\documentclass[reprint]{revtex4-2}

\usepackage[utf8]{inputenc}
\usepackage[british]{babel}

\usepackage{amsfonts}
\usepackage{amssymb}
\usepackage{amsmath}
\usepackage{amsthm}

\usepackage{bbold}
\usepackage{bm}
\usepackage{graphicx}
\usepackage{color}
\usepackage{multirow}
\usepackage{array}
\usepackage{csquotes}
\usepackage{subcaption}
\usepackage{enumitem}

\usepackage{accents}
\usepackage{braket}
\usepackage{mathrsfs}

\usepackage{tensor}

\usepackage{tikz}
\usetikzlibrary{3d, perspective, calc, arrows, arrows.meta}
\usepackage{xcolor}

\usepackage{orcidlink}
\hypersetup{hidelinks}

\usepackage[nolist,nohyperlinks]{acronym}

\graphicspath{ {./figures/} }

\definecolor{oran}{RGB}{222, 131, 11}
\definecolor{oi_blue}{HTML}{0072B2}

\newcommand{\curvi}{\tilde{\imath}}
\newcommand{\hati}{\hat{\imath}}

\newcommand{\curvj}{\tilde{\jmath}}
\newcommand{\hatj}{\hat{\jmath}}

\newcommand{\ddiff}{\mathscr{D}}
\newcommand{\hmu}{\hat{\mu}}
\newcommand{\hnu}{\hat{\nu}}
\newcommand{\hrho}{\hat{\rho}}
\newcommand{\halp}{\hat{\alpha}}
\newcommand{\bmu}{\bar{\mu}}

\newcommand{\ho}{\hat{0}}
\newcommand{\po}{p^{\hat{0}}}
\newcommand{\exppar}[2]{{#1}^{(#2)}}
\newcommand{\hyperind}[1]{\mathbb{#1}}
\newcommand{\lmax}{\ell_\mathrm{max}}
\newcommand{\diag}[1]{\mathrm{diag}\left( #1 \right)}
\renewcommand{\vec}[1]{\mathbf{#1}}
\newcommand{\vm}[1]{\mathbf{#1}}
\newcommand{\symvm}[1]{\boldsymbol{#1}}
\newcommand{\bvm}[1]{\widetilde{\vm{#1}}}

\definecolor{posvec}{HTML}{56B4E9}   % sky blue — position vector (r, R, x, y, z)
\definecolor{momvec}{HTML}{D55E00}   % vermillion — momentum vector p
\definecolor{anglearc}{gray}{0.0}   % neutral gray — construction angle arcs

\makeatletter
\tikzoption{canvas is plane}[]{\@setOxy#1}
\def\@setOxy O(#1,#2,#3)x(#4,#5,#6)y(#7,#8,#9)%
  {\def\tikz@plane@origin{\pgfpointxyz{#1}{#2}{#3}}%
   \def\tikz@plane@x{\pgfpointxyz{#4}{#5}{#6}}%
   \def\tikz@plane@y{\pgfpointxyz{#7}{#8}{#9}}%
   \tikz@canvas@is@plane
  }
\makeatother

\pgfkeys{
    /cone/.is family,
    /cone,
    theta/.store in=\coneTheta,
    phi/.store in=\conePhi,
    ptheta/.store in=\conePtheta,
    pphi/.store in=\conePphi,
    r/.store in=\coneR,
    mag/.store in=\coneMag,
    tilt/.store in=\coneTilt,
    rotation/.store in=\coneRotation,
    viewx/.store in=\coneViewx,
    viewy/.store in=\coneViewy,
    viewz/.store in=\coneViewz,
    phizero/.store in=\conePhizero,
    thanchor/.store in=\coneThAnchor,
    phianchor/.store in=\conePhiAnchor,
    theta=40,
    phi=130,
    ptheta=40,
    pphi=30,
    r=1.8,
    mag=1.5,
    tilt=0,
    rotation=0,
    viewx=1,
    viewy=1,
    viewz=0.6,
    phizero=0,
    thanchor=east,
    phianchor=south west,
}

\newcommand{\drawcone}[1][]{
    \pgfkeys{/cone,#1}

    \def \r {\coneR}              % distance of particle
    \def \angth {\coneTheta}      % angle for position of the particle
    \def \angphi {\conePhi}       % angle for position of the particle
    
	\def \rx { (sin(\angth) * cos(\angphi)) }
	\def \ry { (sin(\angth) * sin(\angphi)) }
	\def \rz { cos(\angth) }
	
	\def \Px { ( \r * \rx ) }
	\def \Py { ( \r * \ry ) }
	\def \Pz { ( \r * \rz ) }
    
    \def \p {\coneMag}              % magnitude of momentum
    \def \angvarth {\conePtheta}  % opening angle of the cone (no tilt)
    \def \angvarphi {\conePphi}   % position on the cone (no tilt)
    
    \def \angalpha {\coneTilt}    % tilt angle alpha between cone axis and r
    \def \angaxphi {\coneRotation} % azimuth (in perp/perpa plane) fixing tilt direction
    
    \def \varphirel { (\angvarphi - \angaxphi) }
    \def \coneth { acos( cos(\angalpha)*cos(\angvarth) + sin(\angalpha)*sin(\angvarth)*cos(\varphirel) ) }
    \def \conephi { atan2(sin(\angvarth)*sin(\varphirel) , -sin(\angalpha)*cos(\angvarth) + cos(\angalpha)*sin(\angvarth)*cos(\varphirel)) }
    
    \def \perpx { sin(\angphi) }
    \def \perpy { - cos(\angphi) }
    \def \perpz { 0 }

    \def \perpax { cos(\angth) * cos(\angphi) }
    \def \perpay { cos(\angth) * sin(\angphi) }
    \def \perpaz { -sin(\angth) }
	
    \def \wx { (cos(\angaxphi)*\perpx + sin(\angaxphi)*\perpax) }
    \def \wy { (cos(\angaxphi)*\perpy + sin(\angaxphi)*\perpay) }
    \def \wz { (cos(\angaxphi)*\perpz + sin(\angaxphi)*\perpaz) }

    \def \axisx { (cos(\angalpha)*\rx + sin(\angalpha)*\wx) }
    \def \axisy { (cos(\angalpha)*\ry + sin(\angalpha)*\wy) }
    \def \axisz { (cos(\angalpha)*\rz + sin(\angalpha)*\wz) }

    \def \eonex { (-sin(\angalpha)*\rx + cos(\angalpha)*\wx) }
    \def \eoney { (-sin(\angalpha)*\ry + cos(\angalpha)*\wy) }
    \def \eonez { (-sin(\angalpha)*\rz + cos(\angalpha)*\wz) }

    \def \etwox { (-sin(\angaxphi)*\perpx + cos(\angaxphi)*\perpax) }
    \def \etwoy { (-sin(\angaxphi)*\perpy + cos(\angaxphi)*\perpay) }
    \def \etwoz { (-sin(\angaxphi)*\perpz + cos(\angaxphi)*\perpaz) }
	
    \def \upx##1 { cos(\coneth)*\axisx + sin(\coneth)*( cos(##1)*\eonex + sin(##1)*\etwox ) }
    \def \upy##1 { cos(\coneth)*\axisy + sin(\coneth)*( cos(##1)*\eoney + sin(##1)*\etwoy ) }
    \def \upz##1 { cos(\coneth)*\axisz + sin(\coneth)*( cos(##1)*\eonez + sin(##1)*\etwoz ) }
    
    \def \dx {\coneViewx}
    \def \dy {\coneViewy}
    \def \dz {\coneViewz}

    \def \du     { (\dx*\axisx + \dy*\axisy + \dz*\axisz) }
    \def \dperp  { (\dx*\eonex + \dy*\eoney + \dz*\eonez) }
    \def \dperpa { (\dx*\etwox + \dy*\etwoy + \dz*\etwoz) }
    \def \dRmag  { sqrt( (\dperp)^2 + (\dperpa)^2 ) }
    \def \phizero{ atan2(\dperpa,\dperp) }
    \def \deltaang{ acos( ( tan(\coneth) * \du ) / \dRmag ) }
    \def \angt   { (\phizero + \deltaang) } % one silhouette generator
    \def \angtt  { (\phizero - \deltaang) } % other silhouette generator
    \def \rcone  { (\p * cos(\coneth) ) } % axial height of the cone
    
    \def \Pcx { ( \Px + \rcone * \axisx ) }
    \def \Pcy { ( \Py + \rcone * \axisy ) }
    \def \Pcz { ( \Pz + \rcone * \axisz ) }
	
	\def \projfac {0.8}  % fraction of radius of projection of P onto xy plane
	\def \projfacz {0.5} % fraction of radius of projection of P onto z axis
	
	\coordinate (P) at ({\Px},{\Py},{\Pz});
	\coordinate (Pc) at ({\Pcx},{\Pcy},{\Pcz});
	\coordinate (p) at ({\Px + \p * ( \upx{\conephi} ) },{\Py + \p * ( \upy{\conephi} ) },{\Pz + \p * ( \upz{\conephi} ) });
	
	\draw[->, >=stealth, very thin] (0,0,0) -- (1.6,0,0);
	\draw[->, >=stealth, very thin] (0,0,0) -- (0,1.6,0);
	\draw[->, >=stealth, very thin] (0,0,0) -- (0,0,1.6);
	
	\fill (P) circle (0.04cm);

    \fill[gray, fill opacity=0.5, draw=none, samples=60, smooth,
          variable=\t, domain=\angtt:\angt]
        (P) -- plot ({\Px + \p * ( \upx{\t} ) },
                     {\Py + \p * ( \upy{\t} ) },
                     {\Pz + \p * ( \upz{\t} ) }) -- cycle;

    \draw[dashed, ultra thin] (P) -- (Pc);
    
	\draw[thin,draw=none,fill=gray, fill opacity=0.25,samples=100,smooth,variable=\t, domain=0:360]
        plot ({\Px + \p * ( \upx{\t} ) },
            { \Py + \p * ( \upy{\t} ) },
        { \Pz + \p * ( \upz{\t} ) });
    
    \draw[thick, ->, >=stealth, momvec] (P) -- (p) node[pos=1,anchor=west]{$\mathbf{p}$};
}

\newcommand{\drawconespherical}[1][]{

    \pgfkeys{/cone,#1}
    \drawcone[#1]
    
    \draw[->, >=stealth, thick, posvec] (0,0,0) -- (P) node[pos=0.5, anchor=north west] {$\mathbf{r}$};
	\draw[ultra thin] (0,0,0) -- ({\Px},{\Py}, 0);
	\draw[ultra thin] ({\Px},{\Py}, 0) -- (P);
    
    \draw[canvas is xy plane at z=0, posvec] ({\projfac * sqrt(\Px^2 + \Py^2)},0) arc (0:{\angphi}:{\projfac * sqrt(\Px^2 + \Py^2)}) node[pos=0.5, anchor=north] {$\boldsymbol{\phi}$};
    \begin{scope}[canvas is plane={O(0,0,0)x({\Px/sqrt(\Px^2+\Py^2)},{\Py/sqrt(\Px^2+\Py^2)},0)y(0,0,1)}]
        \draw[posvec] (0,{\projfacz * \Pz}) arc (90:{90-\angth}:{\projfacz * \Pz}) node[pos=0.5, anchor=south] {$\boldsymbol{\theta}$};
    \end{scope}
    
    \draw[ultra thin, dashed] (Pc) -- ({\Px + \p * ( \upx{0} ) },{\Py + \p * ( \upy{0} ) },{\Pz + \p * ( \upz{0} ) });
    \draw[ultra thin, dashed] (Pc) -- (p);
    
    \begin{scope}[canvas is plane={O({\Pcx},{\Pcy},{\Pcz})x({\Pcx+\eonex},{\Pcy+\eoney},{\Pcz+\eonez})y({\Pcx+\etwox},{\Pcy+\etwoy},{\Pcz+\etwoz})}]
        \draw[momvec] ({0.2*cos(\conePhizero)},{0.2*sin(\conePhizero)}) arc ({\conePhizero}:{\conephi}:{0.2}) node[pos=0.5, anchor=south west] {$\boldsymbol{\varphi}$};
    \end{scope}
    \begin{scope}[canvas is plane={O({\Px},{\Py},{\Pz})x({\Px + \upx{\conephi} },{\Py + \upy{\conephi} },{\Pz + \upz{\conephi} })y({\Px+\axisx},{\Py+\axisy},{\Pz+\axisz})}]
        \draw[momvec] ({0.},{0.5}) arc (90:0:{0.5}) node[pos=0.0, anchor=east] {$\boldsymbol{\vartheta}$};
    \end{scope}
}

\newcommand{\drawconecylindrical}[1][]{

    \pgfkeys{/cone,#1}
    \drawcone[#1]
    
	\draw[thick, ->, >=stealth, posvec] (0,0,0) -- ({\Px},{\Py}, 0) node[pos=0.5,anchor=south east]{$\mathbf{R}$};
	\draw[thick, ->, >=stealth, posvec] ({\Px},{\Py}, 0) -- (P) node[pos=0.5,anchor=west]{$\mathbf{z}$};
	\draw[canvas is xy plane at z=0, posvec] ({\projfac * sqrt(\Px^2 + \Py^2)},0) arc (0:{\angphi}:{\projfac * sqrt(\Px^2 + \Py^2)}) node[pos=0.5, anchor=north] {$\boldsymbol{\phi}$};
    
    \draw[ultra thin, dashed] (Pc) -- ({\Px + \p * ( \upx{\conePhizero} ) },{\Py + \p * ( \upy{\conePhizero} ) },{\Pz + \p * ( \upz{\conePhizero} ) });
    \draw[ultra thin, dashed] (Pc) -- (p);
    
    \begin{scope}[canvas is plane={O({\Pcx},{\Pcy},{\Pcz})x({\Pcx+\eonex},{\Pcy+\eoney},{\Pcz+\eonez})y({\Pcx+\etwox},{\Pcy+\etwoy},{\Pcz+\etwoz})}]
        \draw[momvec] ({0.5*cos(\conePhizero)},{0.5*sin(\conePhizero)}) arc ({\conePhizero}:{\conephi}:{0.5}) node[pos=0.5, anchor=\conePhiAnchor] {$\boldsymbol{\varphi}$};
    \end{scope}
    \begin{scope}[canvas is plane={O({\Px},{\Py},{\Pz})x({\Px + \upx{\conephi} },{\Py + \upy{\conephi} },{\Pz + \upz{\conephi} })y({\Px+\axisx},{\Py+\axisy},{\Pz+\axisz})}]
        \draw[momvec] ({0.},{0.5}) arc (90:0:{0.5}) node[pos=0.0, anchor=\coneThAnchor] {$\boldsymbol{\vartheta}$};
    \end{scope}
}

\newcommand{\drawconecartesian}[1][]{

    \pgfkeys{/cone,#1}
    \drawcone[#1]
    
	\draw[thick, ->, >=stealth, posvec] (0,0,0) -- ({\Px},0, 0) node[pos=0.5,anchor=south]{$\mathbf{x}$};
	\draw[thick, ->, >=stealth, posvec] ({\Px},0, 0) -- ({\Px},{\Py}, 0) node[pos=0.5,anchor=south]{$\mathbf{y}$};
	\draw[thick, ->, >=stealth, posvec] ({\Px},{\Py}, 0) -- (P) node[pos=0.5,anchor=west]{$\mathbf{z}$};
    
    \draw[ultra thin, dashed] (Pc) -- ({\Px + \p * ( \upx{\conePhizero} ) },{\Py + \p * ( \upy{\conePhizero} ) },{\Pz + \p * ( \upz{\conePhizero} ) });
    \draw[ultra thin, dashed] (Pc) -- (p);
    
    \begin{scope}[canvas is plane={O({\Pcx},{\Pcy},{\Pcz})x({\Pcx+\eonex},{\Pcy+\eoney},{\Pcz+\eonez})y({\Pcx+\etwox},{\Pcy+\etwoy},{\Pcz+\etwoz})}]
        \draw[momvec] ({0.8*cos(\conePhizero)},{0.8*sin(\conePhizero)}) arc ({\conePhizero}:{\conephi}:{0.8}) node[pos=0., anchor=east,yshift=0.1cm] {$\boldsymbol{\varphi}$};
    \end{scope}
    \begin{scope}[canvas is plane={O({\Px},{\Py},{\Pz})x({\Px + \upx{\conephi} },{\Py + \upy{\conephi} },{\Pz + \upz{\conephi} })y({\Px+\axisx},{\Py+\axisy},{\Pz+\axisz})}]
        \draw[momvec] ({0.},{0.5}) arc (90:0:{0.5}) node[pos=0.0, anchor=east] {$\boldsymbol{\vartheta}$};
    \end{scope}
}

\pgfkeys{
    /hollow_arrow/.is family,
    /hollow_arrow,
    height/.store in=\haHeight,
    head width/.store in=\haHwidth,
    head height/.store in=\haHheight,
    fill/.store in=\haFill,
    height=1,
    head width=0.2,
    head height=0.5,
    fill=none
}

\definecolor{fluxout}{HTML}{E69F00}  % orange — outgoing flux
\definecolor{fluxin}{HTML}{56B4E9}   % sky blue — incoming flux
\definecolor{residual}{HTML}{D55E00} % vermillion — nonzero residual / imbalance

\makeatletter
\tikzoption{canvas is plane}[]{\@setOxy#1}
\def\@setOxy O(#1,#2,#3)x(#4,#5,#6)y(#7,#8,#9)%
  {\def\tikz@plane@origin{\pgfpointxyz{#1}{#2}{#3}}%
   \def\tikz@plane@x{\pgfpointxyz{#4}{#5}{#6}}%
   \def\tikz@plane@y{\pgfpointxyz{#7}{#8}{#9}}%
   \tikz@canvas@is@plane
  }
\makeatother

\def\parsecoordstart(#1,#2){%
  \def\startx{#1}%
  \def\starty{#2}%
}

\def\parsecoordend(#1,#2){%
  \def\endx{#1}%
  \def\endy{#2}%
}

\newcommand{\drawarrow}[3][]{
    \pgfkeys{/hollow_arrow,#1}
    
    \parsecoordstart#2
    \parsecoordend#3

    \pgfmathsetmacro{\h}{\haHeight}
    \pgfmathsetmacro{\w}{sqrt((\endx-\startx)^2 + (\endy-\starty)^2)}
    \pgfmathsetmacro{\headw}{\haHwidth}
    \pgfmathsetmacro{\headh}{\haHheight}
    \pgfmathsetmacro{\ang}{atan2((\endy-\starty),(\endx-\startx))}
    
    \pgfmathsetmacro{\reculx}{\startx - \h/2*sin(\ang)} % x-coord of upper left corner of the rectangle
    \pgfmathsetmacro{\recllx}{\startx + \h/2*sin(\ang)} % x-coord of lower left corner of the rectangle
    \pgfmathsetmacro{\reculy}{\starty + \h/2*cos(\ang)} % y-coord of upper left corner of the rectangle
    \pgfmathsetmacro{\reclly}{\starty - \h/2*cos(\ang)} % y-coord of lower right corner of the rectangle
    \pgfmathsetmacro{\recurx}{\reculx + \w * (1-\headw)* cos(\ang)} % x-coord of upper right corner of the rectangle
    \pgfmathsetmacro{\reclrx}{\recurx + \h * sin(\ang)}             % x-coord of lower right corner of the rectangle
    \pgfmathsetmacro{\recury}{\reculy + \w * (1-\headw)* sin(\ang)} % y-coord of upper right corner of the rectangle
    \pgfmathsetmacro{\reclry}{\recury - \h * cos(\ang)}             % y-coord of lower right corner of the rectangle
    
    \pgfmathsetmacro{\headux}{\recurx - \h * \headh/2 * sin(\ang)}  % x-coord of upper tip of the head
    \pgfmathsetmacro{\headdx}{\reclrx + \h * \headh/2 * sin(\ang)}  % x-coord of lower tip of the head
    \pgfmathsetmacro{\headuy}{\recury + \h * \headh/2 * cos(\ang)}  % y-coord of upper tip of the head
    \pgfmathsetmacro{\headdy}{\reclry - \h * \headh/2 * cos(\ang)}  % y-coord of lower tip of the head
    \draw[thick,black,fill=\haFill] ({\recllx},{\reclly}) -- ({\reculx},{\reculy}) -- ({\recurx},{\recury}) -- ({\headux},{\headuy}) -- ({\endx},{\endy}) -- ({\headdx},{\headdy}) -- ({\reclrx},{\reclry}) -- cycle;    
    
}

\newcommand{\drawbkggrid}{\fill[fill=gray!12] (0,0) rectangle (6,6);}
\newcommand{\drawaxes}{
    \draw[thick, ->, >=stealth] (6.5,0) -- (6.5,4) node[pos=0.5,anchor=north]{\large $r$};
    \draw[thick, ->, >=stealth] (6,-0.3) -- (2,-0.3) node[pos=0.5,anchor=east]{\large $\vartheta$};
}

\newcommand{\drawarrows}[4]{
    \draw[step=2cm] (0,0) grid (6,6);
    \drawarrow[height=#1, head height=0.7, head width=0.3, fill=fluxout]{(3,2.2)}{(3,1.2)}
    \drawarrow[height=#2, head height=0.7, head width=0.3, fill=fluxin]{(3,4.9)}{(3,3.5)}
    \drawarrow[height=#3, head height=0.7, head width=0.3, fill=fluxin]{(1.5,3)}{(2.5,3)}
    \drawarrow[height=#4, head height=0.8, head width=0.35, fill=fluxout]{(3.6,3)}{(5.3,3)}
    \node[rotate={\roth}] at (3,4.3) {\footnotesize $\frac{F^r\Delta A_r}{\Delta V}$};
    \node[rotate={\rotv}] at (4.4,3) {\footnotesize $\frac{\mathcal{F}^\vartheta\Delta A_\vartheta}{\Delta V}$};
}

\newcommand{\drawinconsistentconservative}[2]{
        \def \angth {#2}
        \def \angphi {#1}
        \def \roth {90-\angphi}
        \def \rotv {-90 + 90-\angphi}
		\begin{scope}[3d view={{\angphi}}{{\angth}}]
            \begin{scope}[canvas is xy plane at z=0]
                \drawbkggrid
                \fill[fill=residual] (2,2) rectangle (4,4);
                \drawarrows{0.4}{0.75}{0.3}{0.55}
                \node at (3,3){$-\mathcal{R}$};
                
                \drawaxes
            \end{scope}
		\end{scope}
}

\newcommand{\drawinconsistentnonconservative}[2]{
        \def \angth {#2}
        \def \angphi {#1}
        \def \roth {90-\angphi}
        \def \rotv {-90 + 90-\angphi}
		\begin{scope}[3d view={{\angphi}}{{\angth}}]
            \begin{scope}[canvas is xy plane at z=0]
                \drawbkggrid
                \drawarrows{0.4}{0.75}{0.3}{0.55}
                \node at (3,3){$+0$};
                
                \drawaxes
            \end{scope}
            
            \begin{scope}[canvas is plane={O(6,0,0)x({6-1},{1},0)y(0,0,0.5)}]
                \drawarrow[height=0.35, head height=0.8, head width=0.4, fill=residual]{(2.5,0.6)}{(2.5,0.15)}
                \node[rotate=-90] at (2.5,0.4) {$+\mathcal{R}$};
                \node at (2.9,0.55){\large $S'$};
            \end{scope}
            
		\end{scope}
}

\newcommand{\drawconsistentconservative}[2]{
        \def \angth {#2}
        \def \angphi {#1}
        \def \roth {90-\angphi}
        \def \rotv {-90 + 90-\angphi}
		\begin{scope}[3d view={{\angphi}}{{\angth}}]
            \begin{scope}[canvas is xy plane at z=0]
                \drawbkggrid
                \drawarrows{0.5}{0.7}{0.4}{0.65}
                \node at (3,3){$+0$};
                
                \drawaxes
            \end{scope}
            
		\end{scope}
}

\begin{document}

\title{A Multidimensional General-Relativistic Boltzmann Solver for Neutrino Transport: Implementation, Discretization and Optimization}
\author{Arthur Offermans\orcidlink{0000-0002-8313-5976}}
\email{arthur.offermans@kuleuven.be}
\affiliation{Department of Physics and Astronomy, KU Leuven, B-3001 Leuven, Belgium}
\affiliation{Leuven Gravity Institute, KU Leuven, Celestijnenlaan 200D box 2415, 3001 Leuven, Belgium}

\author{Harry Ho-Yin Ng\orcidlink{0000-0003-3453-7394}}
\affiliation{TAPIR, Mailcode 350-17, California Institute of Technology, Pasadena, CA 91125, USA}
%\email{}

\author{Patrick Chi-Kit Cheong\orcidlink{0000-0003-1449-3363}}
\affiliation{Department of Physics, University of California, Berkeley, Berkeley, CA 94720, USA}
%\email{}

\author{Anthony Mezzacappa\orcidlink{0000-0001-9816-9741}}
\affiliation{Department of Physics and Astronomy, University of Tennessee, Knoxville, Knoxville, TN 37996-1200, USA}
%\email{}

\author{Tjonnie Guang Feng Li\orcidlink{0000-0003-4297-7365}}
\affiliation{Department of Physics and Astronomy, KU Leuven, B-3001 Leuven, Belgium}
\affiliation{Leuven Gravity Institute, KU Leuven, Celestijnenlaan 200D box 2415, 3001 Leuven, Belgium}
\affiliation{KU Leuven, Department of Electrical Engineering (ESAT), STADIUS Center for Dynamical Systems, Signal Processing and Data Analytics, B-3001 Leuven, Belgium }
%\email{}

\begin{abstract}
    We present the implementation of a multidimensional general-relativistic Boltzmann solver for neutrino transport using the finite volume method. We extend the general-relativistic magnetohydrodynamics solver \texttt{Gmunu} to discretize the full $6$D phase space. We discretize the momentum space in spherical coordinates in the comoving frame, and the position space in the lab frame in Cartesian, cylindrical or spherical coordinates. As in the M1 scheme, we expand the interaction kernels up to first order in a Legendre series. We present a discretization scheme that ensures consistency between the number-conservative and non-conservative formulations in empty flat spacetime, multiple dimensions and coordinate systems,  conserves energy to $\sim 1\%$ with $20$ energy bins in 1D tests. To address the prohibitive computational cost of a 6D method, we discuss optimizations of the implicit solver for stiff source terms to minimize its memory requirements and its computational cost. In particular, we introduce a method that leverages the Legendre expansion of the kernels to reduce the dimensionality of the problem. In a 1D core-collapse supernova snapshot, our method yields a speed-up factor of $300$ compared to a full-matrix LU method for $14$ angular bins when including energy- and species-coupling interactions. Finally, we validate our implementation on standard test cases and report quadratic convergence in space, energy and propagation angles. We compare our solver to the M1 scheme in simplified 1D configurations and find differences of about $10\%$ in the free-streaming luminosities in relaxation test cases, which we attribute primarily to the M1 closure relation. The average energies, on the other hand, agree between the two methods. We compare both methods on a core-collapse supernova test case and find an overall good agreement in the fluid variables' profiles at the time of core bounce for $20$ energy bins.
\end{abstract}

\maketitle

\section{Introduction}

\noindent Neutrinos have been shown to play an important role in multiple astrophysical
systems. In \acp{CCSN}, the core of a massive star collapses until it reaches nuclear saturation densities. When the contraction of the core halts, a shock forms and propagates outwards. However, this shock loses energy and stalls before reaching the surface. A proposed scenario to revive the shock and produce the explosion is neutrino reheating~\citep{Wilson_1985,Bethe_1985}. Neutrinos can interact with the fluid close to the shock and deposit enough energy to revive the shock. This scenario has since been extensively studied (for reviews, see Refs.~\cite{Muller_2020,Mezzacappa_2020,Burrows_2020,Yamada_2024,Janka_2025,Mezzacappa_2026,Raffelt_2026}). In binary neutron-star mergers, the interaction of neutrinos with the fluid alters the proportion of neutrons available in the ejecta and therefore affects the heavy-element nucleosynthesis through r-processes and the composition of the ejecta~\citep{Fujibayashi:2017xsz,Radice_2022,Ng_2024b,Foucart_2023}. Such astrophysical systems are also unique laboratories to study neutrino physics in extreme conditions of density and temperature, such as instabilities in neutrino flavour oscillations (see~\citep{Tamborra_2021,Volpe_2024,Johns_2025} for reviews).

To explain and analyse observations of these astrophysical systems, our models must therefore include the effects of neutrinos, whose evolution is classically described by the Boltzmann equation. The Boltzmann equation is a $3+3+1$D --- position, momentum and time --- integro-differential equation that does not have analytical solutions for such astrophysical systems. Numerical integration is therefore required to evolve neutrinos in these systems. Neutrino transport is thus a key component of these models, but also extremely computationally demanding. Indeed, the dimensionality of this problem is larger than that of radiation-free \ac{GRMHD} models. The higher dimensionality of the problem is one of the main factors that make the computational cost of full neutrino transport outweigh that of fluid evolution.

To mitigate the curse of dimensionality, lower-dimensional approximate schemes have been designed. One of the main approximate schemes used in the literature is the truncated moment scheme. The truncated moment formalism~\citep{Thorne_1981} decomposes the distribution function, or intensity, into its angular moments. The expansion is then truncated to the first $N$ moments. Examples of truncated moment schemes are flux-limited diffusion~\citep{Rahman_2019,Bruenn_2020} and M0~\citep{Radice_2016}, where only the zeroth-order moment is evolved, and M1~\citep{Shibata_2011, Cardall_moments_2013,O'Connor_2015,Foucart_2015,Just_2015,Kuroda2016,Skinner:2018iti,Radice_2022,Cheong_2023,Musolino_2024,Schianchi_2024,Endeve_2026,Daszuta:2026szb,Kuroda_2026}, where the first two moments are evolved. For the latter, both energy-dependent (spectral) and energy-integrated (grey) schemes exist. The truncated moment scheme makes two main approximations. First, higher-order moments are discarded, which may smooth out anisotropies in the distribution function. Second, the use of an approximate closure relation (\textit{e.g.},~\cite{Murchikova_2017}) is required to close the system of equations after truncation. The equation for a moment $m$ requires the knowledge of the moments $m+1$ and $m+2$, which are not evolved. The closure relation provides a prescription to compute these higher-order moments.

Several methods have also been developed to solve the full Boltzmann equation.
The discrete-ordinate method ($S_N$) directly discretizes the full 6D phase space and uses the \ac{FD} method to solve the radiation transport equation~\citep{Mezzacappa_1993,Mezzacappa_1999,Yamada_1999,Livne_2004,Liebendorfer_2004,Ott_2008,Sumiyoshi_2012,Nagakura_2014,Nagakura_2017,Chan_2020,Akaho_2021}.
The Filtered Spherical Harmonics method ($FP_N$; \cite{McClarren_2010, Radice_2013}) is a spectral method that expands the angular part of the distribution function in momentum space into spherical harmonics up to $l=N$. Filters are then applied to the expansion to limit the effect of oscillations in the solution, and limiters were designed to ensure positivity of the distribution function~\citep{Laiu_2018}. An advantage of this method is that it does not suffer from ray effects. Another spectral method by Ref.~\cite{Peres_2014} decomposes the distribution function into basis functions over the whole phase space.
The Finite Element Method ($FEM_N$; \cite{Bhattacharyya_2023}) uses finite elements in the angular part of the momentum space.
\ac{MC} methods (\textit{e.g.}~\cite{Richers_2015,Miller_2019,Foucart_2020,Kawaguchi_2023}) represent the distribution function as a sum of packets of particles, each corresponding to a given number of particles. This method is difficult to apply in optically thick regimes, as the characteristic time of interaction becomes very small. To overcome the difficulties of \ac{MC} in optically thick regimes and the limitation of the M1 closure, the guided-moment scheme was introduced by Ref.~\cite{Izquierdo_2024}. This technique interpolates moments between the \ac{MC} method, efficient in the optically thin regime, and the M1 method, accurate in the optically thick regime. Finally, the lattice Boltzmann method has also been applied to the relativistic Boltzmann equation~\cite{Weih_2020,Olsen_2025}.

In the context of general-relativistic multidimensional \ac{CCSN} simulations with neutrino Boltzmann transport, however, only the solver from Ref.~\cite{Akaho_2021} has been used~\citep{Akaho_bgk_2026,Akaho_2026} to our knowledge. This solver discretizes momentum in the orthonormal frame, where advection is simpler, and complements it with another grid for the source terms~\citep{Nagakura_2014}. Another approach consists in discretizing momentum in the fluid rest-frame (or comoving frame) and has been used in 1D solvers~\citep{Mezzacappa_1993,Mezzacappa_1999,Yamada_1999,Liebendorfer_2004}. The comoving frame offers a more natural description because the source terms are expressed in the comoving frame. Therefore, it removes the need for the two separate grids. On the other hand, the advection in momentum space is coupled to the velocity background in the comoving frame, which makes its numerical treatment and discretization more difficult in the presence of strong velocity gradients. In this paper, we introduce a multidimensional Boltzmann solver with the alternative approach based on the comoving frame momentum discretization, and discuss the aforementioned challenges associated with the advection. Moreover, the development of separate approaches enables cross-validation of methods and simulations. 

Comparisons have already been performed between different approximate schemes and full treatments to evaluate the performance of approximate schemes~\citep{Yamada_1999,Liebendorfer_2005,Richers_2017,O'Connor_2018,Foucart_2024}. For \acp{CCSN} in a general-relativistic framework, such comparisons are however limited to 1D and a thorough multidimensional comparison between M1 schemes and full treatments is still lacking. Special relativistic simulations of \acp{CCSN} with the $S_N$ method in 2D and 3D have found that the M1 closure struggles to reconstruct the off-diagonal elements of the Eddington tensor in the semi-transparent regime at early post-bounce times~\citep{Nagakura_2018,Iwakami_2020}. However, these observations are made with a limited momentum-space resolution, which may affect quantitative comparisons. It has indeed been shown that a higher resolution in momentum space is required for convergence~\citep{Akaho_2026}. The computational cost associated with higher resolutions however makes these comparisons difficult to achieve. As a result, the exact effect of the M1 approximation on the evolution of \acp{CCSN} remains uncertain.

The use of approximate formulations of the Boltzmann equation is not the only way the accuracy of neutrino transport may be affected. Indeed, multiple works have highlighted the complexity of discretizing the Boltzmann equation, including in approximate schemes such as the M1 (see Ref.~\cite{Mezzacappa_2020} for a review). Such challenges include enforcing both number and energy conservation (M1:~\cite{Muller_2010,Cardall_moments_2013}; Boltzmann:~\cite{Cardall_2013,Liebendorfer_2004}), retrieving the diffusive limit in scattering media (M1:~\cite{Audit_2002}; Boltzmann:~\cite{Mezzacappa_1993}) and maintaining the realizability of $f$, \textit{i.e.} the fermionic distribution function between $0$ and $1$~\citep{Chu_2019,Laiu_2025,Hunter_2025}. Finding a discretization that conserves both number and energy has not yet been achieved for the full Boltzmann equation in a multidimensional and general-relativistic framework.

Our discussion has highlighted three main challenges or limitations that our work aims to address: the uncertainty related to the effect of approximations in the neutrino treatment from approximate schemes, the discretization of the equation and the prohibitive cost of a 6D Boltzmann solver. Thus, we extend the code \texttt{Gmunu}~\citep{Cheong_2020,Cheong_2021,Cheong_2022,Cheong_2023,Ng_2024} with a multidimensional general-relativistic Boltzmann solver to reduce approximations to the Boltzmann equation and study the role of neutrinos in astrophysical systems with less uncertainty. The implementation of this Boltzmann solver also enables comparisons between different methods and the assessment of their validity. Our Boltzmann solver applies the \ac{FV} method, already used for the \ac{GRMHD} equations, on a fully discretized phase space. In addition, we present a discretization scheme that ensures consistency between the conservative and non-conservative formulations in flat spacetime. We also address the cost of the solver and introduce a method that leverages approximations to the source terms to reduce the time and memory complexity of the implicit evolution of stiff source terms. The approximation of the source terms consists in expanding the interaction kernels up to first order in a Legendre series, as in the M1 scheme.

The paper is organized as follows. First, we introduce the general formalism on which our solver is based in Sec.~\ref{sec:GRBE}. Then, we focus on the left-hand side of the equation, advection. We describe the implementation of the advection term and generalizations needed to discretize a 6D phase space, as opposed to a 3D space, in Sec.~\ref{sec:phase_space}. In Sec.~\ref{sec:importance_discretization}, we detail our discretization scheme for the momentum-space flux and study the conservation properties of our solver. We also briefly discuss realizability, maintaining the fermionic distribution function between $0$ and $1$. After that, we discuss the right-hand side of the equation, interactions. We introduce the expression of the source terms, discuss the diffusive limit and describe the framework of our implicit solver in Sec.~\ref{sec:interaction_and_implicit}. We discuss the implicit solver further in Sec.~\ref{sec:optimization_implicit}, where we introduce an optimization to reduce the memory requirements and computational cost of the implicit solver. Finally, we test our implementation on standard test cases in Sec.~\ref{sec:tests} and compare our Boltzmann transport scheme to \texttt{Gmunu}'s M1 transport scheme in Sec.~\ref{sec:comparison_M1} for simple configurations. To close our discussion, we summarize our work and discuss future improvements in Sec.~\ref{sec:conclusion}.

Throughout the paper, we use the metric signature $-+++$ and work in units where the speed of light $c$, the gravitational constant $G$, the solar mass $M_{\odot}$ and the Boltzmann constant are equal to $1$, \textit{i.e.} $c=G=M_{\odot}=k_B=1$. These units correspond to the code units used in \texttt{Gmunu}. Greek indices run from $0$ to $3$ and Latin indices run from $1$ to $3$. Double-struck Latin indices like $\hyperind{i}$ range from $1$ to $6$ and represent the spatial components of a $7$D space. We use unadorned indices to denote components measured in the lab frame, hatted indices in the orthonormal comoving frame, indices adorned with a bar in the local orthonormal frame, and indices adorned with a tilde in a curvilinear momentum space. Unless stated otherwise, we will use the terms ``comoving frame'' and ``orthonormal frame'' to refer to the reference frame used for the momentum space only.

\section{General-relativistic Boltzmann equation\label{sec:GRBE}}

\noindent The evolution of neutrinos is described by the Boltzmann equation. Under the strong-gravity conditions encountered in astrophysical systems, a relativistic description is required. The relativistic Boltzmann equation states that the change in $f$, the distribution function of a given particle, along a geodesic is determined by the invariant collision integral $C[f]$. The latter term accounts for the interaction between the radiation and the fluid, such as absorption, emission and scattering. The relativistic Boltzmann equation reads~\citep{Lindquist_1966,Ehlers_1971,Israel_1972}
\begin{equation}
    \frac{df}{d\xi} = C[f], \label{eq:boltzmann_equation}
\end{equation}
where $\xi$ is the affine parameter describing the trajectory of the particle in phase space. This equation thus requires solving a $6+1$D equation, which includes a $3$D position space and a $3$D momentum space.

In this section, we introduce the formalism of the Boltzmann equation used for our solver. We also introduce some useful quantities related to the equation, such as characteristic speeds, and to the distribution function, such as its moments.

\subsection{Coordinates and reference frames \label{sec:coordinates}}

\noindent The formulation of the Boltzmann equation presented in Eq.~\eqref{eq:boltzmann_equation} is not convenient to solve.
Thus, we introduce phase space coordinates $(x^\mu, p^i)$, where $x^\mu$ are the coordinates in position space and $p^i$ in momentum space, both in the coordinate basis. We choose to solve the equation in a coordinate system where the momentum is taken in the fluid rest frame, the comoving frame. The interaction terms between the fluid and the radiation are easier to compute in the comoving frame, but the advection part becomes fluid-dependent and must thus be treated more carefully. This reference frame was also chosen for the 1D Lagrangian code \texttt{AGILE-BOLTZTRAN}~\citep{Liebendorfer_2004}. Another approach, taken in Refs.~\cite{Nagakura_2014} and \cite{Akaho_2021}, consists in using the local orthonormal frame. In that case, the advection and collision terms must be computed in separate reference frames. The two different choices of reference frame will be discussed in more detail in Sec.~\ref{sec:ref_frame}. It is also desirable to solve the equation with the momentum in curvilinear coordinates $p^{\curvi}$, typically spherical polar coordinates. We follow the formalism and notations of Ref.~\cite{Cardall_2013}.

Therefore, we solve the Boltzmann equation in coordinates $(x^\mu, p^{\curvi})$. In spherical polar coordinates, the momentum coordinates are $(\varepsilon, \vartheta, \varphi)$, where $\varepsilon$ is the magnitude of the particle's momentum (or its energy for a massless particle) and $\vartheta$ and $\varphi$ are the propagation angles. The transformation from the lab frame coordinate basis to the orthonormal comoving frame coordinate basis for the momentum space components is performed using the tetrad ${L^\mu}_{\hmu}$ such that
\begin{align}
    &{L^\mu}_{\hmu} \, {L^\nu}_{\hnu} \, g_{\mu\nu} = \eta_{\hmu\hnu} \label{eq:tetrad_eta}\\
    &{L^\mu}_{\hmu} \, {L^\nu}_{\hnu} \, \eta^{\hmu\hnu} = g^{\mu\nu} \label{eq:tetrad_g}
\end{align}
where $g_{\mu\nu}$ is the metric in the lab frame coordinate basis and $\eta_{\hmu\hnu}$ is the Minkowski metric in the orthonormal comoving frame coordinate basis. We can thus express any tensor $A^{\mu_1 \dots \mu_N}_{\nu_1 \dots \nu_M}$ in the comoving frame as 
\begin{equation}
    A^{\hmu_1 \dots \hmu_N}_{\hnu_1 \dots \hnu_M} = {L^{\hmu_1}}_{\mu_1} \cdots {L^{\hmu_N}}_{\mu_N} {L^{\nu_1}}_{\hnu_1} \cdots {L^{\nu_M}}_{\hnu_M} \, A^{\mu_1 \dots \mu_N}_{\nu_1 \dots \nu_M}.
\end{equation}
Christoffel symbols are however not tensors and transform according to
\begin{equation}
    \begin{aligned}
        {\Gamma^{\hrho}}_{\hnu\hmu} &= {L^{\hrho}}_{\nu}  \, {L^{\mu}}_{\hmu} \, \nabla_\mu{L^{\nu}}_{\hnu} \\
        &= {L^{\hrho}}_{\rho} \, {L^{\nu}}_{\hnu} \, {L^{\mu}}_{\hmu} {\Gamma^{\rho}}_{\nu\mu} + {L^{\hrho}}_{\nu}  \, {L^{\mu}}_{\hmu} \, \partial_\mu{L^{\nu}}_{\hnu}.
    \end{aligned}\label{eq:com_Gamma}
\end{equation}

A similar transformation is used in momentum space to introduce curvilinear coordinates. Following Ref.~\cite{Cardall_2013}, we introduce 
\begin{equation}
    {P^{\curvi}}_{\hati} = \frac{\partial p^{\curvi}}{\partial p^{\hati}}
\end{equation}
that satisfies the relation
\begin{equation}
    {P^{\curvi}}_{\hati} \, {P^{\curvj}}_{\hatj} \, \lambda_{\curvi \curvj} = \delta_{\hati \hatj} \, ,
\end{equation}
where $\lambda_{\curvi \curvj}$ is the metric in momentum space defining the momentum space interval $d\Phi^2 = \lambda_{\curvi \curvj} dp^{\curvi} dp^{\curvj}$. We also introduce an invariant volume element in momentum space
\begin{equation}
    dV_p = \frac{\sqrt{\lambda}}{h^3 \, \po} dp^{\tilde{1}}dp^{\tilde{2}}dp^{\tilde{3}} \label{eq:dV_p}
\end{equation}
where $\lambda$ is the determinant of the momentum-space metric and $\po$ is constrained by the mass shell relation $p_{\hmu} p^{\hmu} = -m^2$, with $m$ the mass of a given neutrino species. We constrain the solution to the positive energy $\po >0$. The particle's energy in spherical coordinates thus reads
\begin{equation}
    p^{\hat{0}} = \sqrt{\varepsilon^2 + m^2}. \label{eq:p_hat0}
\end{equation}

Finally, we introduce the angular moments of the distribution function $f$ in these coordinates for massless particles. The first and second moments represent the spectral number flux $\mathcal{N}^{\mu}$ and the spectral stress-energy tensor $\mathcal{T}_{\mathrm{rad}}^{\mu\nu}$ of the radiation, respectively. They are defined similarly to Ref.~\cite{Cardall_moments_2013} as
\begin{align}
    & \mathcal{N}^{\mu}(x^\alpha,\varepsilon) = \frac{1}{4\pi h^3 \varepsilon}\int d\Omega_p~f(x^\alpha,\varepsilon,\vartheta,\varphi)~p^{\mu},\\
    & \mathcal{T}_{\mathrm{rad}}^{\mu\nu}(x^\alpha,\varepsilon) = \frac{1}{4\pi h^3 \varepsilon} \int d\Omega_p~f(x^\alpha,\varepsilon,\vartheta,\varphi)~p^{\mu}p^{\nu},
\end{align}
where $d\Omega_p = \sin\vartheta \, d\vartheta d\varphi$ is the solid angle in momentum space and the integration is performed on the unit sphere. Note that the moments depend on the momentum/energy $\varepsilon$, despite being themselves expressed in the lab frame coordinate basis. We can nonetheless express them straightforwardly in the comoving frame. Since ${L^\mu}_{\hmu}$ does not depend on momentum, the moments in the comoving frame, \textit{i.e.} $\mathcal{N}^{\hmu}$ and $\mathcal{T}_{\mathrm{rad}}^{\hmu\hnu}$, are obtained by replacing $p^{\mu}$ by $p^{\hmu}$ in the last two equations. The particle number four-flux $N^{\mu}$ and stress-energy tensor $T_{\mathrm{rad}}^{\mu\nu}$ are the energy-integrated spectral moments, \textit{i.e.}
\begin{align}
    & N^{\mu}(x^\alpha) = \int d\varepsilon~\varepsilon^2 \mathcal{N}^{\mu}(x^\alpha,\varepsilon), \label{eq:Nmu_rad}\\
    & T_{\mathrm{rad}}^{\mu\nu}(x^\alpha) = \int d\varepsilon~\varepsilon^2 \mathcal{T}_{\mathrm{rad}}^{\mu\nu}(x^\alpha,\varepsilon). \label{eq:Tmunu_rad} 
\end{align}
These moments are however not the ones that are evolved in the truncated moment scheme, because of redundancy with the higher-order moments~\cite{Thorne_1981}. Instead, $p^\mu$ is decomposed into $p^\mu = \varepsilon \left(u^\mu + l^\mu \right)$, where $l^\mu$ is a four-vector orthogonal to the fluid four-velocity $u^\mu$, \textit{i.e.} $u^\mu l_\mu = 0$. We then introduce the following angular moments
\begin{align}
    & \mathcal{J}(x^\alpha,\varepsilon) \equiv \frac{\varepsilon}{4\pi h^3}\int d\Omega_p~f(x^\alpha,\varepsilon,\vartheta,\varphi) \\
    & \mathcal{H}^\mu(x^\alpha,\varepsilon) \equiv \frac{\varepsilon}{4\pi h^3}\int d\Omega_p~f(x^\alpha,\varepsilon,\vartheta,\varphi) l^\mu\\
    & \mathcal{K}^{\mu\nu}(x^\alpha,\varepsilon) \equiv \frac{\varepsilon}{4\pi h^3}\int d\Omega_p~f(x^\alpha,\varepsilon,\vartheta,\varphi) l^\mu l^\nu \\
    & \mathcal{L}^{\mu\nu\rho}(x^\alpha,\varepsilon) \equiv \frac{\varepsilon}{4\pi h^3}\int d\Omega_p~f(x^\alpha,\varepsilon,\vartheta,\varphi) l^\mu l^\nu l^\rho
\end{align}
The moment $\mathcal{J}$ is interpreted as the spectral energy density, $\mathcal{H}^i$ as the spectral momentum density and $\mathcal{K}^{ij}$ as the spectral stress, each measured in the fluid comoving frame. These moments are more conveniently expressed in the comoving frame, where $l^{\hmu}$ is spacelike and can be written~\citep{Cardall_moments_2013}
\begin{equation}
    l^{\hmu} = (0, \cos\vartheta, \sin\vartheta\cos\varphi,\sin\vartheta\sin\varphi)^T.
\end{equation}
The general expression of $l^\mu$ can then be obtained from $l^\mu = \tensor{L}{^\mu_{\hmu}}l^{\hmu}$.

\subsection{Conservative formulation of the Boltzmann equation\label{sec:cons_BE}}

\noindent We continue with the formalism developed in Ref.~\cite{Cardall_2013}, where the conservative formulation of the Boltzmann equation is derived in the same coordinates and reference frame as those introduced in the previous section. In this formalism, the conservative Boltzmann equation can be written in the form 
\begin{equation}
    \nabla_\mu \left( F^\mu[f] \right) + \po \mathcal{D}_{\curvi}\left( \frac{\mathcal{F}^{\curvi}[f]}{\po}\right) = C[f], \label{eq:cons_boltzmann}
\end{equation}
where $f(x^\mu, p^{\curvi})$ is the particle distribution function, $F^0$ is the conserved variable, $F^i$ and $\mathcal{F}^{\curvi}$ are the fluxes in position and momentum space, respectively, $\nabla_{\mu}$ denotes the covariant derivative in spacetime and $\mathcal{D}_{\curvi}$ the covariant derivative in momentum space, and $C$ is the invariant collision integral related to interactions with the fluid. The conserved variable and fluxes are
\begin{align}
    &q \equiv F^0 = \tensor{L}{^0_\hmu} p^{\hmu} f \\
    &F^i = \tensor{L}{^i_\hmu} p^{\hmu} f \\
    &\mathcal{F}^{\curvi} = -\tensor{P}{^\curvi_\hati} \tensor{\Gamma}{^\hati_{\hmu\hnu}} p^{\hmu} p^{\hnu} f
\end{align}
We can also write $C[f] = p^{\mu} \hat{u}_{\mu} B[f]$, where $\hat{u}^\mu$ is a timelike four-vector 
and $B[f]$ is the collision integral.
Following Ref.~\cite{Shibata_2014}, we choose $\hat{u}_{\mu} = {L^{\ho}}_{\mu}$, which is equivalent to $C[f] = \po B[f]$ in the comoving frame. Eq.~\eqref{eq:cons_boltzmann} represents the conservation of the total number of particles. This equation is however not in its final form and must still be expressed in the formalism used in \texttt{Gmunu}.

\subsection{Conservative formulation in \texttt{Gmunu} \label{sec:gmunu_BE}}

\noindent We use the 3+1 formalism \citep{Gourgoulhon_2007, Alcubierre_2008} to rewrite Eq.~\eqref{eq:cons_boltzmann}. The position space metric is then expressed as
\begin{equation}
    ds^2 = -\alpha^2 dt^2 + \gamma_{ij} (dx^i + \beta^i dt) (dx^j + \beta^j dt) \, ,
\end{equation}
with $\alpha$ the lapse function, $\beta^i$ the shift vector and $\gamma_{ij}$ the spatial metric. We also introduce the normal to the spacelike hypersurface $n^\mu = (1, - \beta^i)^T/\alpha$.
In this framework, it is useful to decompose both the fluid four-velocity $u^\mu$ and ${L^{\mu}}_{\hmu}$ into parts tangent and perpendicular to the spacelike hypersurface
\begin{align}
    &u^\mu = W \left( n^\mu + v^\mu \right),\\
    &{L^{\mu}}_{\hmu} = n^\mu \mathcal{L}_{\hmu} + \tensor{\ell}{^\mu_\hmu},
\end{align}
where $W$ is the Lorentz factor and $v^\mu$ is the Eulerian velocity with $n_\mu v^\mu = 0$. Similarly, $\tensor{\ell}{^\mu_\hmu}$ is the tangential component of ${L^{\mu}}_{\hmu}$ with $n_\mu \tensor{\ell}{^\mu_\hmu} = 0 $.
Combining these decompositions with Eq.~\eqref{eq:tetrad_eta}, we can find the components of $\tensor{L}{^\mu_\hmu}$. Since these computations are already shown in Ref.~\cite{Cardall_2013}, we will outline the main results here. We can first use the fact that the observer is at rest relative to the fluid in the comoving frame, such that the fluid four-velocity in that frame is $u^{\hmu} = (1,0,0,0)^T$. We then find that the fluid four-velocity in the coordinate basis is
\begin{equation}
    u^\mu = {L^\mu}_{\hmu} u^{\hmu} = {L^\mu}_{\ho} \, ,\label{eq:fluid_comoving}
\end{equation}
thus setting some components of our tetrad. We can then find $\tensor{\ell}{^\mu_\hmu} = (Wv^\mu, a^\mu, b^\mu, c^\mu)$, where each four-vector is spacelike, and $\mathcal{L}_{\hmu} = (W,A,B,C)^T$  with $A = v_\mu a^\mu$, $B = v_\mu b^\mu$ and $C = v_\mu c^\mu$.
The derivations of each component $a^i$, $b^i$ and $c^i$ can be found in Appendix~\ref{app:comoving_tetrad}. The case where the fluid is at rest reduces to the local orthonormal frame tetrad.

We can now use these decompositions and the $3+1$ formalism to rewrite Eq.~\eqref{eq:cons_boltzmann}. \texttt{Gmunu} uses a conformal decomposition of the spatial metric
\begin{equation}
    \gamma_{ij} = \psi^4 \bar{\gamma}_{ij},
\end{equation}
where $\psi$ is the conformal factor and $\bar{\gamma}_{ij}$ the conformally related metric. Following \texttt{Gmunu}'s framework~\citep{Cheong_2021}, we use the reference-metric formalism~\citep{Montero_2014,Mewes_2020,Baumgarte_2020}. In this formalism, $\bar{\gamma}_{ij}$ is expressed as the sum of a time-independent reference metric $\hat{\gamma}_{ij}$ and deviations $h_{ij}$. We adopt the conformally flat approximation in which the deviations vanish and the conformally related metric is thus equal to the reference metric. 
The reference metric allows us to express the equation in terms of covariant derivatives with respect to the reference metric. As a result, we can use the connection coefficients and volume elements related to the reference metric that are known analytically. 
Using the relation $\sqrt{-g} = \alpha \sqrt{\gamma} = \alpha \psi^6 \sqrt{\bar{\gamma}}$, we first have
\begin{equation}
    \nabla_\mu F^\mu = \frac{1}{\alpha\psi^6\sqrt{\Bar{\gamma}}}\partial_\mu\left(\sqrt{\hat{\gamma}}~\alpha \psi^6 \sqrt{\frac{\bar{\gamma}}{\hat{\gamma}}} F^\mu \right). \label{eq:ref_divergence}
\end{equation}
Accounting for the fact that the reference metric is time-independent, we can rewrite Eq.~\eqref{eq:cons_boltzmann} for a particle $s$ as
\begin{equation}
    \partial_t q_s + \frac{1}{\sqrt{\hat{\gamma}}}\partial_i \left(\sqrt{\hat{\gamma}} F_s^i \right)  + \frac{p_s^{\ho}}{\sqrt{\lambda}} \frac{\partial}{\partial p^{\curvi}} \left(\frac{\sqrt{\lambda}}{p_s^{\ho}} \mathcal{F}_s^{\curvi} \right) = s_{\mathrm{rad},s},\label{eq:cons_bol_ref}
\end{equation}
where the difference between particles comes from the fact that they can have different interactions and different masses. Hence, $p_s^{\hmu}(p_s)_{\hmu} = -m_s^2$ results in different $p_s^{\ho}$ for different particles. Note however that the spatial components of the momentum are the same for all species, \textit{i.e.} $p_s^{\hati} \equiv p^{\hati}$. The conserved quantity $q_s$, the position-space flux $F_s^i$, the momentum-space flux $\mathcal{F}_s^{\curvi}$ and the source term $s_{\mathrm{rad},s}$ are
\begin{align}
    &q_s = \psi^6 \sqrt{\frac{\bar{\gamma}}{\hat{\gamma}}} \, \mathcal{L}_{\hmu} p_s^{\hmu} \, f_s \label{eq:q}\\[1.5ex]
    &F_s^i = \psi^6 \sqrt{\frac{\bar{\gamma}}{\hat{\gamma}}} \, \big(\alpha \, {\ell^{i}}_{\hmu} - \beta^i \mathcal{L}_{\hmu} \big) p_s^{\hmu} \, f_s \label{eq:flux_x} \\[1.5ex]
    &\mathcal{F}_s^{\curvi} = -\psi^6 \sqrt{\frac{\bar{\gamma}}{\hat{\gamma}}} \, \alpha \, {P^{\curvi}}_{\hati} \, {\Gamma^{\hati}}_{\hmu\hnu} \,  p_s^{\hmu} \, p_s^{\hnu} \, f_s \label{eq:flux_p}\\[1.5ex]
    &s_{\mathrm{rad},s} = \psi^6 \sqrt{\frac{\bar{\gamma}}{\hat{\gamma}}} \, \alpha \, C_s[f] = \psi^6 \sqrt{\frac{\bar{\gamma}}{\hat{\gamma}}} \, \alpha \, p_s^{\ho} B_{s} \label{eq:source}
\end{align}

An extended expression for ${\Gamma^{\hati}}_{\hmu\hnu}$ in terms of $ \mathcal{L}_{\hmu}$ and ${\ell^{i}}_{\hmu}$ can be found in Ref.~\cite{Cardall_2013}, where the contributions from gravity and the observer acceleration are made explicit. We however found that computing ${\Gamma^{\hati}}_{\hmu\hnu}$ from Eq.~\eqref{eq:com_Gamma} was in practice more convenient, as we shall explain in Sec.~\ref{sec:discr_tetrad_diff}. Note also that the conservative-to-primitive transformation is straightforward, contrary to the case of magnetohydrodynamics. Once the primitive $v^i$ is recovered, one can indeed compute $\mathcal{L}_{\hmu}$ and recover the primitive $f$ from Eq.~\eqref{eq:q}.  

\subsection{Characteristic speed \label{sec:char_speed}}

\noindent The characteristic speeds of the Boltzmann equation are obtained from Eqs.~\eqref{eq:q}~to~\eqref{eq:flux_p}. From the definition of the characteristic speed, we have
\begin{align}
    &c_{x,s}^i = \frac{\partial F_s^i}{\partial q_s} = \frac{1}{\mathcal{L}_{\hmu} p^{\hmu}_s}\left(\alpha {\ell^i}_{\hmu} - \beta^i \mathcal{L}_{\hmu}\right) p^{\hmu}_s, \\[1.5ex]
    &c_{p,s}^{\curvi} = \frac{\partial \mathcal{F}_s^{\curvi}}{\partial q_s} = - \frac{1}{\mathcal{L}_{\hmu} p^{\hmu}_s}\alpha \, {P^{\curvi}}_{\hati} \, {\Gamma^{\hati}}_{\hmu\hnu} \,  p_s^{\hmu} \, p_s^{\hnu}, \label{eq:cs_p}
\end{align}
where $c_{x,s}^i$ and $c_{p,s}^{\curvi}$ are the characteristic speeds in position and momentum space, respectively, of particle $s$.
The characteristic speed in momentum space does not depend on $f$ at all, nor on any variable that could be discontinuous in the momentum space. As a result, the Harten--Lax--van Leer (HLL) Riemann solver~\citep{Harten_1983} reduces to, for example,
\begin{equation}
    \mathcal{F}^{\varepsilon}_{l+1/2,m,n} = \mathcal{F}^{\varepsilon}_{l,m,n}
\end{equation}
if $c^\varepsilon$ is positive or 
\begin{equation}
    \mathcal{F}^{\varepsilon}_{l+1/2,m,n} = \mathcal{F}^{\varepsilon}_{l+1,m,n}
\end{equation}
if $c^\varepsilon$ is negative, where we omitted the position space indices for simplicity. These relations are also true for the other components of the flux in momentum space.
As for the characteristic speeds in position space, they depend on the velocity (through the comoving frame tetrad) and are hence prone to significant discontinuities in case of shocks.

It is important to note that, although the characteristic speed in position space is bounded by the speed of light, the characteristic speed in the comoving momentum space is not. Larger characteristic speeds may result in smaller time steps, as we shall explain in Sec.~\ref{sec:ref_frame}.

\subsection{Coupling to the full system of equations \label{sec:fluid_coupling_eq}}
\noindent The interaction of particles with the fluid results in an exchange of momentum and energy. In the case of neutrinos, this will also alter the lepton fractions of the fluid. Source terms must therefore be included in the evolution of these three quantities. These terms are computed from the contribution of the radiation to the energy-momentum tensor $T^{\mu \nu} = T_{\mathrm{fluid}}^{\mu \nu}  + T_{\mathrm{rad}}^{\mu \nu}$, with
\begin{equation}
    T_{\mathrm{rad}}^{\mu\nu} = \sum_s T_{\mathrm{rad},s}^{\mu\nu},
\end{equation}
where $T_{\mathrm{rad},s}^{\mu\nu}$ is given by Eq.~\eqref{eq:Tmunu_rad}.
The source terms for the energy and momentum of the fluid are given by the projections of $\nabla_\mu T_{\mathrm{rad}}^{\mu\nu}$ onto $n_\nu$ and $\gamma_{i\nu}$. They become, respectively,
\begin{align}
    &s_\tau \longrightarrow s_\tau - \sum_s \int \mathcal{L}_{\hmu} \, p_s^{\hmu} \, s_{\mathrm{rad},s} \, dV_{p,s} \, , \label{eq:coupling_tau}\\
    &s_{S_i} \longrightarrow s_{S_i} - \sum_s \int \ell_{i\hmu} \, p_s^{\hmu} \, s_{\mathrm{rad},s} \, dV_{p,s} \:, \label{eq:coupling_momentum}
\end{align}
where we used $p^\mu = \tensor{L}{^\mu_\hmu} \, p^{\hmu}$.
If the particles $s$ refer to electron neutrinos and antineutrinos, the source term for the electron fraction becomes
\begin{equation}
    s_{Y_e} \longrightarrow s_{Y_e} - \int \Big(s_{\mathrm{rad},\nu_e} -s_{\mathrm{rad},\bar{\nu}_e} \Big) \, dV_{p, \nu_e}, \label{eq:coupling_ye}
\end{equation}
where $\nu_e$ and $\bar{\nu}_e$ are the electron neutrino and the electron antineutrino, respectively.

Finally, the contributions to the energy-momentum tensor also affect the metric equations. We can then compute the corresponding source terms as
\begin{align}
    U_{\mathrm{rad}} &= \sum_s n_\mu n_\nu T^{\mu\nu}_{\mathrm{rad},s} = \sum_s \int \Big( \mathcal{L}_{\hmu} \, p_s^{\hmu} \Big)^2  f_s\, dV_{p,s} \\
    S^j_{\mathrm{rad}} &= -\sum_s n_\mu {\gamma^j}_\nu T^{\mu\nu}_{\mathrm{rad},s} \nonumber \\
    &= \sum_s \int \mathcal{L}_{\hmu} \, p_s^{\hmu} \, {\ell^j}_{\hnu} \, p_s^{\hnu} \,  f_s\, dV_{p,s} \\
    S_{\mathrm{rad}} &= \gamma_{ij} \sum_s {\gamma^i}_\mu {\gamma^j}_\nu T^{\mu\nu}_{\mathrm{rad},s} \nonumber \\
     &= \sum_s \int {\ell^i}_{\hmu} \, p_s^{\hmu} \, {\ell_i}_{\hnu} \, p_s^{\hnu} \,  f_s\, dV_{p,s} \nonumber \\
    &= \sum_s \int \Big( \big( \mathcal{L}_{\hmu} \, p_s^{\hmu} \big)^2 - m_s^2 \Big) \,  f_s\, dV_{p,s}
\end{align}

\section{From spacetime to phase space \label{sec:phase_space}}
\noindent Although the formalism we presented is effectively equivalent to solving a $7$D equation, as opposed to the usual $4$D, the addition of the momentum dimensions is a more complex process than naively extending the \ac{FV} method to a $7$D space. The associated challenges have been addressed in multiple works under various assumptions and in different frameworks, as described in the introduction.

One of the main sources of these challenges is that momentum dimensions cannot be treated independently of the spatial dimensions. In Sec.~\ref{sec:importance_discretization}, we will indeed see that the discretizations of the position-space and momentum-space fluxes are not independent. There are however other factors that relate the position and momentum spaces.
Momentum is measured in directions defined by the position space basis vectors. This adds the degree of freedom of choosing the reference axis from which the propagation angles are measured. The choice of a coordinate system for the position space will also have an effect on the treatment of the momentum space. 
Finally, the particle velocity, which is represented as a vector for the fluid, is represented by a point in phase space in a kinetic description. As a result, the boundary conditions applied in position space on the fluid velocity must be generalized to a phase space description. More precisely, the boundary conditions at a fixed spatial position must connect different points in momentum space.

In this section, we shall focus on the left-hand side of the Boltzmann equation. We will detail our implementation of the solver and expand on all the intricacies of discretizing a 6D phase space, instead of a $3$D space, in the context of the conservative general-relativistic Boltzmann equation.

\subsection{Choice of reference frame}\label{sec:ref_frame}

\noindent Before diving into any of the practical aspects of the implementation, we must decide in which reference frame the momentum space will be discretized. This choice of reference frame was already mentioned in Sec.~\ref{sec:coordinates}, where we opted for the comoving frame. We will discuss here how this choice impacts the simulation and compare it to the local orthonormal frame treatment.

We start with the choice of the comoving frame. As mentioned in Sec.~\ref{sec:coordinates}, the comoving frame is a more natural choice in the sense that the interaction terms already require momentum components to be expressed in that frame. Discretizing momentum in the comoving frame would thus allow one to solve the whole equation on the same grid. This may also facilitate the development of techniques preserving both number and energy. This choice however comes at the price of several challenges.

First, it affects the constraint on the time step. Indeed, the momentum-space flux then depends on the velocity. In the presence of a shock, as expected in systems like core-collapse supernovae, there are sharp velocity gradients. We see from Eq.~\eqref{eq:com_Gamma} that ${\Gamma^{\hrho}}_{\hmu\hnu}$ depends on the tetrad derivative, hence containing the derivative of the velocity. From Eq.~\eqref{eq:cs_p}, we can see that the associated characteristic speed in momentum space will contain these large derivatives. The time steps determined through the \ac{CFL} condition will therefore be accordingly small.

The presence of shocks has another consequence. The discretization of the velocity inevitably leads to a smoothing of the velocity gradient, which is contained in the momentum-space flux. An underestimation of the gradient then induces a weaker momentum advection that prevents the distribution function from evolving correctly in momentum space. The momentum-space flux therefore requires a high resolution close to large velocity gradients. On the other hand, a higher resolution is also required for the evolution of hydrodynamical variables. The resolution might therefore not need to be increased for the Boltzmann solver. This momentum advection problem is also present in the M1 scheme, since the advection in energy space contains the covariant derivative of the four-velocity~\citep{Shibata_2011,Cardall_moments_2013}.

Finally, it is important to stress that the angles in momentum space are measured in the comoving frame and do not necessarily correspond to the angles in the lab frame. For example, a particle could be radially outgoing while having $\cos\vartheta < 0$. Let us consider a Minkowski metric in spherical coordinates. In a spherically symmetric configuration, the angle $\hat{\vartheta}$ in the comoving frame can be related to the angle in the orthonormal frame $\Bar{\vartheta}$. We denote by $\hat{\mu}$ ($\Bar{\mu}$) the cosine of $\hat{\vartheta}$ ($\Bar{\vartheta}$). It can be shown that, in this system, $\Bar{\mu}$ will always have the same sign as the cosine of the angle in the lab frame (for $\beta^i = 0$). This means that if $\Bar{\mu} < 0$, the particle will propagate towards smaller values of $r$. We can also show (see Appendix~\ref{app:analytic_sol}) that the cosines in the orthonormal and comoving frames are related in flat space through
\begin{equation}
    \hat{\mu} = \frac{v - \Bar{\mu}}{v \, \Bar{\mu} - 1}, \label{eq:mu_com_ortho}
\end{equation}
where $v$ is the radial velocity of the fluid. We can see from that equation that if a particle is propagating outwards radially ($\Bar{\mu} > 0$), then $\hat{\mu} < 0 $ for values of $v$ such that $v>\Bar{\mu}$. Similarly, an ingoing particle ($\Bar{\mu} < 0$) will be described by $\hat{\mu} > 0$ for values of $v$ that satisfy $v < \Bar{\mu}$.

The fact that the cosine of the propagation angle in the comoving frame may have a different sign than that in the lab frame is important to keep in mind when discussing shocks in the fluid. Indeed, as a particle propagates through a shock, $\hat{\mu}$ will change discontinuously according to Eq.~\eqref{eq:mu_com_ortho}. On the other hand, the cosine $\Bar{\mu}$ does not change, since it is independent of the fluid variables. If the change in $\hat{\mu}$ is not properly handled numerically, it may lead to particles having a $\hat{\mu}$ corresponding to a different $\Bar{\mu}$ than the initial one. If that $\Bar{\mu}$ changes sign across the shock because of this numerical inaccuracy, particles might be artificially trapped around the shock region. However, we did not observe this behaviour in our tests (see Sec.~\ref{sec:advection_collisions_curved}).

The local orthonormal frame does not suffer from these shock-related limitations, since the velocity and its derivatives do not enter the momentum-space flux. Momentum advection is thus unaffected by shocks. On the other hand, the right-hand side of the equation still requires comoving-frame momentum quantities. The approach presented in Ref.~\cite{Nagakura_2014} consists in using two different grids. One grid is constructed from a Lorentz-transformed comoving-frame grid and is used to compute the source terms. A second, temporary, grid is used to compute the advection term in the full phase space. The grid mappings are designed to conserve neutrino number. Energy conservation through the grid mapping is however not discussed.

Even though the formulations in the comoving frame and the orthonormal frame are in theory equivalent, a thorough comparison between the two approaches is still needed to ensure that this is also true in the discrete case.

\subsection{Momentum discretization}\label{sec:mom_discretization}

\noindent Now that our framework has been fully established, we turn our attention to the implementation of our solver.
The first addition to consider to extend our code to phase space is the momentum space grid, using the comoving frame as discussed in the previous section.
Both position and momentum spaces are discretized according to the cell-centred \ac{FV} method. The two spaces are however kept independent, in the sense that the momentum space grid is the same at all position space points. The position-space grid can also be dynamically modified throughout the simulation with \ac{AMR}, whereas the momentum-space grid is fixed.
The position space can be discretized in Cartesian, cylindrical or spherical coordinates in the lab frame, in 1, 2 or 3 dimensions, whereas the momentum space is discretized in spherical coordinates in the orthonormal comoving frame. 

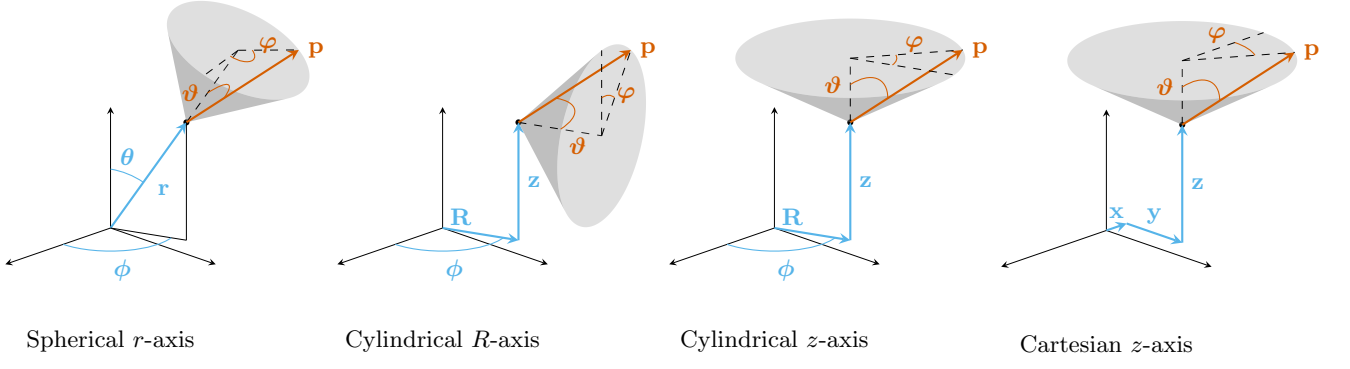
\begin{figure*}
    \centering
    \begin{subfigure}[b]{0.245\linewidth}
		\centering
		\begin{tikzpicture}[x={(-0.87cm,-0.3cm)}, y={(0.87cm,-0.3cm)}, z={(0cm,1cm)}]
            \drawconespherical[phi=110,theta=30,ptheta=35,pphi=150, tilt=0, rotation=0]
            \node at (0,0,-1.5) {Spherical $r$-axis};
		\end{tikzpicture}
	\end{subfigure}%
	\begin{subfigure}[b]{0.245\linewidth}
		\centering
		\begin{tikzpicture}[x={(-0.87cm,-0.3cm)}, y={(0.87cm,-0.3cm)}, z={(0cm,1cm)}]
            \drawconecylindrical[phi=110,theta=30,ptheta=35,pphi=150, tilt=60, rotation=90, phizero=180, thanchor=north west, phianchor=west]
			\node at (0,0,-1.5) {Cylindrical $R$-axis};
		\end{tikzpicture}
	\end{subfigure}%
	\begin{subfigure}[b]{0.245\linewidth}
		\centering
		\begin{tikzpicture}[x={(-0.87cm,-0.3cm)}, y={(0.87cm,-0.3cm)}, z={(0cm,1cm)}]
            \drawconecylindrical[phi=110,theta=30,ptheta=35,pphi=150, tilt=-30, rotation=90]
            \node at (0,0,-1.5) {Cylindrical $z$-axis};
		\end{tikzpicture}
	\end{subfigure}%
	\begin{subfigure}[b]{0.245\linewidth}
		\centering
		\begin{tikzpicture}[x={(-0.87cm,-0.3cm)}, y={(0.87cm,-0.3cm)}, z={(0cm,1cm)}]
            \drawconecartesian[phi=110,theta=30,ptheta=35,pphi=150, tilt=-30, rotation=90, phizero=70]
			\node at (0,0,-1.5) {Cartesian $z$-axis};
		\end{tikzpicture}
	\end{subfigure}
    \caption{Illustration of the same position and momentum coordinates described with different coordinate systems in position space and different reference axes. In all cases, $\varepsilon = |\vec{p}|$, and $\vartheta$ describes the motion along the reference axis, whereas $\varphi$ is related to the motion in the plane perpendicular to that axis. In the schematics, the cone axis corresponds to the reference axis, $\vartheta$ is the opening angle and $\varphi$ indicates the position of $\vec{p}$ on the cone. The same momentum vector is described with different angles depending on the coordinate system and the reference axis.}
    \label{fig:axes}
\end{figure*}

Even though the grids are indeed independent of each other, the choice of coordinate system in position space has an influence on the momentum space. Indeed, the momentum space coordinates $(\varepsilon,\vartheta,\varphi)$ introduced in Sec.~\ref{sec:coordinates} are defined such that
\begin{align}
    &p^{\hat{1}} = \varepsilon \cos\vartheta, \label{eq:p_hat1}\\
    &p^{\hat{2}} = \varepsilon \sin\vartheta \cos\varphi,\label{eq:p_hat2}\\
    &p^{\hat{3}} = \varepsilon \sin\vartheta \sin\varphi, \label{eq:p_hat3}
\end{align}
meaning that the $z_p$-axis (reference axis for measuring $\vartheta$) is aligned with the $x^1$-axis in the absence of fluid. The spatial axes and corresponding momentum axes are illustrated in Fig.~\ref{fig:axes} for spherical, cylindrical and Cartesian coordinates in position space. Note that we can change the reference axis by changing the upper indices of $p$ in the definitions of the momentum coordinates. The $z_p$-axis will always correspond to the axis $\hat{a}$ for which $p^{\hat{a}}=\varepsilon \cos\vartheta$. The consequences of this degree of freedom will be discussed further in Sec.~\ref{sec:coord}.

\subsection{Finite-volume discretization}\label{sec:FV_discretization}
\noindent The next step is to generalize our $3$D \ac{FV} method to evolve conserved variables defined on our phase-space grid.
Their time evolution with the \ac{FV} method is obtained by integrating Eq.~\eqref{eq:cons_bol_ref} over a 6D volume. Before applying the method to the Boltzmann equation, we describe a general 6D \ac{FV} method. We introduce a 6D metric $\Upsilon_{\hyperind{i}\hyperind{j}}$ and 6D volume element $d\hyperind{V} = \sqrt{\Upsilon} dx^{\hyperind{1}}dx^{\hyperind{2}}dx^{\hyperind{3}}dx^{\hyperind{4}}dx^{\hyperind{5}}dx^{\hyperind{6}}$, where $\Upsilon$ is the determinant of the 6D metric. We also introduce the volume and area elements
\begin{align}
    &\Delta \hyperind{V} = \int d\hyperind{V}, \\
    &\Delta A^{\hyperind{i}} = \int \sqrt{\Upsilon} dx^{\hyperind{j}\neq \hyperind{i}}. \label{eq:area_6D}
\end{align}
The general 6D \ac{FV} discretization then reads
\begin{align}
    &\frac{d}{dt} \braket{q}_{i,j,k,l,m,n} = \frac{-1}{\Delta \hyperind{V}}\nonumber\\
    &\Big\{\left[ (\braket{{F}}^{\hyperind{1}} \Delta A^{\hyperind{1}})_{i+1/2,j,k,l,m,n} - (\braket{{F}}^{\hyperind{1}} \Delta A^{\hyperind{1}})_{i-1/2,j,k,l,m,n} \right] \nonumber \\
    &+ \left[ (\braket{{F}}^{\hyperind{2}} \Delta A^{\hyperind{2}})_{i,j+1/2,k,l,m,n} - (\braket{{F}}^{\hyperind{2}} \Delta A^{\hyperind{2}})_{i,j-1/2,k,l,m,n} \right] \nonumber \\
    &+ \left[ (\braket{{F}}^{\hyperind{3}} \Delta A^{\hyperind{3}})_{i,j,k+1/2,l,m,n} - (\braket{{F}}^{\hyperind{3}} \Delta A^{\hyperind{3}})_{i,j,k-1/2,l,m,n} \right] \nonumber \\
    &+ \left[ (\braket{\mathcal{F}}^{\hyperind{4}} \Delta A^{\hyperind{4}})_{i,j,k,l+1/2,m,n} - (\braket{\mathcal{F}}^{\hyperind{4}} \Delta A^{\hyperind{4}})_{i,j,k,l-1/2,m,n} \right] \nonumber \\
    &+ \left[ (\braket{\mathcal{F}}^{\hyperind{5}} \Delta A^{\hyperind{5}})_{i,j,k,l,m+1/2,n} - (\braket{\mathcal{F}}^{\hyperind{5}} \Delta A^{\hyperind{5}})_{i,j,k,l,m-1/2,n} \right] \nonumber \\
    &+ \left[ (\braket{\mathcal{F}}^{\hyperind{6}} \Delta A^{\hyperind{6}})_{i,j,k,l,m,n+1/2} - (\braket{\mathcal{F}}^{\hyperind{6}} \Delta A^{\hyperind{6}})_{i,j,k,l,m,n-1/2} \right] \Big\} \nonumber \\
    &+ \braket{s_{rad,s}}_{i,j,k,l,m,n}, \label{eq:FV_6D}
\end{align}
where the notation $\braket{~\cdot~}$ represents the average over the 6D volume and $\braket{~\cdot~}^{\hyperind{i}}$ over the 5D area of the cell surface perpendicular to the $\hyperind{i}$th axis, \textit{i.e.}
\begin{align}
    &\braket{~\cdot~} = \frac{1}{\Delta \hyperind{V}} \int \cdot~d\hyperind{V},\\
    &\braket{~\cdot~}^{\hyperind{i}} = \frac{1}{\Delta A^{\hyperind{i}}} \int \cdot~\sqrt{\Upsilon} dx^{\hyperind{j}\neq\hyperind{i}} .
\end{align}

Let us now apply the method to the Boltzmann equation. The volume elements can be split into the position and momentum space volumes, since
\begin{equation}
    d\hyperind{V}_s = \sqrt{\hat{\gamma}} \, \frac{\sqrt{\lambda}}{h^3 \, p_s^{\ho}} dx^1 dx^2 dx^3 dp^{\tilde{1}}dp^{\tilde{2}}dp^{\tilde{3}} = dV_x \, dV_{p,s}, \label{eq:phase_space_volume}
\end{equation}
where $dV_x \equiv\sqrt{\hat{\gamma}} dx^1 dx^2 dx^3$ only depends on the position coordinates and $dV_{p,s}$ on the momentum coordinates (since $\lambda$ is independent of $x^\mu$).
As a result, we can indeed split the position and momentum space contributions and use the volume and area elements of each space separately. Therefore, we introduce the volume and area elements in position space
\begin{align}
    &\Delta V_{x} \equiv  \int_{\mathrm{cell,x}} \sqrt{\hat{\gamma}} \, dx^{1}dx^{2}dx^{3},\label{eq:Vx}\\
    &\Delta A_{x}^{i} \equiv \int_{\mathrm{surface}} \sqrt{\hat{\gamma}} \, dx^{j\neq i},\label{eq:Ax}
\end{align}
and in momentum space
\begin{align}
    &\Delta V_{p,s} \equiv  \int_{\mathrm{cell,p}} \frac{\sqrt{\lambda}}{h^3 \,p_s^{\ho}} \, dp^{\tilde{1}}dp^{\tilde{2}}dp^{\tilde{3}},\label{eq:V}\\
    &\Delta A_{p,s}^{\curvi} \equiv \int_{\mathrm{surface}} \frac{\sqrt{\lambda}}{h^3 \, p_s^{\ho}} \, dp^{\curvj \neq \curvi} \label{eq:Ap}.
\end{align}
The full expression of the volume and area elements in position space can be found in the Appendix of Ref.~\cite{Cheong_2021}, and those in momentum space can be found in Appendix~\ref{app:area_and_volume}.

From Eq.~\eqref{eq:phase_space_volume}, we can then write the integrated volume element
\begin{equation}
    \Delta \mathbb{V}_s = \Delta V_x \, \Delta V_{p,s}
\end{equation}
Similarly, the area elements can also be split into components containing areas in position space $\Delta A_x^i$, where $\hyperind{i}\leq 3$ and is represented by the index $i$, and areas in momentum space $\Delta A_p^{\curvi}$ for $\hyperind{i} > 3$ denoted with the index $\curvi$. We therefore rewrite the area elements in the 6D space as
\begin{equation}
    \Delta A^{\hyperind{i}} = \begin{cases}
        \Delta V_{p,s}~\Delta A_x^i ,& \textrm{if } \hyperind{i} \leq 3\\
        \Delta V_x~\Delta A_{p,s}^{\curvi} ,& \textrm{otherwise}
    \end{cases}
\end{equation}
We use the same notation to rewrite the operator $\braket{~\cdot~}^{\hyperind{i}}$ as
\begin{equation}
    \braket{~\cdot~}^{\hyperind{i}} = \begin{cases}
        \braket{~\cdot~}^{i} ,& \textrm{if } \hyperind{i} \leq 3\\
        \braket{~\cdot~}^{\curvi} ,& \textrm{otherwise}
    \end{cases}
\end{equation}
where we introduced
\begin{align}
    &\braket{~\cdot~}^{i} =\frac{1}{\Delta A^{i}~\Delta V_{p,s}} \int \cdot~\sqrt{\hat{\gamma}} dx^{j \neq i}~dV_{p,s}\label{eq:area_avg_x}\\
    &\braket{~\cdot~}^{\curvi} =\frac{1}{\Delta V_x~\Delta A^{\curvi}} \int \cdot~\frac{\sqrt{\hat{\lambda}}}{h^3 p_s^{\hat{0}}} dp^{\curvj \neq \curvi}~dV_x \label{eq:area_avg_p}
\end{align}
Note that the surface average operator in Eq.~\eqref{eq:area_avg_x} differs from the one defined in pure position space in Ref.~\cite{Cheong_2021} through the additional integral over the momentum space volume.

We can then use Eqs.~\eqref{eq:Vx} to \eqref{eq:area_avg_p} and substitute them into Eq.~\eqref{eq:FV_6D} to find the final \ac{FV} discretization. The conserved variables are evolved as
\begin{align}
    &\frac{d}{dt} \braket{q_s}_{i,j,k,l,m,n} = \frac{-1}{\Delta V_{x;\,i,j,k}}\nonumber\\
    &\times \Big\{\left[ (\braket{{F_s}}^1 \Delta A_x^1)_{i+1/2,j,k} - (\braket{{F_s}}^1 \Delta A_x^1)_{i-1/2,j,k} \right] \nonumber \\
    &+ \left[ (\braket{{F_s}}^2 \Delta A_x^2)_{i,j+1/2,k} - (\braket{{F_s}}^2 \Delta A_x^2)_{i,j-1/2,k} \right] \nonumber \\
    &+ \left[ (\braket{{F_s}}^3 \Delta A_x^3)_{i,j,k+1/2} - (\braket{{F_s}}^3 \Delta A_x^3)_{i,j,k-1/2} \right] \Big\}_{l,m,n} \nonumber \\
    &-\frac{1}{\Delta V_{p,s;\,l,m,n}}\nonumber\\
    &\times \Big\{ \left[ (\braket{\mathcal{F}_s}^{\tilde{1}} \Delta A_p^{\tilde{1}})_{l+1/2,m,n} - (\braket{\mathcal{F}_s}^{\tilde{1}} \Delta A_p^{\tilde{1}})_{l-1/2,m,n} \right] \nonumber \\
    &+ \left[ (\braket{\mathcal{F}_s}^{\tilde{2}} \Delta A_p^{\tilde{2}})_{l,m+1/2,n} - (\braket{\mathcal{F}_s}^{\tilde{2}} \Delta A_p^{\tilde{2}})_{l,m-1/2,n} \right] \nonumber \\
    &+ \left[ (\braket{\mathcal{F}_s}^{\tilde{3}} \Delta A_p^{\tilde{3}})_{l,m,n+1/2} - (\braket{\mathcal{F}_s}^{\tilde{3}} \Delta A_p^{\tilde{3}})_{l,m,n-1/2} \right] \Big\}_{i,j,k} \nonumber \\
    &+ \braket{s_{rad,s}}_{i,j,k,l,m,n}, \label{eq:FV}
\end{align}
where $i,j,k$ denote the indices in position space and $l,m,n$ in momentum space. We see that our \ac{FV} method reduces to applying the \ac{FV} method in position space and momentum space separately, although the quantities involved still correspond to averages over the full phase space.

Finally, note that in spherical symmetry where $f = f(t,r,\varepsilon,\vartheta)$, only $p^{\hat{0}}$ and $p^{\hat{1}}$ contribute to the conserved variables and position-space fluxes among all the components of $p^{\hmu}$ defined in Eqs.~\eqref{eq:p_hat0}, \eqref{eq:p_hat1}, \eqref{eq:p_hat2} and \eqref{eq:p_hat3}. The integrals over the momentum space volume and areas indeed cause the $\sin\varphi$ and $\cos\varphi$ contributions to vanish. For the momentum-space flux, several elements of ${P^{\curvi}}_{\hati}p^{\hmu}p^{\hnu}$ vanish because of the integral over $\varphi$, which simplifies the expression of the flux.

\subsection{Boundary conditions \label{sec:BC}}
\noindent Boundary conditions are the last feature to extend to our phase space representation. Indeed, in the fluid formalism, the velocity of the fluid is represented by a vector. The boundary condition accounts for the change in orientation of the axis by changing the orientation of this velocity vector. In the full kinetic description, however, the momentum of particles is represented by a point in momentum space. Thus, the change in orientation of the velocity vector corresponds to a change in the location of the point in momentum space. Similarly, special care is also required for outer boundary conditions, even though the axis orientation remains the same. Spatial boundary conditions must therefore relate different points in momentum space across the spatial boundary. In this section, we will describe how the spatial boundary conditions are treated.

We can more easily discuss the mapping of the particle propagation angles across a boundary in the coordinate basis. In the local comoving frame, on the other hand, the different spatial basis vectors are mixed. To simplify the discussion, we will therefore introduce the boundary condition in the coordinate basis. We can then map the momentum coordinates in the lab frame to those in the comoving frame using
\begin{equation}
    p^{\hatj} = \tensor{L}{^\hatj_\mu} p^\mu.\label{eq:bc_phat}
\end{equation}
We can then invert Eqs.~\eqref{eq:p_hat1}, \eqref{eq:p_hat2} and \eqref{eq:p_hat3} to find the corresponding coordinates in the comoving frame, which are
\begin{align}
    &\varepsilon = \sqrt{{ \sum_{\hatj}} \left( p^{\hatj} \right)^2} = \sqrt{\left(\po\right)^2 - m^2}, \label{eq:vareps}\\
    &\vartheta = \arccos\left( \frac{p^{\hat{1}}}{\sqrt{\left(\po\right)^2 - m^2}} \right), \label{eq:vartheta}\\
    &\varphi = \begin{cases}
        \arctan\left(\frac{p^{\hat{3}}}{p^{\hat{2}}} \right) & \text{, if } p^{\hat{2}} > 0 \text{ and } p^{\hat{3}} > 0\\
        \arctan\left(\frac{p^{\hat{3}}}{p^{\hat{2}}} \right) + \pi & \text{, if } p^{\hat{2}} < 0 \\
        \arctan\left(\frac{p^{\hat{3}}}{p^{\hat{2}}} \right) + 2\pi & \text{, if } p^{\hat{2}} > 0 \text{ and } p^{\hat{3}} < 0
    \end{cases}\label{eq:varphi}
\end{align}
Note that these relations may change depending on the choice of the reference axis. We can also express the boundary conditions without going through the coordinate basis if the tetrad is diagonal, which is only the case if the metric is diagonal and the velocities are all null. The latter condition is not needed if one works in the local orthonormal frame.

\begin{table}
    \renewcommand{\arraystretch}{1.15}
    \centering
    \begin{tabular}{|c|c|c|c|}
        \hline
        BC & $p^1$ & $p^2$ & $p^3$ \\
        \hline
        \multicolumn{4}{|c|}{Spherical coordinates}\\
        \hline
        $r=0$ & $-$ & $+$ & $-$\\
        \hline
        $\theta=0$ & $+$ & $-$ & $-$ \\
        \hline
        $xy-$plane & $+$ & $-$ & $+$\\
        \hline
        $xz-$, $yz-$plane & $+$ & $+$ & $-$ \\
        \hline
        \multicolumn{4}{|c|}{Cylindrical coordinates}\\
        \hline
        $R=0$ & $-$ & $+$ & $-$ \\
        \hline
        $xy-$plane & $+$ & $-$ & $+$\\
        \hline
        $xz-$, $yz-$plane & $+$ & $+$ & $-$ \\
        \hline
        \multicolumn{4}{|c|}{Cartesian coordinates}\\
        \hline
        $yz-$plane & $-$ & $+$ & $+$ \\
        \hline
        $xz-$plane & $+$ & $-$ & $+$ \\
        \hline
        $xy-$plane & $+$ & $+$ & $-$ \\
        \hline
    \end{tabular}
    \caption{Summary of the change of signs of the momentum components in the lab frame. A $+$ indicates no change of sign, whereas $-$ indicates a change of sign across the boundary. The boundary condition ``$-$plane'' indicates a reflection symmetry around the corresponding plane.}
    \label{tab:bc}
\end{table}

\begin{figure}
    \centering
    \begin{tikzpicture}
        \node at (0.75,-0.5) {domain};
        \node at (-2.25, -0.5) {ghost cell};
		\fill[draw=white, fill=oi_blue, opacity=0.15] (-3,-4) rectangle (-1.5,-1);
		\draw[xstep=1.5, ystep=0.5] (-3,-4) grid (3,-1);
		\node at (-0.75,-3.25) {$f(\vartheta_{\mathrm{in}})$};
		\node at (-0.75,-1.75) {\small$f(\pi-\vartheta_{\mathrm{in}})$};
		\node at (-2.25,-3.25) {\small$f(\vartheta_{\mathrm{out}})$};
		\draw[draw=oi_blue, ultra thick] (-3,-3.5) rectangle (-1.5,-3);
		\draw[draw=oi_blue, ultra thick] (-1.5,-2) rectangle (0,-1.5);

		\draw[->, >=stealth,thick] (-0.5,-4.5) -- (0.5,-4.5) node[pos=0.5, anchor=north] {\large $r$};
		\draw[->, >=stealth,thick] (-3.3,-3.) -- (-3.3,-2) node[pos=0.5, anchor=east] {\large $\vartheta$};

	\end{tikzpicture}
	\caption{Example of symmetric spatial boundary conditions in an empty Minkowski space in spherical coordinates. Using the procedure explained in the main text, we find that the corresponding $\vartheta$ coordinate outside the boundary is given by $\vartheta_{\mathrm{out}}=\pi-\vartheta_{\mathrm{in}}$. We then impose $f=f(\pi-\vartheta_{\mathrm{in}})$ in the ghost cell corresponding to the coordinate $\vartheta_{\mathrm{in}}$.}
	\label{fig:boundary}
\end{figure}
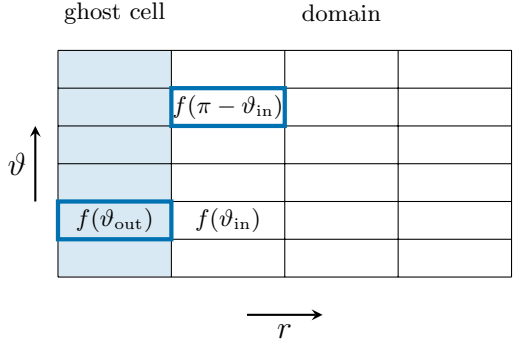

In Tab.~\ref{tab:bc}, we indicate how each component of the momentum in the lab frame changes for different coordinates, boundaries and symmetries. The components $1$, $2$ and $3$ correspond to the following axes:
\begin{align*}
    \text{Cartesian coordinates:} && 1 &\rightarrow x, & 2&\rightarrow y, & 3 &\rightarrow z ;\\
    \text{Cylindrical coordinates:} && 1 &\rightarrow R, & 2&\rightarrow z, & 3 &\rightarrow \phi;\\
    \text{Spherical coordinates:} && 1 &\rightarrow r, & 2&\rightarrow \theta, & 3 &\rightarrow \phi.
\end{align*}
A $+$ sign indicates that the component does not change, whereas a $-$ sign indicates a change of sign across the boundary.
In Tab.~\ref{tab:bc}, we indicate reflection symmetries around a given plane with the notation ``$-$plane''.
For each point outside of the domain, we match $p^i_{\mathrm{out}}$ to $\pm p^i_{\mathrm{in}}$, where in and out refer to inside and outside the domain, and the sign is chosen according to Tab.~\ref{tab:bc}. The resulting $p^i_{\mathrm{out}}$ is then used to compute $p^{\hati}_{\mathrm{out}}$ and finally the coordinates $(\varepsilon_{\mathrm{out}},\vartheta_{\mathrm{out}},\varphi_{\mathrm{out}})$ with the method described above. We can then apply the desired boundary condition as in a position-space-only grid, but relating the values of $f$ at $(\varepsilon_{\mathrm{out}},\vartheta_{\mathrm{out}},\varphi_{\mathrm{out}})$ and $(\varepsilon_{\mathrm{in}},\vartheta_{\mathrm{in}},\varphi_{\mathrm{in}})$. For example, let us consider an empty Minkowski space in spherical coordinates as illustrated in Fig.~\ref{fig:boundary}. At the boundary $r=0$, $p^r$ and $p^{\phi}$ change sign. As a result, $\vartheta_{\mathrm{out}} = \pi - \vartheta_{\mathrm{in}}$ and $\varphi_{\mathrm{out}} = 2\pi - \varphi_{\mathrm{in}}$. A symmetric boundary condition around $r=0$ in spherical coordinates would thus read
\begin{equation*}
    \begin{split}
    &f(r=0^{-}, \theta, \phi, \varepsilon, \vartheta, \varphi) \\
    &= f(r=0^{+},\pi-\theta, \phi+\pi,\varepsilon, \pi-\vartheta, 2\pi-\varphi).
    \end{split} 
\end{equation*}

For a ``no inflow'' boundary condition along the axis $j$, preventing quantities from entering the domain, we only compute $p^j_{\mathrm{out}}$. If it is negative (positive) for the outer (inner) boundary, \textit{i.e.} if matter would enter the domain, we set the value of $f$ to $0$ outside of the domain. This is equivalent to the position-space-only case where the velocity is set to $0$ depending on its sign. Note that we use the lab frame momentum because, as mentioned in Sec.~\ref{sec:ref_frame}, a comoving-frame momentum pointing inwards does not necessarily correspond to one pointing inwards in the lab frame.

In momentum space, we implemented the boundary conditions that are generally used in $3+1$ simulations, namely periodic (for $\varphi$), symmetry across the boundary, zero-flux and the no-inflow boundary condition. For the latter, we impose $f=0$ if the characteristic speed $c^{\curvi}$ is negative (positive) for the outer (inner) boundary. Unlike the point $r=0$ in position space, particles cannot cross the point $\varepsilon = 0$. In position space, points at negative radii can effectively be represented by rotated $\theta$ and $\phi$ angles. This is however not the case for negative energies, which are unphysical. Therefore, we impose the zero-flux boundary condition at the boundary $\varepsilon = 0$.

\subsection{Choice of the position space coordinate system}\label{sec:coord}
\noindent The previous sections have highlighted the relation between the position and momentum spaces. In this last section on the phase space extension, we want to draw attention to the effect that the choice of the position space coordinates has on the computational performance of a simulation.

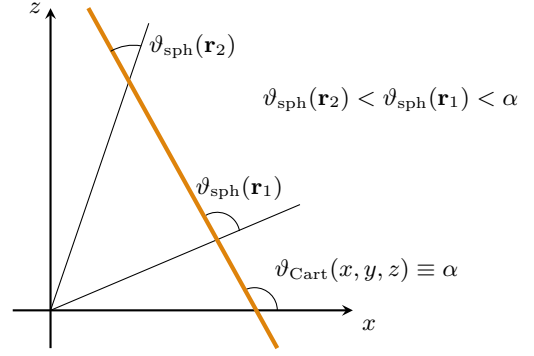
\begin{figure}
    \centering
    \begin{tikzpicture}
		\draw[thick,->, >=stealth] (-0.5,0) -- (4,0) node[pos=1,anchor=north west]{$x$};
		\draw[thick,->, >=stealth] (0,-0.5) -- (0,4) node[pos=1,anchor=east]{$z$};
		\draw (3,0) arc (0:120:0.3) node[pos=0.5,anchor=south west] {$ \vartheta_{\mathrm{Cart}}(x,y,z)\equiv \alpha $};
		\draw[thin] (0,0) -- (3.3, 1.4);
		\draw[thin] (0,0) -- (1.3, 3.8);
		\draw (2.5,1.05) arc (10:120:0.3) node[pos=0.,anchor=south,yshift=0.3cm] {$ \vartheta_{\mathrm{sph}}(\vec{r}_1)$};
		\draw[oran, ultra thick] (3,-0.5) -- (0.5, 4);
		\draw (1.2,3.5) arc (80:120:0.6) node[pos=0.5,anchor=west,xshift=0.2cm] {$ \vartheta_{\mathrm{sph}}(\vec{r}_2)$};
		\node at (4.5,2.8) {$\vartheta_{\mathrm{sph}}(\vec{r}_2) < \vartheta_{\mathrm{sph}}(\vec{r}_1) < \alpha$};
	\end{tikzpicture}
	\caption{Illustration of the change in $\vartheta$ with the change of the reference axis $\vec{r}$ in spherical coordinates. In Cartesian coordinates, the measured angle $\vartheta_{\mathrm{Cart}}$ is constant.}
	\label{fig:ref_axis_change}
\end{figure}

The choice of coordinate system in position space affects the momentum space through the choice of the reference axis for measuring the propagation angles. This axis may indeed be position-dependent, as is the case for all axes in spherical coordinates and for the $R-$ and $\phi-$axes in cylindrical coordinates. In these cases, a change of the axis induces a change in the propagation angles. This is illustrated in Fig.~\ref{fig:ref_axis_change} for spherical coordinates. Even though the particle moves in a straight line, \textit{i.e.} at a constant angle in Cartesian coordinates, the value of $\vartheta$ measured at different positions does change. This behaviour translates into a purely geometrical contribution to the momentum-space flux in both spherical and cylindrical coordinates. These contributions arise from the Christoffel symbols of the reference metric.

More concretely, this geometrical advection impacts the maximum time step allowed. The maximum characteristic speeds in momentum space in a Minkowski space and on a uniform grid read
\begin{align}
    &c_{\mathrm{max},\vartheta} = \max \frac{\sin{\vartheta}}{r} = \frac{1}{r_{\mathrm{min}}} = \frac{2}{\Delta r} \label{eq:c_sph_theta}\\
    &c_{\mathrm{max},\varphi} = \max \frac{\sin{\varphi}\sin{\vartheta}}{r \tan{\theta}} = \frac{2}{\Delta r \, \tan{(\Delta \theta/2)}} \label{eq:c_sph_phi}
\end{align}
in the spherical coordinate system $(r,\theta,\phi)$ with reference axis $\vec{r}$,
\begin{align}
    &c_{\mathrm{max},\vartheta} = \max \frac{\sin^2{\varphi}\sin{\vartheta}}{R} = \frac{1}{R_{\mathrm{min}}} = \frac{2}{\Delta R} \label{eq:c_cyl_theta}\\
    &c_{\mathrm{max},\varphi} = \max \frac{2\sin{(2\varphi)}\cos{\vartheta}}{R} = \frac{4}{\Delta R} \label{eq:c_cyl_phi}
\end{align}
in the cylindrical coordinate system $(R,\phi,z)$ with reference axis $\vec{R}$, and
\begin{align}
    &c_{\mathrm{max},\vartheta} = 0  \label{eq:c_cart_theta}\\
    &c_{\mathrm{max},\varphi} = 0  \label{eq:c_cart_phi}
\end{align}
in Cartesian coordinates with reference axis $\vec{x}$.

A direct consequence of these characteristic speeds is that the maximum time step, determined through the \ac{CFL} condition, may be less constrained in Cartesian coordinates than in spherical or cylindrical coordinates. Indeed, the maximum time step that can be used given the \ac{CFL} condition and Eqs.~\eqref{eq:c_sph_theta} to \eqref{eq:c_cart_phi} is
\begin{align}
    &\Delta t_{\mathrm{max,sph,r}} \propto \min\left(\Delta t_{\vec{x}},\:\frac{\Delta r \Delta \vartheta}{2},\: \frac{\Delta r \tan(\frac{\Delta \theta}{2}) \Delta \varphi}{2} \right),\\
    &\Delta t_{\mathrm{max,cyl,R}} \propto \min\left(\Delta t_{\vec{x}},\:\frac{\Delta R \Delta \vartheta}{2},\: \frac{\Delta R}{4} \Delta \varphi \right),\\
    &\Delta t_{\mathrm{max,Cart,x}} \propto \Delta t_{\vec{x}},
\end{align}
where $\Delta t_\vec{x}$ is the largest time step allowed by the spatial advection and the last index in $\Delta t_{\mathrm{max}}$ represents the reference axis. Note that in cylindrical coordinates, the constraint on the time step can be modified to
\begin{equation}
    \Delta t_{\mathrm{max,cyl,z}} \propto \min\left(\Delta t_{\vec{x}},\: \frac{\Delta R}{4} \Delta \varphi \right)
\end{equation}
if one uses the $z-$axis as the reference axis. In that case, there is no geometric contribution to the momentum advection for the $\vartheta-$axis, since it is measured from a fixed axis.

The influence of the spatial resolution on the time step becomes apparent. The worst case occurs for $2+3$ and $3+3$D spherical coordinates, in which the maximum time step depends on the resolution along $3$ dimensions. Even though fluid velocities and gravity would induce additional contributions to the advection, including in the energy space, we can see that spherical and cylindrical coordinates contain additional intrinsic geometrical advection terms that could result in more stringent constraints on the time step than Cartesian coordinates.

Another consequence is that the choice of coordinate system in position space and its resolution may also modify the momentum space resolution required for convergence. Our method is indeed prone to ray effects if the angular resolution in momentum space is lower than that in position space. An extended Cartesian grid may have a very small angle resolution, increasing the number of points needed for the momentum space grid. This could thus have a significant impact on the computational cost and memory requirements. Note that a larger number of points would also imply smaller time steps.

Finally, the choice of reference axis to measure $\vartheta$ and $\varphi$ is only natural in spherical coordinates, where it is chosen to be the radial axis. Indeed, the reference axis then depends on the location in space and does not introduce a preferred direction of propagation. In Cartesian coordinates, however, none of the three axes should be preferred over the others. By using one of them as a reference axis, we may introduce differences in how the propagation along each axis is treated. For rotating objects, however, the $z-$axis may be a natural choice and may benefit, in cylindrical coordinates, from looser constraints on the time steps.

\section{Discretizing the Boltzmann equation \label{sec:importance_discretization}}
\noindent From the presented formalism and implementation, it may still not be clear why the discretization of the Boltzmann equation is as challenging as mentioned in the introduction. We shall therefore now clarify this point and discuss three main discretization issues that impact the physical realism of our model. First, a naive discretization does not ensure consistency between the conservative and non-conservative formulations of the Boltzmann equation. As a consequence, the solver's prediction may not converge towards the true solution locally. In Sec.~\ref{sec:consistency}, we detail a discretization scheme that ensures consistency in flat spacetime. Second, a direct discretization of the number conservative equation does not ensure energy and momentum conservation. Since our discretization is not designed to mitigate this problem, we study the solver's energy conservation properties in Sec.~\ref{sec:conservation_laws} and suggest improvements. Third, maintaining the realizability of the fermionic distribution function, \textit{i.e.} enforcing $0 \leq f \leq 1$, is not trivially achieved. We briefly discuss this issue in Sec.~\ref{sec:realizability}.

\subsection{Consistency \label{sec:consistency}}
\noindent We start our discussion with the investigation of the consistency between the discrete conservative and non-conservative formulations of the Boltzmann equation. A discretization is consistent if the discretization of the conservative (non-conservative) equation yields a valid discretization for the non-conservative (conservative) equation. This is a known problem and solutions have been proposed~\cite{Mezzacappa_1993,Liebendorfer_2004}. These adaptations to the discretizations are however targeted to one-dimensional codes in spherical coordinates and must still be generalized to asymmetrical configurations and other coordinate systems. In this section, we will present the first steps towards such a generalization. Even though the discretization scheme we suggest is tailored to our formalism, it can in principle be extended to methods other than \ac{FV} and does not require working in the conformally flat approximation or with the reference metric. We shall also highlight the limitations of our discretization scheme and point to future improvements.

\subsubsection{From continuous to discrete formulations}

\noindent Before deriving our solution, we introduce the consistency problem in more detail. To understand the source of the problem, we must look into the derivation of the conservative formulation of the Boltzmann equation. The derivation that we use here is detailed in Ref.~\cite{Cardall_2013}. We shall thus only outline the important points. 
We start from the conservative form
\begin{equation}
    \nabla_\mu\left( \tensor{L}{^\mu_\hmu} p^{\hmu} f \right) - \po \mathcal{D}_{\curvi}\left(\frac{1}{\po}\tensor{P}{^\curvi_\hati} {\Gamma^{\hati}}_{\hmu\hnu} \,  p^{\hmu} \, p^{\hnu} f \right) = C[f].
\end{equation}
To retrieve the non-conservative formulation, we use the product rule
\begin{equation}
    \begin{split}
        &f \nabla_\mu\left( \tensor{L}{^\mu_\hmu} p^{\hmu} \right) + \tensor{L}{^\mu_\hmu} p^{\hmu} \nabla_\mu f \\
        &- f \po \mathcal{D}_{\curvi}\left(\frac{1}{\po}\tensor{P}{^\curvi_\hati} {\Gamma^{\hati}}_{\hmu\hnu} \,  p^{\hmu} \, p^{\hnu}  \right) - \tensor{P}{^\curvi_\hati} {\Gamma^{\hati}}_{\hmu\hnu} \,  p^{\hmu} \, p^{\hnu}  \mathcal{D}_{\curvi}f = C[f].\label{eq:cons_to_non_cons_bol}
    \end{split}
\end{equation}
The resulting equation must then be the same as the non-conservative formulation
\begin{equation}
    \tensor{L}{^\mu_\hmu} p^{\hmu} \partial_\mu f - \tensor{P}{^\curvi_\hati} {\Gamma^{\hati}}_{\hmu\hnu} \, p^{\hmu} \, p^{\hnu}  \frac{\partial f}{\partial p^{\curvi}} = C[f]. \label{eq:non_cons_bol}
\end{equation}
This condition makes it clear that the equality
\begin{equation}
    f \nabla_\mu \tensor{L}{^\mu_\hmu} p^{\hmu} = f \po \mathcal{D}_{\curvi}\left(\frac{1}{\po}\tensor{P}{^\curvi_\hati} {\Gamma^{\hati}}_{\hmu\hnu} \,  p^{\hmu} \, p^{\hnu}  \right) \label{eq:cancel}
\end{equation}
is key to ensure the correspondence between the conservative and non-conservative formulations. Any discretization of the Boltzmann equation must therefore ensure that this equality holds in the discrete limit. If this relation is violated, Eq.~\eqref{eq:cons_to_non_cons_bol} does not reduce to the non-conservative form, introducing local inconsistencies between the conservative and non-conservative forms. We shall return to this point in the next section.

An important property of the discrete conservative Boltzmann equation then stems from Eq.~\eqref{eq:cancel}: the discretization of the flux in momentum space is not independent from that in position space. Indeed, the terms on the left-hand side of Eq.~\eqref{eq:cancel} arise from the four-divergence of the particle number four-flux $F^\mu$, whereas the right-hand side comes from the divergence in momentum space of the momentum-space flux $\mathcal{F}^{\curvi}$ defined in Eq.~\eqref{eq:cons_bol_ref}. More precisely, the derivative of the tetrad contained in $\tensor{\Gamma}{^{\hrho}_{\hnu\hmu}}$ must be discretized consistently with the discretization of the four-divergence $\ddiff_{x;\mu} \tensor{L}{^\mu_\hmu}$.

To discuss the issue more formally, we introduce the operators $\ddiff_{x;\mu}$ and $\ddiff_{p;\curvi}$ as being the discretized divergence operator with respect to spatial and momentum space variables, respectively. More precisely, we discretize the operators as
\begin{align}
    &\nabla_\mu F^\mu = \frac{1}{\alpha \sqrt{\gamma}} \partial_\mu (\alpha \sqrt{\gamma} F^\mu) \longrightarrow \ddiff_{x;\mu} F^\mu, \\
    &\po \mathcal{D}_{\curvi} ( \frac{1}{\po} \mathcal{F}^{\curvi} ) \longrightarrow \ddiff_{p;\curvi} \mathcal{F}^{\curvi}
\end{align} 
These discrete operators are determined by the method employed to solve the Boltzmann equation, the \ac{FV} method for our solver. At this point, we do not provide any expression for these operators and remain agnostic about the exact scheme used to solve the equation. Note that we include the $\po$ factors inside and outside $\mathcal{D}_{\curvi}$ in the discretization $\ddiff_{p;\curvi}$ to be consistent with the formalism presented in the previous sections. To be more general, we also include $\alpha \sqrt{\gamma}$ in the discretization of $\ddiff_{x;\mu}$. In our formalism, only $\sqrt{\hat\gamma}$ would be included, whereas a factor $\alpha \psi^6$ would be absorbed in $F^\mu$.

Then, we write the product rule for the discrete spatial operator as
\begin{equation}
    \ddiff_{x,\mu} (F^\mu) = \widehat{c}^\mu \widehat{\partial}_\mu f + \widetilde{f} \, \widehat{\nabla}_\mu c^\mu, \label{eq:discrete_product_rule}
\end{equation}
where $c^\mu$ is such that $F^\mu = c^\mu f$, $\widehat{\cdot}$ and $\widetilde{\cdot}$ are specific discretizations of the terms themselves, and $\widehat{\partial} \cdot$ and $\widehat{\nabla} \cdot$ specific discretizations of the partial derivative and covariant derivative. For example, for the \ac{FV} method, one can write along the dimension $i$ indexed by $j$ (no sum)
\begin{equation*}
    \begin{split}
    &\ddiff_{x,i} (F^i) = \\ 
    &\frac{(c^i \Delta A_i)_{j+1/2} + (c^i \Delta A_i)_{j-1/2}}{2\Delta V_j} \left[ (f)_{j+1/2} - (f)_{j-1/2}\right] \\
    & + \frac{(f)_{j+1/2} + (f)_{j-1/2}}{2} \, \ddiff_{x,i} c^i.
    \end{split}
\end{equation*}
As we can see, the discretizations $\widehat{c}^\mu$ and $\widetilde{f}$ depend on the axis considered. We shall omit this dependence in the notation to avoid confusion in the indices. We only indicate an index $x$ or $p$ to differentiate between the terms arising from position and momentum space derivatives. The two previous expressions are also valid for the momentum space operators.

To ensure that Eq.~\eqref{eq:cancel} is also true in the discrete limit, our discretization must then satisfy
\begin{equation}
    \widetilde{f}_p \widehat{\mathcal{D}}_{\curvi}\left( \tensor{\Gamma}{^{\hati}_{\hmu\hnu}} {P^{\curvi}}_{\hati} \, p^{\hmu}p^{\hnu} \right) = \widetilde{f}_x \, \widehat{\nabla}_\mu \left( {L^\mu}_{\hmu}p^{\hmu}  \right),
\end{equation}
or, similarly, enforce that the residuals
\begin{equation}
    \mathcal{R} = \widetilde{f}_x \, \widehat{\nabla}_\mu \left( {L^\mu}_{\hmu}p^{\hmu}  \right) - \widetilde{f}_p \widehat{\mathcal{D}}_{\curvi}\left( \tensor{\Gamma}{^{\hati}_{\hmu\hnu}} {P^{\curvi}}_{\hati} \, p^{\hmu}p^{\hnu} \right) \label{eq:full_residuals}
\end{equation}
vanish. Using the discrete product rule on the conservative equation, we thus find
\begin{equation}
    \ddiff_{x;\mu}\left( c_x^\mu f \right) - \ddiff_{p;\curvi}\left( c_p^{\curvi} f \right) = \widehat{c}_x^{\mu} \, \widehat{\partial}_\mu f - \widehat{c}_p^{\curvi} \widehat{\partial}_{\curvi} f + \mathcal{R} = C[f].
\end{equation}
If the residuals do not vanish, the discretization of the conservative equation does not reduce to a direct discretization of the non-conservative equation. In other words, solving the discrete conservative equation with $\mathcal{R} \neq 0$ is not equivalent to solving the discrete non-conservative equation. Making the residuals vanish is then the only way to ensure consistency between the two formulations. 

In this work, we shall only consider the homogeneous case where $f$ is constant, which neglects the effect of the discrete product rule. Small derivatives of $f$ are indeed expected at the centre of (proto-)neutron stars, where the largest residuals are expected in spherical coordinates. We note that the corrections in Ref.~\cite{Mezzacappa_1993} were also derived within that assumption. We can then rewrite our condition and residuals in the homogeneous case as
\begin{align}
    &\ddiff_{p;\curvi}\left( \tensor{\Gamma}{^{\hati}_{\hmu\hnu}} {P^{\curvi}}_{\hati} \, p^{\hmu}p^{\hnu} \right) = \ddiff_{x;\mu}\left( {L^\mu}_{\hmu}p^{\hmu}  \right), \label{eq:discr}\\
    &\mathcal{R}_{h} = \ddiff_{x;\mu}\left( {L^\mu}_{\hmu}p^{\hmu}  \right) - \ddiff_{p;\curvi}\left(\tensor{\Gamma}{^{\hati}_{\hmu\hnu}} {P^{\curvi}}_{\hati} \, p^{\hmu}p^{\hnu} \right). \label{eq:residuals}
\end{align}
Our goal in the next sections is thus to find a discretization that makes the residuals $\mathcal{R}_h$ vanish.

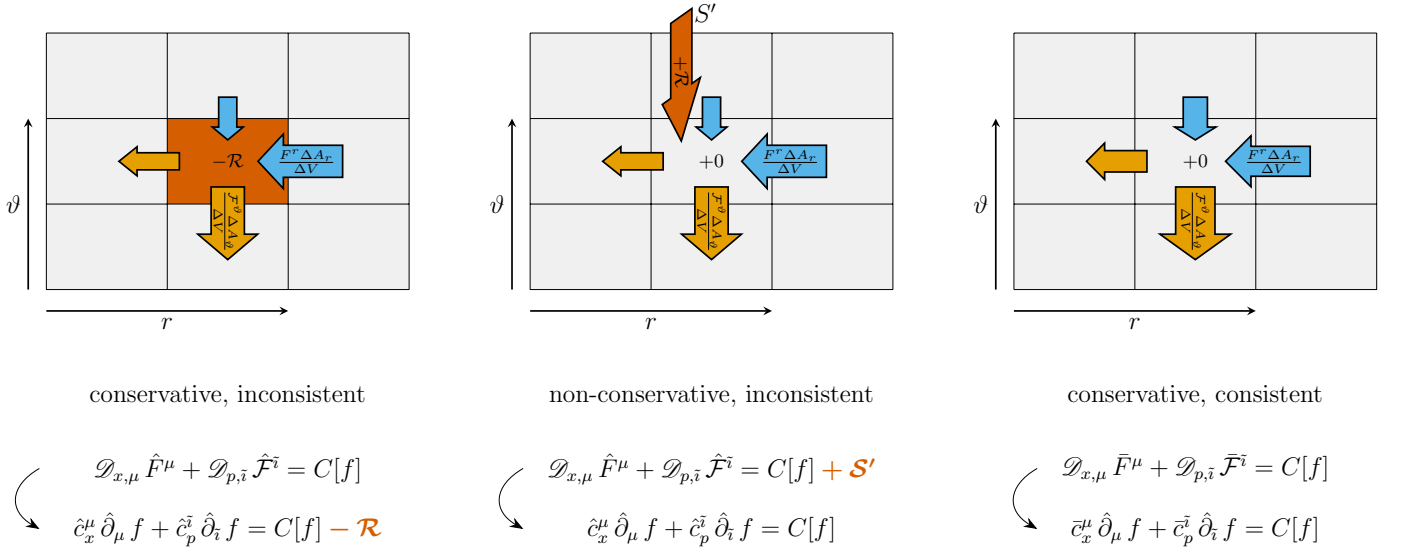
\begin{figure*}
    \centering
    \scalebox{0.8}{
	\begin{tikzpicture}
		\def \th {45}
		\def \ph {90}
		\def \yshift {1.2}
		\begin{scope}[shift={(-8,0)}]
    		\drawinconsistentconservative{\ph}{\th}
    		\node at (3,-6) {\large conservative, inconsistent};
            \node (cons) at (3,{-6-\yshift}) {\parbox{6cm}{\centering\large $ \ddiff_{x,\mu} \, \hat{F}^{\mu} + \ddiff_{p,\curvi} \, \hat{\mathcal{F}}^{\curvi} = C[f]$}};
            \node (notcons) at (3,{-6-1-\yshift}) {\parbox{6cm}{\centering\large $ \hat{c}_{x}^{\mu} \, \hat{\partial}_{\mu} \, f + \hat{c}_{p}^{\curvi} \, \hat{\partial}_{\curvi} \, f = C[f] \mathcolor{residual}{\bm{-\mathcal{R}}}$}};
            \draw [-{Stealth[scale=1.3]}] (cons.west) to [out=210,in=150,looseness=1.5] (notcons.west);
        \end{scope}
        \begin{scope}[shift={(0,0)}]
    		\drawinconsistentnonconservative{\ph}{\th}
    		\node at (3,-6) {\large non-conservative, inconsistent};
            \node (cons) at (3,{-6-\yshift}) {\parbox{6cm}{\centering\large $ \ddiff_{x,\mu} \, \hat{F}^{\mu} + \ddiff_{p,\curvi} \, \hat{\mathcal{F}}^{\curvi} = C[f] \mathcolor{residual}{\bm{+\mathcal{S'}}}$}};
            \node (notcons) at (3,{-6-1-\yshift}) {\parbox{6cm}{\centering\large $ \hat{c}_{x}^{\mu} \, \hat{\partial}_{\mu} \, f + \hat{c}_{p}^{\curvi} \, \hat{\partial}_{\curvi} \, f = C[f]$}};
            \draw [-{Stealth[scale=1.3]}] (cons.west) to [out=210,in=150,looseness=1.5] (notcons.west);
        \end{scope}

        \begin{scope}[shift={(8,0)}]
    		\drawconsistentconservative{\ph}{\th}
    		\node at (3,-6) {\large conservative, consistent};
            \node (cons) at (3,{-6-\yshift}) {\parbox{5cm}{\centering\large $ \ddiff_{x,\mu} \, \Bar{F}^{\mu} + \ddiff_{p,\curvi} \, \Bar{\mathcal{F}}^{\curvi} = C[f]$}};
            \node (notcons) at (3,{-6-1-\yshift}) {\parbox{5cm}{\centering\large $ \Bar{c}_{x}^{\mu} \, \hat{\partial}_{\mu} \, f + \Bar{c}_{p}^{\curvi} \, \hat{\partial}_{\curvi} \, f = C[f]$}};
            \draw [-{Stealth[scale=1.3]}] (cons.west) to [out=210,in=150,looseness=1.5] (notcons.west);
        \end{scope}
	\end{tikzpicture}
	}
    \caption{Illustration of different discretizations assuming finite-volume discretization of the conservative equation. The diagram on the left shows a given discretization $\hat{F}$ of the fluxes that yields residuals $\mathcal{R}$ in the non-conservative formulation. The middle diagram shows a possible solution to remove the residuals using source terms. Although $\mathcal{R}$ does vanish from the non-conservative equation, the added source terms induce non-conservation of neutrino number. On the right, the ideal discretization $\bar{F}$ makes the residuals vanish and conserves neutrino number. The discretization is then consistent, since both discrete equations correspond to their continuous counterparts; they do not contain additional isolated terms.}
    \label{fig:consistency}
\end{figure*}

\subsubsection{Conservative versus consistent \label{sec:conservative_vs_consistent}}

\noindent Before diving into the details of the discretization, we shall first explain in more detail how to interpret the residuals, what their effect is and how they can be removed. In particular, we will see that an inconsistent discretization is not necessarily non-conservative, \textit{i.e.} that the residuals do not prevent conservation of the conserved variable.

Let us consider the \ac{FV} method. By construction, this discretization is conservative. The change in the global conserved quantity is purely determined by the fluxes at the boundary and the source terms. Indeed, the total conserved quantity $Q = \int q dV$ would be expressed, in 1D with no source terms, as
\begin{equation*}
    \begin{split}
        \frac{dQ}{dt} &= \sum_{i=1}^N \frac{\partial q_i}{\partial t} \Delta V_i \\
        &= -\sum_{i=1}^N (F_{i+1/2} \Delta A_{i+1/2} - F_{i-1/2} \Delta A_{i-1/2}) \\
        &= F_{1/2} \Delta A_{1/2}-F_{N+1/2} \Delta A_{N+1/2}.
    \end{split}
\end{equation*}
In the absence of source terms, the only way for $Q$ to change is that the flux be non-zero at the boundaries, \textit{i.e.} $q$ would flow outside or inside the domain. The residuals do not play any role in the conservation, such that conservation is still guaranteed by construction.

This discretization is represented in Fig.~\ref{fig:consistency}. On the left-most diagram, we discretize the conservative Boltzmann equation with a discretization with non-vanishing residuals $\mathcal{R}$. When using the discrete equivalent of the product rule, we find a discrete analog of the non-conservative Boltzmann equation. As discussed in the previous section, the residuals appear on the right-hand side. As a result, the discrete solution may locally diverge from the true solution, since the discrete equation contains an additional term compared to the true equation. This divergence depends on the amplitude of the residuals, which vanish in the continuous limit.

A possible (\textit{ad hoc}) solution to retrieve the proper non-conservative formulation is to remove the residuals from $q$. This effectively corresponds to adding a source term $S'$ in the conservative equation, as shown in the middle diagram in Fig.~\ref{fig:consistency}. This new source term allows one to find the true solution, since the corresponding discrete non-conservative equation is of the same form as the continuous one. On the other hand, we introduce a source term into the conservative equation, which induces non-conservation of the discrete conserved quantity. Therefore, the discretizations are still inconsistent, since the discrete conservative equation now differs from its continuous counterpart. Note that this can alternatively be viewed as trying to discretize the integrated non-conservative equation and use integration by parts (see also~\cite{Liebendorfer_2004} and~\cite{Cardall_moments_2013}). Indeed, using integration by parts, one would obtain the integral of the conservative equation, which is what we discretize with  the \ac{FV} method, while the second term of integration by parts is equivalent to the residuals and vanishes in the continuous limit. If these residuals do not vanish in the discrete limit, they can be interpreted as an artificial source term in the conservative equation, hence violating the actual particle number conservation.

One way to ensure both conservation and consistency is to modify the discretization of the fluxes used in the conservative equation. This is shown on the right-most diagram in Fig.~\ref{fig:consistency}, where we changed the discretization from $\hat{F}$ to $\bar{F}$. This in turn results in a different discretization for the non-conservative formulation. Our constraints are then that the new discretization make the residuals vanish and that it still be a valid discretization of the Boltzmann equation, \textit{i.e.} we retrieve the continuous equation in the continuous limit.

\subsubsection{Sources of inconsistency}

\noindent From the last section, we find that we must modify the fluxes to make the residuals vanish. To achieve this goal, there are several continuous identities that must be satisfied in the discrete limit. To identify all of these identities, we shall work further on the right-hand side of Eq.~\eqref{eq:cancel}. From the right-hand side of Eq.~\eqref{eq:cancel}, we can show that
\begin{equation}
     \po\mathcal{D}_{\curvi}\left(\frac{1}{\po} {\Gamma^{\hati}}_{\hmu\hnu} {P^{\curvi}}_{\hati} \, p^{\hmu}p^{\hnu} \right) = {\Gamma^{\hnu}}_{\hmu\hnu} \, p^{\hmu}. \label{eq:mom_div}
\end{equation}
The derivation of the last equality highlights several terms that are relevant to our problem.
We can thus further detail the left-hand side of Eq.~\eqref{eq:mom_div}, without the $\po$ factors, as
\begin{equation}
    \begin{aligned}
        \mathcal{D}_{\curvi}\left({\Gamma^{\hati}}_{\hmu\hnu} {P^{\curvi}}_{\hati} \, p^{\hmu}p^{\hnu} \right) &= -\po{\Gamma^{\hati}}_{\hmu\hnu} {P^{\curvi}}_{\hati} \, p^{\hmu}p^{\hnu} \mathcal{D}_{\curvi}\frac{1}{\po}\\
        & + {\Gamma^{\hnu}}_{\hnu\hmu} \, p^{\hmu}+ {\Gamma^{\hnu}}_{\hmu\hnu} \, p^{\hmu}\\
        & + \frac{1}{\po}{\Gamma^{\hmu}}_{\hnu\hat{0}} \, p_{\hmu}p^{\hnu}.
    \end{aligned}\label{eq:detail_dp}
\end{equation}
The second and fourth terms of the right-hand side are trivially zero, since $\Gamma_{\hrho\hnu\hmu}$ is antisymmetric over its first two indices, as seen from Eq.~\eqref{eq:com_Gamma}. The third term is given by 
\begin{equation}
    {\Gamma^{\hnu}}_{\hmu\hnu} = \nabla_\mu \tensor{L}{^\mu_\hmu}, \label{eq:tr13_Gamma}
\end{equation}
and the first term cancels with the additional derivative of $(\po)^{-1}$ coming from the full derivative in the left-hand side of Eq.~\eqref{eq:mom_div}.

To express the discrete equivalents of these previous relations, we must introduce another discrete derivative operator, $\hat{\ddiff}_{x;\mu}$, which is the discretization of the covariant derivative in position space. We stress that, in general, $\hat{\ddiff}_{x;\mu} A^\mu \neq \ddiff_{x;\mu} A^\mu$, since the four-divergence may be discretized differently from the more general covariant derivative. This is for example the case in the \ac{FV} method, where the four-divergence directly discretizes $\sqrt{-g}^{-1} \partial_\mu ( \sqrt{-g}~\cdot~)$, which cannot be directly used for the more general covariant derivative. We introduce the discretization of the Ricci rotation coefficients from Eq.~\eqref{eq:com_Gamma} as
\begin{equation}
    \tensor{\Gamma}{^{\hrho}_{\hnu\hmu}} \longrightarrow \tensor{L}{^\hrho_\nu}\tensor{L}{^\mu_\hmu} \hat{\ddiff}_{x;\mu} \tensor{L}{^\nu_\hnu} \label{eq:discr_ricci_coeff}
\end{equation}
One of the key relations to ensure vanishing residuals is the discrete equivalent of Eq.~\eqref{eq:tr13_Gamma}, \textit{i.e.}
\begin{equation}
    \tensor{\Gamma}{^{\hnu}_{\hmu\hnu}} = \hat{\ddiff}_{x;\mu} \tensor{L}{^\mu_\hmu} = \ddiff_{x;\mu}\tensor{L}{^\mu_\hmu} . \label{eq:tr13_Gamma_discr}
\end{equation}
Similarly, the discrete version of Eq.~\eqref{eq:mom_div} reads
\begin{equation}
    \ddiff_{p;\curvi}\left( {\Gamma^{\hati}}_{\hmu\hnu} {P^{\curvi}}_{\hati} \, p^{\hmu}p^{\hnu} \right) = {\Gamma^{\hnu}}_{\hmu\hnu} \, p^{\hmu} . \label{eq:mom_div_discr}
\end{equation}
We stress that the last two equations are conditions to enforce and are not true for any discretization of the derivative operators and of $\tensor{\Gamma}{^{\hrho}_{\hmu\hnu}}$.

From our discussion, we identify three main ways in which the discretization of the equation may lead to $\mathcal{R}_h$ being non-zero.
\begin{enumerate}
    \item ${\Gamma^{\hrho}}_{\hnu\hmu}$ discretization: in the continuous limit, Eq.~\eqref{eq:tr13_Gamma} is satisfied, but it is not always true for the discrete case in Eq.~\eqref{eq:tr13_Gamma_discr}. If we assume that $\ddiff_{p;\curvi} = \po \mathcal{D}_{\curvi}((\po)^{-1}~\cdot~)$, then the residuals are simply given by $\mathcal{R}_h = \ddiff_{x;\mu}\left( {L^\mu}_{\hmu}p^{\hmu} \right) - \hat{\ddiff}_{x;\mu}\left( {L^\mu}_{\hmu}p^{\hmu}\right)$. Thus, the residuals arise from a discrepancy between the different discretizations of the derivative operators. Note that this means that even if $\tensor{\Gamma}{^{\hrho}_{\hmu\hnu}}$ could be computed analytically, the residuals would still not vanish in the case where $\ddiff_{x;\mu}$ is not exact. Moreover, the antisymmetry over the first two indices must be preserved to ensure ${\Gamma^{\hnu}}_{\hnu\hmu} = 0$. If not preserved, additional terms from Eq.~\eqref{eq:detail_dp} will contribute to the residuals. Note that this consistency of discretization of derivatives was already pointed out in Ref.~\cite{Cardall_moments_2013}.
    \item Discretization $\ddiff_{p;\curvi}\left({P^{\curvi}}_{\hati} \, p^{\hmu}p^{\hnu} \right)$: in this case, the discrete condition from Eq.~\eqref{eq:mom_div_discr} may be violated, independently of the discretization of $\tensor{\Gamma}{^{\hrho}_{\hmu\hnu}}$. Since both the continuous and discrete derivative of this term can be computed, we can express their difference 
    \begin{equation}
        \tensor{\mathscr{R}}{_{\hati}^{\hmu\hnu}} = \po \mathcal{D}_{\curvi}\left(\frac{{P^{\curvi}}_{\hati} \, p^{\hmu}p^{\hnu}}{\po} \right) - \ddiff_{p;\curvi}\left({P^{\curvi}}_{\hati} \, p^{\hmu}p^{\hnu}\right). \label{eq:mom_residuals}
    \end{equation}
    As a result, assuming point 1 is solved, the residuals would take the form $\mathcal{R}_h = \tensor{\mathscr{R}}{_{\hati}^{\hmu\hnu}} \tensor{\Gamma}{^{\hati}_{\hnu\hmu}}$. Thus, contrary to point 1, the residuals arise here from a discrepancy between the discretized and continuous derivative operator.
    \item Flux reconstruction: the use of an approximate Riemann solver modifies how the flux is reconstructed at the cell interfaces. These reconstructions change the discretization of the derivative of the tetrad in the position space term, but not in ${\Gamma^{\hrho}}_{\hmu\hnu}$. Similarly to point 1, the residuals would then not vanish. This situation could also arise from the modified reconstruction needed for the diffusive limit (see Sec.~\ref{sec:diffusive_limit}). If we encapsulate the flux reconstruction in the definition of $\ddiff_{x;\mu}$, the issue becomes the same as point 1, \textit{i.e.} the residuals arise from a discrepancy with $\hat{\ddiff}_{x;\mu}$.
\end{enumerate}
In this work, we will only discuss points 1 and 2. Since, in our case, the operators $\ddiff_{x;\mu}$ and $\ddiff_{p;\curvi}$ are fixed by the \ac{FV} method to ensure number conservation, we will use the freedom of defining a discretization for $\hat{\ddiff}_{x;\mu}$ and for the momentum-space flux to ensure that the contribution to the residuals from points 1 and 2 vanishes. Throughout the rest of the paper, we shall refer to the vanilla, or naive, discretization as being $\hat{\ddiff}_{x;\mu}$ discretized using \ac{FD} for both the partial derivative and the connection coefficients.

\subsubsection{Reformulating the momentum-space flux \label{sec:mom_flux_discretization}}
\noindent In the previous section, we mentioned that the discretized Ricci rotation coefficients must preserve the properties of their continuous counterparts. With this idea in mind, we find it more convenient to re-express $\tensor{\Gamma}{^\hrho_{\hmu\hnu}}$ in a form that intrinsically enforces the vanishing trace ${\Gamma^{\hnu}}_{\hnu\hmu}$. In other words, we look for a formulation of the Ricci rotation coefficients that would be intrinsically antisymmetric over the first two indices independently of the discretization of the derivative. Such an expression can be derived by substituting Eq.~\eqref{eq:tetrad_g} into the definition of the Christoffel symbols and Eq.~\eqref{eq:com_Gamma}. We then find
\begin{align}
    \tensor{\Gamma}{^\hrho_{\hmu\hnu}} &= {L^{\hrho}}_{\mu} \, {L^{\nu}}_{\hmu} \, {L^{\rho}}_{\hnu} {\Gamma^{\mu}}_{\nu\rho} + {L^{\hrho}}_{\mu}  \, {L^{\rho}}_{\hnu} \, \partial_\rho{L^{\mu}}_{\hmu} \nonumber\\
    &= \tensor{\bar\Gamma}{^\hrho_{\hmu\hnu}} - \tensor{\bar\Gamma}{^\hrho_{\hnu\hmu}} - \tensor{\bar\Gamma}{_\hmu^\hrho_\hnu} - \tensor{\bar\Gamma}{_\hnu^\hrho_\hmu} + \tensor{\bar\Gamma}{_\hmu_\hnu^\hrho} + \tensor{\bar\Gamma}{_\hnu_\hmu^\hrho}, \label{eq:def_Gamma}
\end{align}
where we introduced
\begin{equation}
    \tensor{\bar\Gamma}{_{\hrho\hmu\hnu}} = -\frac{1}{2} \tensor{L}{_{\hrho\nu}} \tensor{L}{^\mu_\hmu} \partial_\mu \tensor{L}{^\nu_\hnu} . \label{eq:Gamma_bar}
\end{equation}
This formulation of $\tensor{\Gamma}{^\hrho_{\hmu\hnu}}$ indeed ensures that the antisymmetry property holds independently of the discretization used for the derivative $\partial_\mu \tensor{L}{^\nu_{\hnu}}$ in Eq.~\eqref{eq:Gamma_bar}, since all $\tensor{\bar\Gamma}{_{\hrho\hmu\hnu}}$ are discretized the same way and thus cancel exactly in the discrete limit. This property is the main advantage compared to computing $\nabla_\mu \tensor{L}{^\nu_\hnu}$ as the partial derivative and Christoffel symbols separately. Modifications to their discretizations separately may not straightforwardly satisfy the antisymmetry property. With the expression~\eqref{eq:def_Gamma}, further discretization corrections to $\tensor{\Gamma}{^\hrho_{\hmu\hnu}}$ must solely focus on the discretization of $\partial_\mu \tensor{L}{^\nu_{\hnu}}$ and the tetrad itself.

\subsubsection{Discretization of the tetrad derivative\label{sec:discr_tetrad_diff}}
\noindent Two key requirements stem from Point 1: the trace of the discrete covariant derivative $\hat{\ddiff}_{x;\mu}\tensor{L}{^\nu_\hnu}$ must match the discrete four-divergence $\ddiff_{x;\mu} \tensor{L}{^\mu_\hnu}$, and one must choose $\hat{\ddiff}_{x;\mu}$ such that $\tensor{\Gamma}{_{\hrho\hnu\hmu}}$ is antisymmetric over its first two indices. The latter property is already satisfied from the expression in Eq.~\eqref{eq:def_Gamma} provided in the previous section. In other words, finding the proper discretization for $\hat{\ddiff}_{x;\mu} \tensor{L}{^\nu_\hnu}$ reduces to finding the appropriate discretization of $\partial_\mu \tensor{L}{^\nu_\hnu}$, which will be the focus of this section. We remind the reader that we are working under the assumption that $f$ is constant. We shall also neglect the effect of the discretization of the momentum space derivative and assume that it is exact.

Before discussing our discretization scheme, we will give a concrete example for the residuals in spherical symmetry and briefly describe the solutions proposed in the literature.
We consider an empty Minkowski space in spherical coordinates. Assuming a uniform grid, the \ac{FV} discretization of the spatial divergence of the tetrad is given by
\begin{equation}
    \ddiff_{x;j}\left( {L^j}_{\hmu}p^{\hmu} \right) = \frac{2 \cos\vartheta}{r_i} \frac{1}{1+a_i^2/3}, \label{eq:discr_mink_spherical}
\end{equation}
where we defined
\begin{equation}
    a_i = \frac{\Delta r}{2 r_i}. \label{eq:a}
\end{equation}
If we use the vanilla discretization of the Ricci rotation coefficients and neglect the boundary treatment, the residuals are
\begin{equation}
    \mathcal{R}_i = \frac{2 \cos\vartheta}{r_i} \frac{4 a_i^2}{(3+a_i^2)(1-a_i^2)} \label{eq:residuals_spherical}
\end{equation}
Ignoring the boundary $r_i = \Delta r /2$ ($a_i=1$), we see that the residuals vanish at large $r$ (or $i$), as $a$ tends to $0$. For the first few points ($a_i \rightarrow 1$), however, the residuals can be significant at small radii. These residuals therefore introduce artificial source terms for $f$ when approaching $r=0$ that are not present in the Boltzmann equation. An illustration of their effect is shown in Sec.~\ref{sec:comparison_vanilla}. Therefore, some modifications to the discretization are needed to suppress the residuals.

Such a modification was proposed in Ref.~\cite{Mezzacappa_1993}, later extended by Ref.~\cite{Liebendorfer_2004} for curved spacetimes, and used in different codes and formalisms~(\textit{e.g.},~\cite{Sumiyoshi_2012}). In Minkowski space, the conservative Boltzmann equation reads
\begin{equation}
    \partial_t f + \frac{1}{r^2}\frac{\partial}{\partial r} (r^2 \cos\vartheta f) - \frac{1}{r \sin\vartheta} \frac{\partial}{\partial\vartheta} \big( \sin^2\vartheta f \big) = 0. \label{eq:mink_spherical}
\end{equation}
The discretization proposed by Ref.~\cite{Mezzacappa_1993} is to discretize the $1/r$ term as 
\begin{equation}
     \frac{1}{r} \longrightarrow \frac{3}{2} \frac{r^2_{i+1/2} - r^2_{i-1/2}}{r^3_{i+1/2} - r^3_{i-1/2}}  \label{eq:mod_1/r}.
\end{equation}
In fact, this corresponds to a discretization of $1/r$ expressed as
\begin{equation*}
    \frac{1}{r} = \frac{1}{2} \frac{1}{r^2}\frac{\partial}{\partial r}(r^2) = \frac{3}{2} \frac{\partial}{\partial r^3}{r^2}.
\end{equation*}
We can show that this discretization is consistent with our requirements of vanishing residuals. The condition given in Eq.~\eqref{eq:cancel} is, in this case,
\begin{equation}
    \frac{1}{r^2}\frac{\partial}{\partial r} (r^2 \cos\vartheta) = \frac{1}{r \sin\vartheta} \frac{\partial}{\partial\vartheta} \big( \sin^2\vartheta \big). \label{eq:cancel_spherical_1d}
\end{equation}
This is true in the continuous limit, and must be enforced in the discrete limit. The \ac{FV} discretization of the left-hand side, neglecting the $\cos\vartheta$ term, is
\begin{equation}
    \frac{1}{r^2}\frac{\partial}{\partial r} r^2 \longrightarrow 3 \frac{r^2_{i+1/2} - r^2_{i-1/2}}{r^3_{i+1/2} - r^3_{i-1/2}} \label{eq:FV_dr}
\end{equation}
Assuming the $\vartheta$ derivative is exact, Eq.~\eqref{eq:cancel_spherical_1d} becomes a discretization rule for $1/r$, which matches Eq.~\eqref{eq:mod_1/r}. Hence, Eq.~\eqref{eq:cancel} is satisfied in the discrete limit and the residuals vanish.

In our case, however, the $1/r$ term does not appear explicitly, but arises as a more complex computation of $\tensor{\Gamma}{^\hrho_{\hmu\hnu}}$. As a consequence, we cannot straightforwardly apply this modification and we must generalize it in terms of the components of $\tensor{\Gamma}{^\hrho_{\hmu\hnu}}$.

Therefore, we will now proceed to the derivation of a discretization scheme that makes the residuals vanish under the considered assumptions.
To remain more general, we do not consider the reference metric formalism and use $\gamma$ instead of $\hat\gamma$. The \ac{FV} discretized operator is (no sum)
\begin{align}
   \ddiff_{j} F^j &=\frac{(F^j)_{i+1/2} (\Delta A_j)_{i+1/2} - (F^j)_{i-1/2} (\Delta A_j)_{i-1/2}}{\Delta V_i} \nonumber\\
   & = \frac{((\Delta A_j)_i)_{+,j} ((F^j)_i)_{-,j}}{\Delta V_i} + \frac{((\Delta A_j)_i)_{-,j} ((F^j)_i)_{+,j}}{\Delta V_i}, \label{eq:discr_diff}
\end{align}
where we introduced the notations
\begin{align}
    &(~Q_i~)_{+,j} \equiv \frac{Q_{i+1/2} + Q_{i-1/2} }{2}, \label{eq:plus_notation}\\
    &(~Q_i~)_{-,j} \equiv Q_{i+1/2} - Q_{i-1/2}.  \label{eq:minus_notation}
\end{align}
The index $i$ represents the index along the axis $j$, whereas all the other indices are omitted for simplicity and correspond to cell centres. All area elements are defined in Eq.~\eqref{eq:Ax} for the position space and Eq.~\eqref{eq:Ap} and Appendix~\ref{app:area_and_volume} for the momentum space.
For the time component, we consider the forward Euler method, \textit{i.e.}
\begin{equation}
    \ddiff_0 q = \frac{q^{(n+1)} - q^{(n)}}{\Delta t},
\end{equation}
where $n$ indicates the time step.
The second equality in Eq.~\eqref{eq:discr_diff} indicates that the first term corresponds to the divergence of the flux $\partial_\mu F^\mu$, whereas the second corresponds to the ``connection term'' $F^i \tensor{{\Gamma}}{^j_i_j}$, where $\tensor{{\Gamma}}{^k_i_j}$ are the Christoffel symbols. The change from the first to the second equality is a discrete equivalent of the product rule.

To extend the solution proposed in the literature to more general cases, we look for a discretization of $\partial_\mu \tensor{L}{^\mu_\hmu}$ in $\tensor{\bar\Gamma}{_{\hrho\hmu\hnu}}$, defined in Eq.~\eqref{eq:Gamma_bar}, that satisfies Eq.~\eqref{eq:tr13_Gamma_discr}. Thus, we express the trace $\tensor{\Gamma}{^\hnu_{\hmu\hnu}}$ as a function of $\tensor{\bar\Gamma}{_{\hrho\hmu\hnu}}$ as
\begin{equation}
    {\Gamma^{\hnu}}_{\hmu\hnu} = 2 \left(\tensor{\bar\Gamma}{^{\hnu}_{\hmu\hnu}} - \tensor{\bar\Gamma}{^{\hnu}_{\hnu\hmu}} \right).\label{eq:tr13_Gamma_detail}
\end{equation}
The first term on the right-hand side
\begin{equation}
    2\tensor{\bar\Gamma}{^{\hnu}_{\hmu\hnu}} = -\tensor{L}{^\mu_\hmu} \tensor{L}{^\hnu_\nu} \partial_\mu \tensor{L}{^\nu_\hnu} = \tensor{L}{^\mu_\hmu} \tensor{\Gamma}{^\nu_{\mu\nu}} \label{eq:tr13_Gamma_bar}
\end{equation}
is the connection term. It is thus related to the derivative of the metric. The second term
\begin{equation}
    2\tensor{\bar\Gamma}{^{\hnu}_{\hnu\hmu}} = -\partial_\mu\tensor{L}{^\mu_\hmu} \label{eq:tr12_Gamma_bar}
\end{equation}
is the divergence term. We see that both terms arise from different traces of $\tensor{\bar\Gamma}{_{\hrho\hnu\hmu}}$. The modification of the discretization of the derivatives $\partial_\mu\tensor{L}{^\nu_\hnu}$ must however ensure that both terms get the appropriate \ac{FV} discretization given in Eq.~\eqref{eq:discr_diff}. We can use the fact that our definition of $\tensor{\bar\Gamma}{_{\hrho\hmu\hnu}}$ makes the divergence $\partial_\mu\tensor{L}{^\mu_\hmu}$ appear exactly in the trace~\eqref{eq:tr12_Gamma_bar} to replace it by the four-divergence $\nabla_\mu \tensor{L}{^\mu_\hmu}$. We thus rewrite the derivatives as
\begin{equation}
    \partial_\mu \tensor{L}{^\nu_\hnu} = \frac{1}{\alpha \sqrt{\gamma}} \partial_\mu\left(\alpha\sqrt{\gamma}\tensor{L}{^\nu_\hnu}\right) - \frac{\tensor{L}{^\nu_\hnu}}{\alpha \sqrt{\gamma}}\partial_\mu\left(\alpha\sqrt{\gamma}\right). \label{eq:modified_discr}
\end{equation}
The first term of the right-hand side of the equation will then indeed be equal to  $\nabla_\mu \tensor{L}{^\mu_\hmu} = \sqrt{-g}^{~-1}\partial_\mu(\sqrt{-g} \tensor{L}{^\mu_\hmu})$ when computing the trace~\eqref{eq:tr12_Gamma_bar}. Therefore, we can discretize the derivative of the diagonal components $\mu = \nu$ using the \ac{FV} method, that is, using $\ddiff_{x;\mu}$. To ensure consistent additions or cancellations in the computation of the connection terms in Eq.~\eqref{eq:tr13_Gamma_bar}, we also discretize the components $\mu \neq \nu$ with an \ac{FV}-like expression. Thus, we write
\begin{equation}
    \frac{1}{\alpha \sqrt{\gamma}} \partial_\mu\left(\alpha\sqrt{\gamma}\tensor{L}{^\nu_\hnu}\right) \rightarrow \ddiff_{x;\mu} \tensor{L}{^\nu_\hnu}, \label{eq:FV_like_discr}
\end{equation}
where we introduced a generalization of Eq.~\eqref{eq:discr_diff} 
\begin{equation}
    \ddiff_{j} F^\nu =\frac{(F^\nu \Delta A_j)_{i+1/2} - (F^\nu \Delta A_j)_{i-1/2}}{\Delta V_i}  \label{eq:generalized_discr_diff} \\
\end{equation}
Note that contrary to Eq.~\eqref{eq:discr_diff} we compute $(F^\nu \Delta A_j)_{i+1/2}$ instead of $(F^\nu)_{i+1/2} (\Delta A_j)_{i+1/2}$. We shall return to this subtlety later and show that the two discretizations are the same for the diagonal elements in our case.

This new discretization in Eq.~\eqref{eq:generalized_discr_diff} satisfies $\tensor{\Gamma}{^\hmu_{\hnu\hmu}} = \ddiff_{x;\mu} \tensor{L}{^\mu_\hnu}$ provided the extra terms vanish. We can then discretize the second term in Eq.~\eqref{eq:modified_discr} to satisfy this condition. First, we insert Eq.~\eqref{eq:modified_discr} into Eq.~\eqref{eq:tr13_Gamma_detail}, which yields
\begin{equation}
    \begin{split}
    {\Gamma^{\hnu}}_{\hmu\hnu} & = -\frac{\tensor{L}{^\mu_\hmu} \tensor{L}{^\hnu_\nu}}{\alpha \sqrt{\gamma}} \partial_\mu\left(\alpha\sqrt{\gamma}\tensor{L}{^\nu_\hnu}\right) + 4\frac{\tensor{L}{^\mu_\hmu}}{\alpha \sqrt{\gamma}}\partial_\mu\left(\alpha\sqrt{\gamma}\right)\\
    &+ \frac{1}{\alpha \sqrt{\gamma}} \partial_\mu\left(\alpha\sqrt{\gamma}\tensor{L}{^\mu_\hmu}\right) - \frac{\tensor{L}{^\mu_\hmu} }{\alpha \sqrt{\gamma}}\partial_\mu\left(\alpha\sqrt{\gamma}\right).
    \end{split} \label{eq:new_tr13_Gamma}
\end{equation}
Because the trace is given by Eq.~\eqref{eq:tr13_Gamma}, we get the equality
\begin{equation}
    \frac{1}{\alpha \sqrt{\gamma}}\partial_\mu\left(\alpha\sqrt{\gamma}\right) = \frac{\tensor{L}{^\hnu_\nu}}{3\alpha \sqrt{\gamma}} \partial_\mu\left(\alpha\sqrt{\gamma}\tensor{L}{^\nu_\hnu}\right). \label{eq:ddiff_alp_psi}
\end{equation}
This relation can then be used to discretize the second term in Eq.~\eqref{eq:modified_discr}. We thus get the discretization
\begin{equation}
    \frac{1}{\alpha \sqrt{\gamma}}\partial_\mu\left(\alpha\sqrt{\gamma}\right) \rightarrow \frac{\tensor{L}{^\hnu_\nu}}{3} \ddiff_{x;\mu}\tensor{L}{^\nu_\hnu}.
\end{equation}
In practice, we define a new variable
\begin{equation}
    \tensor{\widetilde{\Gamma}}{_{\hrho\hmu\hnu}} = -\frac{1}{2} \frac{\tensor{L}{_{\hrho\nu}} \tensor{L}{^\mu_\hmu}} {\alpha\sqrt{\gamma}}\partial_\mu \left( \alpha \sqrt{\gamma} \tensor{L}{^\nu_\hnu} \right). \label{eq:Gamma_tilde}
\end{equation}
Using Eqs.~\eqref{eq:Gamma_bar}, \eqref{eq:modified_discr} and \eqref{eq:ddiff_alp_psi}, we can then express $\tensor{\bar{\Gamma}}{_{\hrho\hmu\hnu}}$ as 
\begin{equation}
    \tensor{\bar{\Gamma}}{_{\hrho\hmu\hnu}} = \tensor{\widetilde{\Gamma}}{_{\hrho\hmu\hnu}} - \frac{1}{3} \eta_{\hrho\hnu} \tensor{\widetilde{\Gamma}}{^{\halp}_{\hmu\halp}} \label{eq:Gamma_bar_function_Gamma_tilde}
\end{equation}
Inserting this definition into Eq.~\eqref{eq:tr13_Gamma_detail}, we find
\begin{equation}
    \frac{\tensor{\Gamma}{^\hnu_{\hmu\hnu}}}{2} = \tensor{\widetilde{\Gamma}}{^{\hnu}_{\hmu\hnu}} - \frac{4}{3} \tensor{\widetilde{\Gamma}}{^{\halp}_{\hmu\halp}} - \tensor{\widetilde{\Gamma}}{^{\hnu}_{\hnu\hmu}} + \frac{1}{3} \tensor{\widetilde{\Gamma}}{^{\halp}_{\hmu\halp}} = -\tensor{\widetilde{\Gamma}}{^{\hnu}_{\hnu\hmu}} \label{eq:tr_13_Gamma_with_Gamma_tilde}
\end{equation}
If we discretize the derivative in $\tensor{\widetilde{\Gamma}}{_{\hrho\hmu\hnu}}$ using Eq.~\eqref{eq:FV_like_discr}, we find the desired relation
\begin{equation*}
    {\Gamma^{\hnu}}_{\hmu\hnu} = -2\tensor{\widetilde{\Gamma}}{^{\hnu}_{\hnu\hmu}} = \ddiff_{x;\mu}\tensor{L}{^\mu_\hmu},
\end{equation*}
which ensures that the residuals vanish. Note that we applied our reasoning to the \ac{FV} method, but it can in principle be extended to other methods. The only part that must be modified is the generalization of the $\ddiff_{x,j}$ operator in Eq.~\eqref{eq:generalized_discr_diff}. Also, we did not make use of the reference metric formalism. This discretization scheme should thus also hold for more general metrics. However, our derivation assumes that Eq.~\eqref{eq:ddiff_alp_psi} holds in the discrete limit, which is not true for our discretization. This discrepancy between the continuous and the discrete limits introduces artificial advection in momentum space if $f$ is not homogeneous.

Note that additional care must be taken in our case in spherical and cylindrical coordinates around irregular points. This is in particular the reason for the discretization of Eq.~\eqref{eq:generalized_discr_diff} using $(F^\nu \Delta A_j)_{i+1/2}$ instead of $(F^\nu)_{i+1/2} (\Delta A_j)_{i+1/2}$.
The comoving tetrad can also be written as a boosted orthonormal tetrad, \textit{i.e.}
\begin{equation}
    \tensor{L}{^\mu_\hmu} = \tensor{e}{^\mu_\bmu}~\tensor{D}{^\bmu_\hmu},
\end{equation}
where indices adorned with a bar correspond to the local orthonormal frame, $\tensor{e}{^\mu_\bmu}$ is the tetrad to the local orthonormal frame and $\tensor{D}{^\bmu_\hmu}$ is the (product of) Lorentz transformation(s) between the orthonormal and fluid rest frames. In the \texttt{Gmunu} framework, the tetrad $\tensor{e}{^\mu_\bmu}$ is given by
\begin{equation}
    \tensor{e}{^\mu_\bmu} = \begin{bmatrix} 
     \frac{1}{\alpha} & 0 & 0 & 0 \\
     -\frac{\beta^1}{\alpha} & \frac{1}{\psi^2 \sqrt{\hat{\gamma}_{11}}} & 0 & 0\\
     -\frac{\beta^2}{\alpha} & 0 & \frac{1}{\psi^2 \sqrt{\hat{\gamma}_{22}}} & 0\\
     -\frac{\beta^3}{\alpha} & 0 & 0 & \frac{1}{\psi^2 \sqrt{\hat{\gamma}_{33}}}
     \end{bmatrix}
\end{equation}
where, as a reminder, $\hat{\gamma}_{ij}$ is the analytical, diagonal and time-independent reference metric. In fact, the largest discrepancy between the discretization $\hat{\ddiff}_{x;\mu}$ and $\ddiff_{x;\mu}$ comes from $\tensor{e}{^\mu_\bmu}$, since $\tensor{D}{^\bmu_\hmu}$ contains velocities and does not contribute to the connection coefficients. The tetrad $\tensor{e}{^\mu_\bmu}$, on the other hand, contains the geometrical factors, such as the $1/r$ factor in spherical coordinates that was discussed at the beginning of this section. At the irregular points, the spatial area elements vanish, which ensures that the value of the flux $F^i$ at these points does not contribute to the evolution. In fact, $\Delta A_i F^i$ at these irregular points is always analytically zero in our case, because the tetrad $\tensor{e}{^i_\bmu}$ in the flux does not contain any factor that diverges in spherical, cylindrical and Cartesian coordinates.
When extending our operator with Eq.~\eqref{eq:generalized_discr_diff}, we are now introducing terms that do not vanish at the boundary and should then be treated appropriately. In cylindrical coordinates, we multiply each component of $\tensor{L}{^\mu_\hmu}$, and hence of $\tensor{e}{^\mu_\bmu}$, by $\Delta A_R \propto R$. The element $\tensor{e}{^\phi_{\bar{\phi}}} = R^{-1}$ would become a constant term $1$, while all the other elements then contain an $R$ factor that vanishes at $R=0$ (provided $\psi$ and $\alpha$ are regular at that point). Therefore, we must discretize the product $\tensor{e}{^\phi_{\bar{\phi}}} \Delta A_R$ as $(\tensor{e}{^\phi_{\bar{\phi}}} \Delta A_R)_{i+1/2}$ to ensure that the discretized value corresponds to the continuous one. This discretization is exactly the same as using $(F^\nu)_{i+1/2} (\Delta A_j)_{i+1/2}$ for all the other elements, since they do not explicitly contain any $R$ factors. A similar reasoning can be applied to the terms with $\Delta A_\theta$ in spherical coordinates. For $\Delta A_r$ in spherical coordinates, all the elements of $(\tensor{e}{^\mu_{\bmu}} \Delta A_r)$ vanish at $r=0$ and the two discretizations are equivalent.

We finish our discussion on the discretization of the tetrad derivative by showing that our discretization retrieves the discretization suggested by Ref.~\cite{Mezzacappa_1993}, as discussed above, for an empty Minkowski space in spherical coordinates. This discretization consisted in discretizing the $2/r$ terms from the momentum-space flux using Eq.~\eqref{eq:mod_1/r}. In our case, this term arises from the trace ${\Gamma^{\hnu}}_{\hat{1}\hnu}$. Using our modified discretization and the fact that the tetrad is diagonal in this case, we can then rewrite the discretization of $2/r$ as
\begin{equation}
    \frac{2}{r} = \tensor{\Gamma}{^{\hnu}_{\hat{1}\hnu}} \longrightarrow \ddiff_{x;\mu} \tensor{L}{^{\mu}_{\hat{1}}} = \ddiff_{x;r} \tensor{L}{^r_{\hat{1}}}
\end{equation}
If we insert Eq.~\eqref{eq:discr_diff} with the area and volume elements associated with $r$ into the last equation, we find
\begin{equation}
    \frac{2}{r} \longrightarrow \ddiff_{x;r} \tensor{L}{^r_{\hat{1}}} = 3 \frac{r^2_{i+1/2} - r^2_{i-1/2}}{r^3_{i+1/2} - r^3_{i-1/2}}.
\end{equation}
This discretization is then the same as the position space one in Eq.~\eqref{eq:FV_dr}, \textit{i.e.} we retrieve the correction suggested in previous works and the residuals vanish.

\subsubsection{Momentum-space derivative of the flux \label{sec:mom_diff_discr}}
\noindent We now turn our attention to the discretization of the momentum space derivative (Point 2). As a reminder, we want to make the residuals $\tensor{\mathscr{R}}{_{\hati}^{\hmu\hnu}}$ defined in Eq.~\eqref{eq:mom_residuals} vanish. As previously mentioned, these residuals can be expressed analytically. Our approach is thus to add second-order corrections $\tensor{\mathscr{C}}{^\curvi_{\hati}^{\hmu\hnu}}$ such that 
\begin{equation}
    \po \mathcal{D}_{\curvi}\left(\frac{{P^{\curvi}}_{\hati} \, p^{\hmu}p^{\hnu}}{\po} \right) - \ddiff_{p;\curvi}\left({P^{\curvi}}_{\hati} \, p^{\hmu}p^{\hnu} +\tensor{\mathscr{C}}{^\curvi_{\hati}^{\hmu\hnu}} \right) = 0.
\end{equation}
After some tedious algebra, an analytical expression for each component of the correction can be found. The current expressions however assume that the grid is uniform. Moreover, even though this approach does make the residuals vanish, artificial momentum advection terms of the order of the correction are introduced. Indeed, the term $\tensor{\mathscr{C}}{^\curvi_{\hati}^{\hmu\hnu}}$ will also multiply the derivative of $f$. We thus shift the problem from having residuals, acting as a source term for $f$, to having spurious momentum advection. An improvement of this discretization correction is left as future work. We also note that no correction must be applied for $\curvi = \varepsilon$ if cell-centroid values are used instead of cell-centre values. The cell centroid $\varepsilon_C$ corresponding to the volume $dV_p = \varepsilon \sin\vartheta d\varepsilon d\vartheta d\varphi$ is given by
\begin{equation}
    \varepsilon_{C,i} = \varepsilon_i + \frac{(\Delta \varepsilon_i)^2}{12 \varepsilon_i}.
\end{equation}
The \ac{FV} discretization of the derivative $\varepsilon^{-1} \partial_\varepsilon \varepsilon^3$ yields $3\varepsilon_C$, as expected in the continuous case.

\subsubsection{Comparison with vanilla discretization \label{sec:comparison_vanilla}}

\begin{figure}
    \centering
    \includegraphics[width=\linewidth]{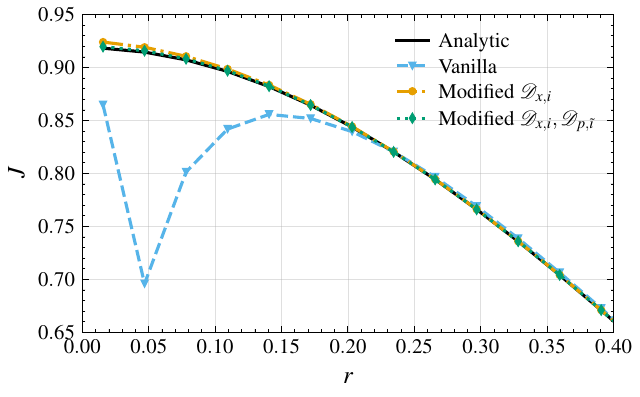}
    \caption{Comparison of the simulation of the diffusion of a Gaussian profile with the vanilla \ac{FV} discretization (blue dashed line), modified discrete spatial covariant derivative $\ddiff_{x,i}$ (orange dash-dotted line), and modified discrete spatial and momentum-space covariant derivatives $\ddiff_{x,i}$ and $\ddiff_{p,i}$ (green dotted line). We can see that the vanilla discretization yields wrong results at small radius, as expected. The corrected discretizations remove the effect of the residuals present in the vanilla discretization.} 
    \label{fig:diffusive_discr}
\end{figure}

\noindent In this section, we illustrate the effect of the non-vanishing residuals in the vanilla discretization and demonstrate the ability of our modified discretization to at least minimize their effect. To this end, we consider the diffusion of a Gaussian profile in a medium with a large scattering opacity in spherical coordinates (see later Sec.~\ref{sec:tests_diffusive}). In Fig.~\ref{fig:diffusive_discr}, we compare the analytical solution to the prediction of our solver with the vanilla \ac{FV} discretization and the modified discretization. We split the discretization modifications into two parts: the modified tetrad derivative only, and the full modification that includes the modified derivative in momentum space. In this case, the residuals are due to a mismatch between the discretizations of the connection terms, since $\partial_r {L^r}_{\hmu} = 0$. We can see that without the modified discretization, the residuals induce a significant error close to $r=0$, as expected from Eq.~\eqref{eq:residuals_spherical}. We also show that the accuracy of the solver is improved when adding a correction to the covariant derivative in momentum space, as explained in Sec.~\ref{sec:mom_diff_discr}. Note that our modifications were designed for homogeneous $f$ and may thus have an effect when $\partial f$ is significant. However, we do not probe this regime in this test because $f$ is isotropic in momentum space and $\partial_r f$ is small close to $r=0$, where the modifications are the largest. Despite the limitations of our corrected discretization, we thus find that it suppresses the artifacts induced by non-vanishing residuals from the vanilla discretization.

\subsubsection{Limitations and improvements of the discretization scheme \label{sec:limitation_and_improvements_discr}}

\noindent To end the discussion on consistency, we summarize and discuss the limitations of our discretization scheme mentioned throughout the previous sections and suggest improvements for future work.

Our discretization scheme was derived using several assumptions. First, we assume $f$ to be constant, at least spatially. This assumption captures optically thick regions, which are typically located at the centre of the domain inside neutron stars. This region is also where the residuals are the largest in spherical coordinates because of geometrical factors ($1/r$). The constant $f$ assumption discards any intricacy related to the discrete product rule, which may only be relevant close to large gradients in the distribution function.

Second, our derivation assumes Eq.~\eqref{eq:ddiff_alp_psi} to hold in the discrete limit, which is not true in practice as mentioned in Sec.~\ref{sec:discr_tetrad_diff}. Even though Eq.~\eqref{eq:tr13_Gamma_discr} is always true with our discretization scheme, its derivation relies on Eq.~\eqref{eq:ddiff_alp_psi} to ensure consistency between the initial formulation of $\tensor{\bar\Gamma}{^\hrho_{\hnu\hmu}}$ and the final formulation in Eq.~\eqref{eq:Gamma_bar_function_Gamma_tilde}. Because Eq.~\eqref{eq:ddiff_alp_psi} does not hold exactly in the discrete limit, the difference produces secondary residuals that do not contribute to the trace $\tensor{\Gamma}{^\hmu_{\hnu\hmu}}$, since our discretization is designed to enforce Eq.~\eqref{eq:tr13_Gamma_discr}. However, these residuals do contribute to the advection in momentum space if $f$ is not constant. The secondary residuals would indeed multiply the term corresponding to the derivative of $f$. This artificial advection is a similar problem to the momentum space derivative correction discussed in Sec.~\ref{sec:mom_diff_discr}. Note that the comments from Sec.~\ref{sec:conservative_vs_consistent} still hold. The artificial advection does not alter number conservation (unless neutrinos exit the domain), but only how number is distributed among cells. However, we do expect this artificial advection to contribute to violations of four-momentum conservation. We stress that such four-momentum violations are also present in the vanilla discretization. To reduce the secondary residuals, one could modify the discretization of the elements of $\ddiff_{x,\mu} \tensor{L}{^\nu_\hnu}$ that do not contribute to the four-divergence, and that of the tetrad itself in $\tensor{\Gamma}{^\hrho_{\hnu\hmu}}$. To be fully consistent, our discretization of $\tensor{\widetilde\Gamma}{^\hrho_{\hnu\hmu}}$ should be expressed as
\begin{equation*}
    \tensor{\widetilde\Gamma}{_{\hrho\hnu\hmu}} \rightarrow -\frac{1}{2} \tensor{\hat L}{_{\hrho\nu}} \tensor{\bar L}{^\mu_\hmu} \ddiff_{x,\mu} \left( \tensor{L}{^\nu_\hnu} \right),
\end{equation*}
where $\hat{L}$ and $\bar{L}$ are discretizations of the tetrad to be determined to satisfy Eq.~\eqref{eq:ddiff_alp_psi} from the trace $\tensor{\widetilde\Gamma}{^{\hnu}_{\hmu\hnu}}$ while preserving $\tensor{\hat L}{_{\hnu\nu}} \tensor{\bar L}{^{\mu\hmu}} = \tensor{\delta}{^\mu_\nu}$, which is implicitly used in Eq.~\eqref{eq:tr_13_Gamma_with_Gamma_tilde}.

Third, our correction to the discretization of the momentum-space flux derivative is only valid for uniform grids and introduces artificial advection similarly to the previous point.

Finally, none of the proposed modifications accounts for the difference in divergence discretization arising from Riemann solvers or from the diffusive limit treatment. The changes are also not designed to satisfy energy conservation. Further improvements to the discretization must address these three points.

\begin{figure*}
    \centering
    \includegraphics[width=\linewidth]{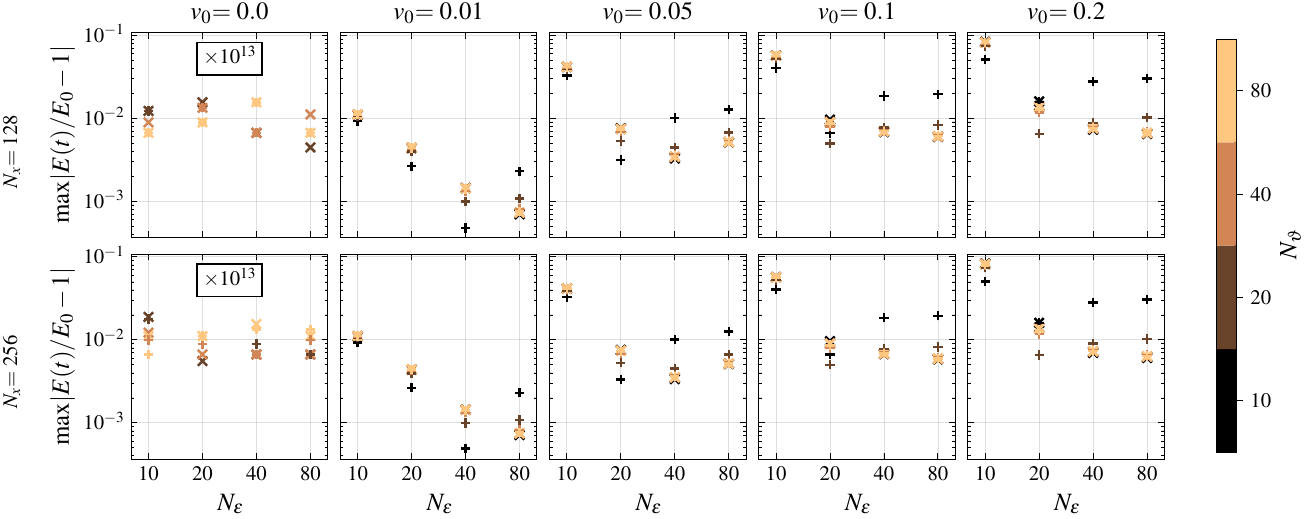}
    \caption{Maximum violation of total energy ($\times$) and momentum ($+$) for a neutrino packet propagating in a 1D Minkowski space in Cartesian coordinates with a velocity background described in Eq.~\eqref{eq:cons_velocity}. The violations are stronger when velocity is larger, inducing stronger momentum advection. Both energy and momentum are conserved in the absence of fluid.}
    \label{fig:cons_cart}
\end{figure*}

\subsection{Conservation laws \label{sec:conservation_laws}}

\noindent  In the previous sections, we have discussed in detail how to discretize our number-conservative formulation of the Boltzmann equation such that it is consistent with the non-conservative formulation. We have however omitted that our discretization should also be compatible with a discrete form of the energy-momentum-conservative formulations of the Boltzmann equation. In the continuous limit, the different conservative formulations are indeed redundant~\cite{Cardall_2013}. In the discrete limit, similarly to the reasoning used in Sec.~\ref{sec:consistency}, some key relations may not hold in the discrete limit. While consistency with the non-conservative formulation ensures correct advective behaviour, consistency with energy-momentum-conservative formulations ensures energy and momentum conservation. Deriving a discretization that would ensure this consistency is beyond the scope of this paper. Nevertheless, we investigate the current performance of the solver in terms of energy and momentum conservation in flat spacetime. Our investigation will thus be limited to simple and controlled test cases to identify possible sources of non-conservation. Note that we expect the particle number to be conserved to machine precision in pure advection since our solver uses the \ac{FV} method on the number-conservative equation. We summarize and discuss our findings in Sec.~\ref{sec:cons_summary}.

\subsubsection{Energy and momentum conservation in Cartesian coordinates}

\noindent  We start our investigation with particles propagating in a $1$D flat spacetime described by the Minkowski metric in Cartesian coordinates. We assume axisymmetry in the momentum space. Then, we add a background fluid velocity field and neglect its action on the metric. In this context, both the neutrino number and energy are constant. This specific setup is used to study the effect of velocity-induced advection without the pollution of geometric momentum advection terms present in spherical and cylindrical coordinates. We set the distribution function to $0$ everywhere, except at the phase-space position of the source $x_s = 10$, $\varepsilon_s = 75$MeV and $\vartheta_s = 0.999\pi$ (bin closest to $\pi$), where we set it to $1$. We consider a velocity profile
\begin{equation}
    v(x) = v_0 \, e^{-\frac{(x-x_0)^2}{\sigma^2}}, \label{eq:cons_velocity}
\end{equation}
where $v_0$, $x_0$ and $\sigma$ are real parameters. In the following tests, we choose $v_0 \in \{0, 0.01, 0.05, 0.1, 0.2\}$, $x_0 = 0$ and $\sigma = 4$. We repeat our test for all the values of $v_0$, as well as for different resolutions. We take $N_{x} \in \{128, 256\}$ and $N_\varepsilon \, ,~N_\vartheta \in \{10, 20, 40, 80\}$. We can then study the dependence of four-momentum conservation on each dimension. The phase-space domain for this test is $[-20,20] \times [0,300]\,\mathrm{MeV} \times [0,\pi]$ (for space, energy and $\vartheta$).

In Fig.~\ref{fig:cons_cart}, we show the maximum deviation from energy and momentum conservation as a function of spatial, energy and angular resolutions for different amplitudes $v_0$ of the velocity profile. A first observation is that our discretization is compatible with energy and momentum conservation in the absence of velocity. Then, we see that the maximum deviation increases with velocity. At all velocities larger than $0.01$, the maximum deviation saturates and does not decrease significantly with higher angular or energy resolution. This highlights the presence of a source of non-conservation that only dominates at high resolutions. Finally, the amplitude of the violations does not depend significantly on the spatial resolution. This suggests that the non-conservation is not dominated by the accuracy of the velocity gradient.
\begin{figure}[h]
    \centering
    \includegraphics[width=\linewidth]{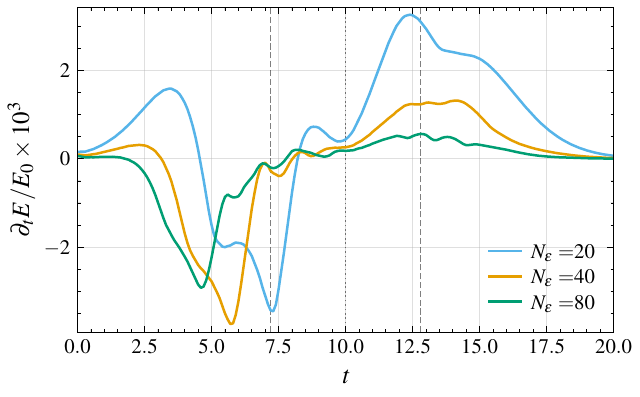}
    \caption{Change in total energy as a function of time. We show the result for $N_x = 256$, $N_\vartheta = 80$ and $v_0 = 0.2$. The dashed vertical lines represent maxima in $|\partial_x v|$ and the vertical dotted line where $\partial_x v = 0$. For $N_\varepsilon=20$ (blue line), the change in energy is correlated with the change in velocity gradient. This suggests that the violations of energy conservation are sourced by the momentum advection induced by velocity gradients. As $N_\varepsilon$ increases (orange and green lines), the first peak shifts away from the maximum velocity gradient. This suggests that the conservation violations are dominated by another phenomenon at higher resolutions.}
    \label{fig:cons_dedt_cart}
\end{figure} 

Because energy is conserved up to $\sim 10^{-15}$ in the absence of fluid, the main source of non-conservation is the advection in momentum space induced by velocity gradients. To support this conclusion, we show the derivative of the total energy as a function of time in Fig.~\ref{fig:cons_dedt_cart} for $N_x = 256$, $N_\vartheta = 80$, $v_0 = 0.2$, and $N_\varepsilon = 20$, $40$ and $80$. We also show the position of the maximum velocity gradients (dashed vertical lines) and of vanishing gradient (dotted vertical line). We can see that, for $N_\varepsilon = 20$ (blue line), the maximum change in energy occurs at the maximum values of the velocity gradient, which corresponds to a maximum in momentum advection. This suggests that momentum advection is one of the main contributors to the non-conservation. The discretization of the energy advection must therefore be further refined to be consistent with energy conservation. As $N_\varepsilon$ increases, however, we see that the first peak in energy change shifts away from the maximum of the velocity gradient. This highlights the presence of another phenomenon that starts to dominate at higher resolution, as also suggested by the saturation of the deviations in Fig.~\ref{fig:cons_cart}.

\begin{figure}[h]
    \centering
    \includegraphics[width=\linewidth]{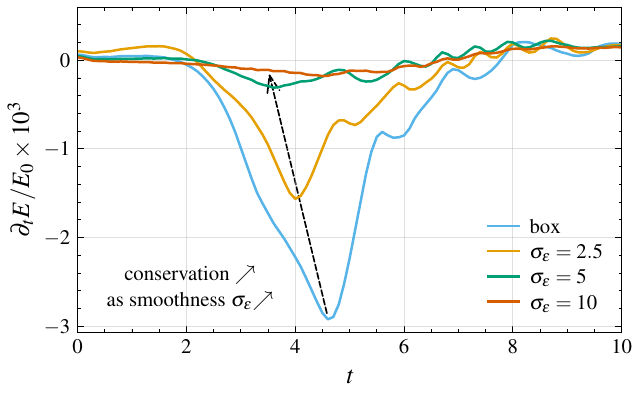}
    \caption{Change in the total energy as a function of time for different initial distribution functions given by Eq.~\eqref{eq:cons_init_profile}. The black dashed vertical lines represent the time at which the neutrino packet encounters the maximum velocity gradients and the dotted line where the velocity gradient vanishes. The stronger energy violation peak observed in Fig.~\ref{fig:cons_dedt_cart} is smoothed out only with a more diffuse initial energy profile. This suggests that the feature is driven by sharp gradients of $f$ in energy space.}
    \label{fig:cons_dedt_cart_gauss}
\end{figure}

As mentioned in Sec.~\ref{sec:limitation_and_improvements_discr}, our discretization scheme only ensures consistency for $f$ constant and may introduce artificial advection and four-momentum conservation violations if that assumption breaks down. Two main terms then contribute. First, the factor $f$ in $f \ddiff_{p,\curvi} \mathcal{F}^{\curvi}$ prevents the residuals from vanishing. The amplitude of these residuals decreases with higher position-space and momentum-space resolution. Second, there is the term $\mathcal{D}_{\curvi} f$, which increases for larger, or better resolved, gradients. Any source of artificial advection or violations present in $\tensor{\Gamma}{^\hrho_{\hmu\hnu}}$ is then amplified (or damped) by the gradient of $f$. The low-resolution violations would then be dominated by the non-vanishing residuals, since the gradient in $f$ is not properly resolved. At higher resolution, the residuals' amplitude decreases whereas the sharp gradient becomes larger (better resolved) and increases the effect of the artificial advection. The latter effect would be the dominating source of non-conservation at high resolution. This explanation is also consistent with the observations in Fig.~\ref{fig:cons_dedt_cart}. As the number of energy bins increases, the gradient in $f$ becomes larger, since it is better resolved, and yields more significant violations for lower values of velocity gradients. This would explain why the first peak shifts to earlier times. To support this hypothesis, we investigate the role of the energy gradient of $f$ on energy conservation. We initialize the distribution function with a smoother profile in energy space. We consider a Gaussian profile
\begin{equation}
    f_0(x_s, \vartheta_s, \varepsilon) = A_0 e^{-\frac{(\varepsilon - \varepsilon_s)^2}{\sigma^2_\varepsilon}}, \label{eq:cons_init_profile}
\end{equation}
where we choose $A_0 = 1$. We choose different values for $\sigma_\varepsilon$ to verify the influence of changing the gradient of $f$. From the sharpest to the smoothest Gaussian profile, we use $\sigma_\varepsilon = 2.5$, $5$, and $10$. In Fig.~\ref{fig:cons_dedt_cart_gauss}, we compare the box initialization to the Gaussian initialization in energy space for $N_\varepsilon = 80 = N_\vartheta$. Even though the profile with Gaussian initialization still shows a more significant change in total energy at early times, the bump is smaller than with a box initialization, and the maximum amplitude of the violations decreases with $\sigma_\varepsilon$. These observations suggest that the dominant effect in energy conservation violations with box initialization is mainly caused by the derivative of $f$ in energy space. This is however coupled to the presence of a significant velocity gradient, since no deviations are observed for $v=0$. The total effect is thus a combination of the velocity gradient and the distribution function gradient in energy space. Note that the effect of the velocity gradient alone is also still visible, mainly when the maximum gradient is negative. 

\begin{figure}[h!]
    \centering
    \includegraphics[width=\linewidth]{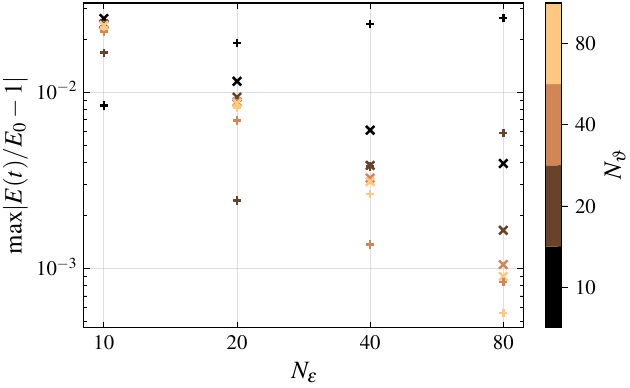}
    \caption{Same as Fig.~\ref{fig:cons_cart}, but for an initial profile given in Eq.~\eqref{eq:cons_init_profile} with $\sigma_\varepsilon = 10$ and only for $N_x = 256$ and $v_0 = 0.2$. Contrary to Fig.~\ref{fig:cons_cart}, the maximum violations do not saturate at high energy and angular resolutions. This suggests that this saturation is caused by the derivative of $f$ in energy space.}
    \label{fig:conservation_gauss_cart}
\end{figure} 

Finally, we plot the maximum deviations from conservation for $\sigma_\varepsilon = 10$, $N_x = 256$ and $v_0 = 0.2$ in Fig.~\ref{fig:conservation_gauss_cart}. Contrary to Fig.~\ref{fig:cons_cart}, where the initial profile is a box in phase space, the deviations do not saturate and decrease with increasing resolution. This suggests that the source of the saturation is also the derivative of $f$ in energy space. This interpretation is consistent with the previous observations that sharper initial profiles cause larger deviations. The saturation would then only occur when the resolution in energy space is sufficient to resolve the large gradient in $f$ and when this contribution outweighs the contribution of the velocity-induced advection. We can see two exceptions for the momentum deviations at $N_\vartheta = 10$ and $20$, where the maximum deviation increases with increasing energy resolution. 

In order to conclude on the exact sources of non-conservation and verify our hypotheses, one would need to have access to a discretization conserving simultaneously number and four-momentum. Therefore, we may only conclude that $f$ inhomogeneity coupled to velocity-induced advection is a plausible dominating source of non-conservation of four-momentum.

\subsubsection{Energy conservation in spherical coordinates \label{sec:conservation_laws_spherical}}
\noindent With the first insights provided by the previous tests, we now repeat the test in spherical coordinates to study whether the geometrical terms in $1/r$ also contribute to the energy conservation violations. The only difference with the previous test setup is that we change the spatial domain to $r\in[0,20]$. In particular, we expect the region at small radii to be more sensitive to non-vanishing residuals (between the energy-conservative and number-conservative formulations) because of $1/r$ terms. Even though the velocity profile is unrealistic because of large velocities at small radii, we reuse this profile to increase advection at small radii. We can then verify whether sources of conservation violations are indeed dominating at small radii.

\begin{figure}[h]
    \centering
    \includegraphics[width=\linewidth]{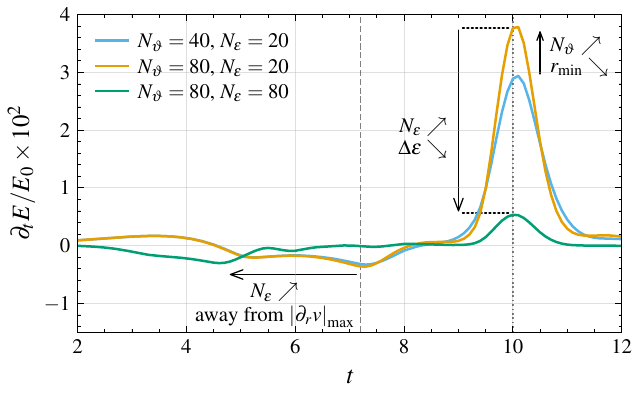}
    \caption{Change in total energy as a function of time for the velocity profile in Eq.~\eqref{eq:cons_velocity} with $N_r = 256$ and $v_0 = 0.2$ and in spherical coordinates. The vertical dashed line represents the velocity gradient maximum, and the vertical dotted line the velocity maximum. The change in energy increases close to the maximum of velocity gradient at lower energy resolution (orange and blue lines), but shifts away from it at higher resolutions (green line). Close to $r=0$ ($t=10$), the changes increase with larger $N_\vartheta$ because the neutrino packet reaches smaller radii. We also see that increasing the energy resolution lowers the dominant peak in energy non-conservation.}
    \label{fig:cons_dedt_spherical}
\end{figure}

In Fig.~\ref{fig:cons_dedt_spherical}, we plot the derivative of the total energy as a function of time for $N_r =256$ and $v_0 = 0.2$. We can see that the dominating pattern is different from Fig.~\ref{fig:cons_dedt_cart} in Cartesian coordinates. Even though changes in energy also increase close to the maxima of velocity gradients (vertical dashed lines), the largest change in energy is obtained when $\partial_r v \sim 0$ and $r \sim 0$ (vertical dotted line). This is expected if the advection term, $Wv\sin^2\vartheta/r$ in this case, is a source of conservation violations, since the term grows with decreasing radius. We also see that increasing $N_\vartheta$ increases the violations. These larger deviations would then correspond to the fact that smaller radii are reached at higher resolution ($\vartheta_0$ closer to $\pi$), where violations are stronger. Increasing the energy resolution (orange line to green line) significantly reduces the height of the peak. This shows a strong dependence of the conservation violations on $\Delta \varepsilon$. This dependence suggests that the residuals between the number-conservative and energy-conservative equations do not vanish because of a factor $\Delta \varepsilon$, which naturally arises from the discretization of the derivative in momentum space. We also note that, similarly to Cartesian coordinates, a first smaller peak is aligned with the maximum of the velocity gradient at lower resolutions (orange and blue lines), but it shifts away from it at a higher resolution. The source of this non-conservation is thus the gradient in $f$ in energy space found in Cartesian coordinates, but it is not dominant in this case. 

Because the point $r=0$ is especially sensitive to the discretization (see Sec.~\ref{sec:discr_tetrad_diff}), we perform an additional test where we use the vanilla \ac{FV} discretization instead of our modified discretization. We find that the relative difference of the maximum violation between the vanilla discretization and the modified discretization is between $\sim 8\%$  and $ \sim -2\%$. This difference tends to $0$ for larger energy resolutions. As a result, our discretization scheme neither significantly improves nor worsens the conservation properties of the vanilla discretization.

\subsubsection{Energy conservation through a shock \label{sec:cons_shock}}

\noindent As a last test of energy conservation, we consider a neutrino packet propagating through a velocity shock in spherical coordinates. This is a first probe of the energy conservation in a supernova-like environment, even though we neglect gravitational effects and interactions here. The velocity is set to $0$ for $r < 8$ and to $-0.2$ otherwise. The neutrino packet is emitted at $r_0 = 10$ and propagates inwards. It will thus cross the shock twice. For this test, we use an energy grid spacing similar to the grid used in opacity tables. We report a maximum violation of energy conservation of the order of $1\%$ for $N_\varepsilon \geq 20$ and $10\%$ for $N_\varepsilon = 10$. We point out that these numbers are possibly increased because of the large gradient of $f$ in energy space.

\subsubsection{Discussion \label{sec:cons_summary}}
\noindent With the tests in Cartesian coordinates, we identified the energy advection induced by velocity gradients to be a source of non-conservation. Part of the violations are embedded in non-vanishing residuals and decrease as the energy resolution increases. Another part multiplies the gradient of $f$ in energy space and becomes non-negligible for a larger number of energy bins.

With the tests in spherical coordinates, we confirmed that the violations are amplified at small radii because of $1/r$ terms and depend strongly on $\Delta \varepsilon$. This suggests that the residuals are non-zero because of a term in $\Delta \varepsilon$ (or some power of it). Note that, even though the radial velocity is not expected to be significant at small radii in astrophysical systems, one may expect violations from curvature effects.

Our observations show that our discretization does not make the discrete energy-conservative and number-conservative equations consistent. Even though this is expected from the fact that our discretization scheme was not designed for this purpose and is designed only for empty flat spacetimes, this suggests that the residuals from Sec.~\ref{sec:consistency} that do not vanish exactly contribute also to the conservation violations, or that terms other than the residuals treated before contribute to the energy non-conservation. Even if these residual terms are small, they can induce substantial conservation violations if they multiply a large term. In our case, such large terms are provided by velocity gradients, distribution function gradients and $1/r$ at small radii. A more detailed comparison between the discretized formulations of the number-conservative and energy-conservative formulations should highlight the exact discretization mismatches responsible for the energy conservation violations. Such a comparison and discretization improvements are left for future work. 

\subsection{Realizability \label{sec:realizability}}
\noindent The distribution function of fermions should be bounded between $0$ and $1$. Our current discretization does not, however, guarantee these bounds to be satisfied. A similar issue is encountered by the M1 scheme, where the moments are not necessarily realizable, \textit{i.e.} not necessarily consistent with a distribution function $ 0\leq f \leq 1$. A realizability-preserving Implicit-Explicit (IMEX) scheme was designed for the two-moment scheme to order $O(1)$~\citep{Chu_2019} and to order $O(v)$~\citep{Laiu_2025}. Solutions for Boltzmann transport have also been proposed in~\citep{Mezzacappa_1993,Liebendorfer_2004}, where the excess of particles in one energy bin is transferred to the higher energy bins while conserving total particle number. In our case, however, neither technique was implemented. Our current approach is to replace the current value at time step $n$ $\exppar{f}{n}$ by $1$ or $0$ if it is above or below the bounds, respectively, before the implicit update. The blocking factors in the expression of the source terms then ensure that the solution $\exppar{f}{n+1}$ for the time step $n+1$ be between $0$ and $1$. None of the tests presented in Sec.~\ref{sec:tests} or Sec.~\ref{sec:comparison_M1} showed values of $f$ outside of the accepted range, except for \acp{CCSN} where $f$ is found above $1$ with violations smaller than $10^{-3}$.

Our approach is sufficient to ensure a physically acceptable solution of the implicit update for collision terms. Our approach however breaks number conservation, in principle guaranteed by the \ac{FV} method, whenever the distribution function takes values outside the physical bounds in the advection part. Even though advection is theoretically expected to keep $f$ within the bounds from the very definition of the Boltzmann equation, numerical inaccuracies, such as the non-vanishing residuals discussed in the previous sections, may induce local errors that produce values of $f$ outside the physical bounds.

Improvements to enforce realizability in advection, such as adapting the methods presented in Refs.~\cite{Chu_2019,Laiu_2025} and improving the discretization scheme, are left for future developments.

\section{Interactions and implicit evolution \label{sec:interaction_and_implicit}}
All the previous sections have focused on advection, the left-hand side of the equation. In this section, we shall focus on the right-hand side of the equation, the collision terms. Unlike the previous sections, which are valid for any (neutral) particle, this section only considers the case of neutrinos. We also assume in this section that neutrinos are massless.

First, we describe how neutrino interactions are implemented. Then, we discuss how the discretization is adapted for the diffusive limit. Finally, we present the implicit treatment of the source terms.

\subsection{Source terms \label{sec:source_terms}}
\noindent The source terms are calculated from the invariant collision integrals. We consider absorption, emission, scattering and pair processes. The difference with the M1 source terms 
is that the latter are the moments of the source terms of the Boltzmann equation. Following Ref.~\cite{Bruenn_1985} and omitting the spatial dependence in the notation, the invariant collision integral for absorption and emission is
\begin{equation}
    C_{\mathrm{E/A}}[f] = \po \Big[ \eta(\varepsilon) \big(1-f(\varepsilon,\Omega_p)\big) - \kappa_a(\varepsilon) f(\varepsilon,\Omega_p) \Big] \label{eq:ae_no_eq}\\
\end{equation}
where $\eta$ is the emissivity and $\kappa_a$ the absorption coefficient. Using detailed balance, we can write the emissivity as $\eta(\varepsilon) = f_{\mathrm{eq}}\left(\varepsilon, T, \mu_\nu^\mathrm{eq}\right) \kappa_a^*(\varepsilon)$. In the latter relation, $f_{\mathrm{eq}}$ is the equilibrium distribution function. For neutrinos, $f_{\mathrm{eq}}$ is given by the Fermi-Dirac statistics
\begin{equation}
    f_{\mathrm{eq}}\left(\varepsilon, T, \mu_\nu^\mathrm{eq}\right) = \frac{1}{e^{(\varepsilon-\mu_\nu^\mathrm{eq})/k_B T} + 1}, \label{eq:f_eq}
\end{equation}
where $T$ is the temperature, $\mu_\nu^\mathrm{eq}$ the chemical potential of neutrinos in $\beta$-equilibrium and $k_B$ is the Boltzmann constant. We can then rewrite Eq.~\eqref{eq:ae_no_eq} as
\begin{equation}
    C_{\mathrm{E/A}}[f] = \po \Big[ \eta(\varepsilon) - \kappa^*_a(\varepsilon) f(\varepsilon,\Omega_p) \Big] \label{eq:ae}
\end{equation}
where we introduced the absorption coefficient corrected for stimulated absorption $\kappa^*_a$ (\textit{e.g.},~\cite{Rampp2002})
\begin{equation}
    \kappa^*_a (\varepsilon) = \frac{\kappa_a}{1-f_{\mathrm{eq}}} = \eta(\varepsilon) + \kappa_a(\varepsilon).\\
\end{equation}

Inelastic scattering is described by the invariant collision integral~\citep{Bruenn_1985}
\begin{align}
    C_{\mathrm{IS}}[f] &= \po \int d\varepsilon' d\Omega_p'\,\varepsilon'^2 \nonumber \\[1.5ex] & R^{\mathrm{in}}_{\mathrm{IS}}(\varepsilon,\varepsilon',\omega) f(\varepsilon',\Omega_p') \Big[1-f(\varepsilon,\Omega_p) \Big] \nonumber \\[1.5ex]
    -& R^{\mathrm{out}}_{\mathrm{IS}}(\varepsilon,\varepsilon',\omega) f(\varepsilon,\Omega_p) \Big[1-f(\varepsilon',\Omega_p') \Big],  \label{eq:is}
\end{align}
where $R^{\mathrm{in/out}}_{\mathrm{IS}}$ are the kernels for in- and outgoing neutrinos, respectively. They depend on the energy of both neutrinos, as well as on $\omega$, the cosine of the angle between the in- and outgoing directions of the neutrino. 
This cosine is defined by
\begin{equation}
    \omega = \cos\vartheta\cos\vartheta' + \sin\vartheta\sin\vartheta' \cos(\varphi-\varphi'), \label{eq:cos_omega}
\end{equation}
where $(\vartheta,\, \varphi)$ and ($\vartheta',\, \varphi'$) are the propagation angles of the two neutrino states.

In the case of elastic scattering (ES), there is no net exchange of energy and the invariant collision integral reads~\citep{Bruenn_1985}
\begin{equation}
    C_{\mathrm{ES}}[f] = \po \varepsilon^2 \int d\Omega_p'\, R_{\mathrm{ES}}(\varepsilon,\omega) \Big[ f(\varepsilon,\Omega_p') - f(\varepsilon,\Omega_p) \Big], \label{eq:es}
\end{equation}
where $R_{\mathrm{ES}}(\varepsilon,\omega)$ is the kernel of ES processes.

Finally, the pair process invariant collision integral reads~\citep{Bruenn_1985}
\begin{align}
    C_{\mathrm{PP}}[f] &= \po \int d\varepsilon' d\Omega_p'\,\varepsilon'^2 \nonumber \\[1.5ex] & R^{\mathrm{pro}}_{\mathrm{PP}}(\varepsilon,\varepsilon',\omega) \Big[1-f(\varepsilon,\Omega_p) \Big] \Big[1-\bar{f}(\varepsilon',\Omega_p') \Big] \nonumber \\[1.5ex]
    -& R^{\mathrm{ann}}_{\mathrm{PP}}(\varepsilon,\varepsilon',\omega) f(\varepsilon,\Omega_p) \bar{f}(\varepsilon',\Omega_p'), \label{eq:pp}
\end{align}
where $R^{\mathrm{pro/ann}}$ are the kernels for pair production and pair annihilation, respectively, and $\bar{f}$ denotes the distribution function of the antiparticle.

In this work, we use the same approach as for the M1 scheme and expand the kernels in a Legendre series up to first order as
\begin{align}
    R\left(\varepsilon, \varepsilon', \omega\right) &= \frac{1}{2} \sum_n(2 n+1) \Phi_{n}\left(\varepsilon,\varepsilon'\right) P_n(\omega) \nonumber \\
     &\approx\frac{1}{2} \Phi_{0}\left(\varepsilon,\varepsilon'\right) + \frac{3}{2} \Phi_{1}\left(\varepsilon,\varepsilon'\right) \omega, \label{eq:kernels_expansion}
\end{align}
where $P_n$ is the Legendre polynomial of degree $n$ and the $\Phi_n$ are the coefficients of the expansion. In the specific case of ES, we have
\begin{align}
R_{\mathrm{ES}}\left(\varepsilon, \omega\right)& \approx R_{\mathrm{ES},0}(\varepsilon) + R_{\mathrm{ES},1}(\varepsilon) \omega \nonumber \\ 
 &=\frac{1}{2} \Phi_{\mathrm{ES},0}(\varepsilon) + \frac{3}{2} \Phi_{\mathrm{ES},1}(\varepsilon) \omega, \label{eq:ES_kernel_expansion}
\end{align}
since the energy does not change through the interaction. The radiation four-force $\mathcal{S}^\mu$ in the M1 source term can then be expressed as~\citep{Shibata_2011}
\begin{equation}
    \mathcal{S}^\mu_{\mathrm{ES}}=-\kappa_{s} \mathcal{H}^\mu, \label{eq:M1_es}
\end{equation}
which only depends on the scattering opacity, and the opacity can be written as a linear combination of kernel coefficients (\textit{e.g.}~\cite{Ng_2024}):
\begin{equation}
\kappa_s(\varepsilon) =4 \pi \varepsilon^2\left[R_{\mathrm{ES},0}(\varepsilon)-\frac{1}{3} R_{\mathrm{ES},1}(\varepsilon)\right] \, . \label{eq:kappa_s}
\end{equation}
Comparing Eqs.~\eqref{eq:es}~and~\eqref{eq:M1_es}, we can see that if only the Legendre coefficients are to be used, the Boltzmann solver requires the expression of the coefficients themselves, whereas the M1 only uses $\kappa_s$. Most interaction tables focus on M1 source terms and thus only provide $\kappa_s$. Libraries generating interaction tables would thus need to be extended to provide directly the kernel, or its Legendre expansion, for the Boltzmann solver. 
Therefore, we have extended the \texttt{Weakhub} library~\citep{Ng_2024} to 
provide $R_{\mathrm{ES},0}$ and $R_{\mathrm{ES},1}$ instead of $\kappa_s$.
More details on how $R_{\mathrm{ES},0}$ and $R_{\mathrm{ES},1}$ are computed in \texttt{Weakhub} can be found in Appendix~\ref{app:kernels_ES}.

In this work, we only use a first-order Legendre expansion of the kernels. As we shall explain in Sec.~\ref{sec:optimization_implicit}, this approximation enables a linear complexity with respect to the number of angular bins for the implicit time step. A full Boltzmann solver will ultimately require the full angular dependence of kernels. Such developments are left for future work.

\subsection{Diffusive limit \label{sec:diffusive_limit}}
\noindent When the scattering opacity is large, the Boltzmann equation behaves as a diffusion equation~(\textit{e.g.} \cite{Bruenn_1985}). We must therefore make sure that our solver behaves properly in both the optically thin and optically thick limits. More precisely, particular care is needed when the mean free path of a particle is not resolved by the grid. Based on previous works, the diffusive limit can be correctly modelled by modifying the flux reconstruction at the cell surface with an opacity-dependent factor. Following Refs.~\cite{Mezzacappa_1993,Sumiyoshi_2012,Akaho_2021}, we define
\begin{equation}
    \delta^a = 1 - \frac{1}{2} \frac{\Delta x^a \bar{\kappa}}{1+ \Delta x^a \bar{\kappa}},\label{eq:diffusive}
\end{equation}
where $\bar{\kappa}$ is an averaged total opacity and $a$ denotes a spatial axis. The parameter $\delta^a$ is close to $1$ when $\bar{\kappa} \Delta x^a \ll 1$, \textit{i.e.} when the mean free path of the particle is resolved, while it tends to $1/2$ when $\bar{\kappa}\Delta x^a \gg 1$. The flux is then reconstructed according to
\begin{equation}
    F^a_{i+1/2,j,k} = \delta^a_{i,j,k} F^a_{i,j,k} + ( 1-\delta^a_{i,j,k}) F^a_{i+1,j,k} 
\end{equation}
if the characteristic speed $c^a_x$ is positive and
\begin{equation}
    F^a_{i+1/2,j,k} = \delta^a_{i,j,k} F^a_{i+1,j,k} + ( 1-\delta^a_{i,j,k}) F^a_{i,j,k} 
\end{equation}
if $c^a_x$ is negative. Note that these expressions may need to be modified when using Riemann solvers. This generalization is left for future work.

\subsection{Implicit evolution of the source terms \label{sec:implicit_evolution}}

\noindent The collision integrals lead to stiff source terms that must be treated implicitly. The equation to solve is of the form
\begin{equation}
    \exppar{\vm{q}}{n+1} = \exppar{\vm{q}}{n} + \Delta t \,\vm{s}_{\mathrm{rad}}\left(\exppar{\vm{q}}{n+1}\right) . \label{eq:implicit}
\end{equation}
We use IMEX schemes to split the advection part, evolved explicitly, from the stiff collision terms, evolved implicitly. This approach, also used for \texttt{Gmunu}'s M1 solver~\citep{Cheong_2023}, allows one to use larger time steps thanks to the implicit solver, while limiting the size of the system of equations to solve implicitly thanks to the explicit solver. Note, however, that the exchange of number and four-momentum with the fluid is performed explicitly to limit the cost of the implicit step.

\subsubsection{Implicit solver equations}

Since the collision terms do not couple terms at different positions, this equation can be solved at each point of the position space independently. It however couples different neutrino species from different locations in momentum space. The components of the vector $\vm{q}$ are therefore $q_I$ with $I=g(l,m,n,s)$, where $l,m,n$ are indices of momentum space bins and $s$ indicates the neutrino species, and $g$ is some function mapping the $4$-dimensional indices into a single one. It should be noted that the collision integrals only depend on $\vm{f}$ and not on $\vm{q}$. Therefore, we solve the equation for $\vm{f}^{n+1}$ instead, using the definition of $q_s$ in Eq.~\eqref{eq:q}.

From Eqs.~\eqref{eq:source}, \eqref{eq:ae}, \eqref{eq:is}, \eqref{eq:es} and \eqref{eq:pp}, we can show that the discretized $\vm{s}_{\mathrm{rad}}$ term can be written in a matrix form as
\begin{equation}
    \vm{s}_{\mathrm{rad}} = \alpha \psi^6 \sqrt{\frac{\bar{\gamma}}{\hat{\gamma}}} \left[ \left( \tilde{\vm{A}} \vm{f} \right) \odot \vm{f} + \tilde{\vm{B}} \vm{f} + \tilde{\vm{c}}\right], \label{eq:discr_source}
\end{equation}
where $\odot$ denotes the component-wise product. The expressions of $\tilde{\vm{A}}$, $\tilde{\vm{B}}$ and $\tilde{\vm{c}}$ are given in Appendix~\ref{app:source_terms_implicit}. Eq.~\eqref{eq:implicit} then becomes a set of coupled quadratic equations that requires a rootfinding method to be solved. It can however reduce to a set of linear equations if only emission, absorption and elastic scattering are considered, since those interactions do not involve any quadratic term and thus $\tilde{\vm{A}} = 0$. In the general case, though, we solve Eq.~\eqref{eq:implicit} rewritten as
\begin{equation}
    \begin{split}
        \vm{h} \equiv & -\exppar{\vm{w}_p}{n+1} \odot \exppar{\vm{f}}{n+1} + \left(\vm{A} \exppar{\vm{f}}{n+1} \right) \odot \exppar{\vm{f}}{n+1} \\
         & + \vm{B} \exppar{\vm{f}}{n+1} + \vm{c} = 0 ,
    \end{split}\label{eq:implicit_f}
\end{equation}
where $f^n$ is in $\vm{c}$ and $\Delta t$ in $\vm{A}$, $\vm{B}$ and $\vm{c}$. In the last equation, we introduced the weights
\begin{equation}
    \exppar{\vm{w}_p}{n} = \frac{\exppar{\mathcal{L}_{\hmu}}{n}\vm{p}^{\hmu}}{\vm{\po}}, \label{eq:w_p}
\end{equation}
where the division is here understood as an element-wise operation.
More details can be found in Appendix~\ref{app:source_terms_implicit}. As already mentioned, a rootfinding method is required to solve this equation. The Jacobian of the equation for the multidimensional \ac{NR} method is directly computed from the different matrices and is (no sum on repeated indices)
\begin{equation}
    J_{IJ} \equiv \frac{\partial h_I}{\partial f_J} = -(w_p)_{I}\delta_{IJ} + \delta_{IJ} \left(\vm{A} \vm{f}\right)_J + A_{IJ} f_I + B_{IJ}.\label{eq:jacobian}
\end{equation}
It becomes apparent that Eq.~\eqref{eq:implicit_f} is expensive to solve. We know that the vector $\vm{f}$ has a size $N = N_{\varepsilon}\times N_{\vartheta} \times N_{\varphi} \times N_{spec}$. The cost of evaluating $\vm{h}$ is $O(N^2)$ since constructing the matrices and applying the matrix-vector product is $O(N^2)$. The computation of the Jacobian is also of the same order, although it can be computed straightforwardly from quantities that have already been computed to evaluate $\vm{h}$. Solving the \ac{NR} iteration with LU decomposition, however, would be of order $O(N^3)$, which rapidly makes the problem intractable even with a $2$D momentum space. It is therefore critical to find a method to accelerate the rootfinding method, either through derivative-free methods or accelerated inversion. The latter will be discussed in Sec.~\ref{sec:optimization_implicit}.

\subsubsection{Implicit solver modes \label{sec:implicit_modes}}
\noindent Another approach to limit the implicit step cost is to limit the size $N$ of the problem. To be more precise, we consider the cases where the matrices $\vm{A}$ and $\vm{B}$ are block-diagonal. If the blocks have a size $N_b$, then the full system of $N$ equations reduces to $N/N_b$ independent systems of $N_b$ coupled equations. The cost of solving the implicit step with \ac{NR} and LU-decomposition of the Jacobian is then $O\big( (N/N_b) N^3_b \big)$. The smaller $N_b$, the larger the acceleration. In practice, the size of the blocks depends on the interactions considered in the model. Thus, similarly to \texttt{Gmunu}'s M1 solver~\citep{Cheong_2023}, we define different implicit-solver modes:
\begin{enumerate}
    \item \textit{\ac{MSMG}}: the matrices involved in the system of equations are not block diagonal. The full system must thus be solved as a single system of size $N = N_{\varepsilon}\times N_{\vartheta} \times N_{\varphi} \times N_{spec}$. This mode contains both energy-coupling and species-coupling interactions.
    \item \textit{\ac{SSMG}}: only energy-coupling interactions are considered for the implicit solve. Each species is then independent from the others, and the system of $N$ equations reduces to $N_{spec}$ independent systems of $N_b = N_{\varepsilon}\times N_{\vartheta} \times N_{\varphi}$ equations. Species-coupling interactions are then treated explicitly, which is less accurate in case of strong coupling.
    \item \textit{\ac{SSSG}}: neither energy-coupling nor species-coupling interactions are considered for the implicit solve. Each species and energy bin is then independent from the others, and the system of $N$ equations reduces to $N_{spec} \times N_{\varepsilon}$ independent systems of $N_b = N_{\vartheta} \times N_{\varphi}$ equations. Energy-coupling and species-coupling interactions are evolved explicitly, which is again more approximate in case of strong coupling. In this case, the resulting systems of equations become linear and do therefore not require a rootfinding method to be solved.
\end{enumerate}

\subsubsection{Coupling to the fluid \label{sec:coupling_to_the_fluid}}
\noindent Even though the evolution of radiation variables is computed implicitly for collision terms, the transfer of number, energy and momentum from neutrinos to the fluid is computed explicitly to reduce the cost of the implicit step. More formally, from Eqs.~\eqref{eq:coupling_tau} to \eqref{eq:coupling_ye}, we update fluid variables as
\begin{align}
    &\exppar{\vm{q}}{n+1}_\tau = \exppar{\vm{q}}{n}_\tau - \Delta t \sum_s \int dV_p~ \vm{s}_\mathrm{rad,s}~ \mathcal{L}_{\hmu} \vm{p}^{\hmu},\\
    &\exppar{\vm{q}}{n+1}_{S_i} = \exppar{\vm{q}}{n}_{S_i} - \Delta t \sum_s \int dV_p~ \vm{s}_\mathrm{rad,s}~ \tensor{\ell}{_{i\hmu}} \vm{p}^{\hmu},\\
    &\exppar{\vm{q}}{n+1}_{Y_e} = \exppar{\vm{q}}{n}_{Y_e} - \Delta t \int dV_p~ (\vm{s}_\mathrm{rad,\nu_e} - \vm{s}_\mathrm{rad,\bar{\nu}_e}).
\end{align}
Extensions of the formalism to evolve these equations jointly with radiative variables in an implicit manner are left for future work.

\section{Optimization of the implicit step \label{sec:optimization_implicit}}

In practice, the multidimensional \ac{NR} iteration for the implicit solver is performed by solving the linear system
\begin{equation}
    \vm{J} \symvm{\Delta} \vm{f} = -\vm{h}(\vm{f}^{(i)}), \label{eq:NR_iter}
\end{equation}
with $\symvm{\Delta} \vm{f} \equiv \vm{f}^{(i+1)} - \vm{f}^{(i)}$. This linear system can typically be solved using LU-decomposition, which is of complexity $O(N^3)$. In our case, we have $N = N_{\text{spec}} N_\varepsilon N_\vartheta N_\varphi$. The cubic complexity thus quickly makes the problem intractable.

On the other hand, the \ac{NR} method is not the only method that can solve Eq.~\eqref{eq:implicit}. Other methods, such as the fixed-point iterative method, can also be used. In our case, each iteration of the fixed-point method is of complexity $O(N^2)$, since the evaluation of Eq.~\eqref{eq:implicit} involves matrix-vector products. However, the order of convergence of the classical fixed-point method is lower than that of the \ac{NR} method and a shorter computation time is thus not guaranteed. In this work, we shall only study the \ac{NR} method.

The computational cost of the \ac{NR} method may however still be too large because of the need for matrix inversion. Therefore, we exploit our formalism to reduce the computational cost. Our optimization relies on two key concepts. The first is that one generally does not need to compute the matrix inverse, but only the product of the inverse with a vector. This is also applicable to the M1 transport schemes. Moreover, we shall use matrix-free methods when possible, \textit{i.e.} avoid storing full matrices and implement matrix-vector products instead. The second concept we use is the low-rank representation of the angular space components of kernels when using a Legendre expansion. In particular, we will show that we can reduce the dimensionality of the problem to that of a truncated moment scheme. We will then show that both the memory requirements and computational costs are accordingly reduced for low-order expansions.

In this section, we present how the \ac{NR} method can be improved and optimized to reduce the complexity and computation time using iterative methods. We also discuss how to leverage the truncated Legendre expansion of the kernels. To discuss complexity more easily, we introduce $N_e \equiv N_{\text{spec}} N_\varepsilon$ and $N_a \equiv N_\vartheta N_\varphi$, which are respectively the total number of energy and species bins per angle and the total number of angular bins per energy bin and species.

\subsection{Iterative methods for matrix inversion \label{sec:iterative}}

\noindent As discussed in previous sections, the \ac{NR} method requires solving a linear system. The matrix representing the system of equations is generally not inverted explicitly. An alternative method to the LU decomposition consists in computing the product of the inverse matrix with a vector, instead of constructing the inverse matrix. Computing matrix-vector products is of complexity $O(N^2)$, whereas matrix inversion (and LU) are of complexity $O(N^3)$. In practice, this can be achieved by solving iteratively the corresponding linear system. Our only requirement for choosing an iterative method is that it must have a complexity at most $O(nN^2)$, where $n$ is the number of iterations before convergence and $O(N^2)$ is the complexity for matrix-vector products. Additionally, each iteration must be fast to compute. The actual efficiency of the iterative method as a replacement to the LU-decomposition will thus depend on the combination of its convergence rate and the computational cost of one iteration. Among the many iterative methods that exist, we shall only consider the Jacobi method. A sufficient condition for convergence is that the matrix be diagonally dominant. Such iterative methods are thus suited to our problem as long as the interactions are not too strong, or the time step sufficiently small. Most matrices to invert in our methods can indeed be expressed as perturbations of order $\Delta t$ of a diagonal matrix. If the iterative method fails to converge, the LU-decomposition is used to solve the linear system.

We use the Jacobi method to solve the \ac{NR} iteration in Eq.~\eqref{eq:NR_iter}. The Jacobi iteration can then be written as
\begin{equation}
    (\symvm{\Delta} \vm{f})^{(k+1)} = -\vm{D}^{-1}\left( (\vm{J} - \vm{D} )(\symvm{\Delta} \vm{f})^{(k)} \right) - \vm{D}^{-1}\vm{h},\label{eq:Jacobi}
\end{equation}
where $k$ denotes the iteration in the Jacobi method, and $D_{IJ} \equiv \delta_{IJ} J_{IJ}$ is the diagonal matrix containing only the diagonal elements of $\vm{J}$. Because $\vm{D}$ is diagonal, both inversion and matrix-vector multiplications are of complexity $O(N)$. The most costly operation is the matrix-vector multiplication of $(\vm{J} - \vm{D})$. The main advantage of the Jacobi method over other methods is thus the simplicity and limited computational cost of a single iteration, despite its possibly slower convergence.

As mentioned above, the efficiency of the iterative method also depends on its convergence rate. It thus depends on the convergence criterion that is used and how accurate the computation must be. Moreover, if the iterative method fails too often or takes many iterations to converge, the cost of the whole implicit time step may, in fact, increase compared to the LU method. Hence, it is important to set appropriate convergence criteria and check the convergence of the method. For the method considered, the error $\symvm\epsilon_k$ at iteration $k$ can be expressed as $\symvm\epsilon_{k+1}$ = $\vm{U}\symvm\epsilon_k$, where $\vm{U}$ is the update matrix (see, \textit{e.g.}, Ref.~\cite{Saad_2003}). The error at the $(k+1)$-th iteration is thus bounded by
\begin{equation*}
    ||\symvm\epsilon_{k+1}||_{\infty} \leq \sigma^{k+1} ||\symvm\epsilon_0||_{\infty}.
\end{equation*}
In practice, we estimate $\sigma$ as the ratio of the errors at the first iteration and compute an estimation of the number of iterations required for convergence. If that number exceeds the maximum number of iterations allowed, the LU-decomposition is used to solve the linear system. This limits the overhead induced by a failed or slow convergence of the iterative method.

\subsection{Low-rank approximation \label{sec:low_rank}}
\noindent In this section, we consider optimizations to the implicit time step when the kernels are expanded in a Legendre series in angles up to order $\lmax$, which induces a low-rank structure in the tensor representing the problem. Before diving into these details, we lay the foundations of our method in the more general case of separable kernels.

\subsubsection{Separable kernels \label{sec:separable_kernels}}
\noindent First, we consider the case where the energy and angular dependences of the kernels are separable, \textit{i.e.} when 
\begin{equation}
    R(\varepsilon, \varepsilon',\omega) = R_\varepsilon(\varepsilon,\varepsilon') R_\omega(\omega),
\end{equation}
where $\omega$ is defined in Eq.~\eqref{eq:cos_omega}, $R_\varepsilon$ only contains the energy dependence and $R_\omega$ the angular dependence.

When discretizing the kernels, we can then express them as a Kronecker product ``$\otimes$'' because they are separable. The discretization of $\omega$ is given by
\begin{equation}
    \begin{aligned}
        \omega_{ij} &= \cos\vartheta_i \cos\vartheta_j + \sin\vartheta_i\cos\varphi_i \sin\vartheta_j\cos\varphi_j \\
        & + \sin\vartheta_i\sin\varphi_i \sin\vartheta_j\sin\varphi_j,
    \end{aligned}
\end{equation}
where the indices range from $1$ to $N_a$. We can see that $\omega$ is low-rank, since it only has rank $3$. Indeed, each of the three terms in the sum can be seen as an outer product of two vectors. It can similarly be seen as a sum of $3$ Kronecker products.
Then, we have
\begin{equation}
    \vm{R} = \vm{R}_\varepsilon \otimes \vm{R}_\omega, \label{eq:R_separable}
\end{equation}
where $(R_\varepsilon)_{ij} = R_\varepsilon(\varepsilon_i,\varepsilon_j)$ is an $N_e \times N_e$ matrix and $(R_\omega)_{ij} = R_\omega(\omega_{ij})$ is an $N_a \times N_a$ matrix.

This structure is leveraged by the property of the Kronecker product $(\vm{R}_\varepsilon \otimes\vm{R}_\omega)\mathrm{vec}(\vm{W}) = \mathrm{vec}(\vm{R}_\omega\vm{W}\vm{R}_{\varepsilon}^T)$, where ``vec'' denotes the vectorization of a matrix and $\vm{W}$ is an $N_a \times N_e$ matrix representation of a vector $\vm{w}$ of size $N$. The complexity of the matrix-vector product reduces from $O(N^2)$ to $O\big( N_e^2 N_a +  N_e N_a^2 \big)$. The final vectorization operation may be skipped if all variables are stored in a matrix representation instead of a vector one. The memory cost of this representation also decreases, since the kernels need not be fully stored, but only $\vm{R}_\varepsilon$ and $\vm{R}_\omega$ separately.

Let us now assume that $\vm{R}_\omega$ is of low rank, \textit{i.e.} of rank $r' \ll N_a$. We can then express $\vm{R}_\omega$ as a sum of $r'$ outer-products of vectors
\begin{equation}
    \vm{R}_\omega = \sum_{i=1}^{r'} \sigma_i \, \vm{u}_i \vm{v}_i^T, \label{eq:R_omega_low_rank}
\end{equation}
where $\sigma_i$ are the singular values of $\vm{R}_\omega$ and $\vm{u}_i$ and $\vm{v}_i$ are vectors of size $N_a$. This expression significantly reduces the number of operations for multiplications if $r' \ll N_a$. The complexity of the matrix-vector multiplication $\mathrm{vec}(\vm{R}_\omega\vm{W}\vm{R}_{\varepsilon}^T)$ thus becomes $O\big(N_e^2 r' +  r' N_e N_a\big)$. This also reduces the memory cost of storing $\vm{R}_\omega$ to $O(2r' N_a)$, since we store $2$ vectors of size $N_a$ for each outer-product.

Finally, we can partially relax the separable condition and consider that the kernels can be expressed as a sum of $Q$ separable kernels
\begin{equation}
    R(\varepsilon, \varepsilon',\omega) = \sum_{i=1}^{Q} R_{\varepsilon,i}(\varepsilon,\varepsilon') R_{\omega,i}(\omega).
\end{equation}
This allows us to generalize Eq.~\eqref{eq:R_separable} to
\begin{equation}
    \vm{R} = \sum_{i=1}^Q \vm{R}_{\varepsilon,i} \otimes \vm{R}_{\omega,i}. \label{eq:R_sum_separable}
\end{equation}
All the operations detailed above would then need to be repeated $Q$ times. If $Q$ is too large, the gain in computational time is thus lost.

\subsubsection{Low-rank approximation of kernels \label{sec:low_rank_kernels}}

\noindent The methods presented in the previous section cannot be straightforwardly applied to our problem. In general, the interaction kernels are indeed not separable. Moreover, the matrices involved in the system of equations we have to solve contain sums of such kernels (see Appendix~\ref{app:source_terms_implicit}). However, we will show that a truncated Legendre series of the kernels, typically used in truncated-moment schemes, enables the use of separable kernels to solve the implicit time step more efficiently.

Similar to Eq.~\eqref{eq:kernels_expansion}, we write the truncated Legendre expansion of the kernels up to order $\lmax$ as
\begin{equation*}
    \begin{split}
        R(\varepsilon, \varepsilon',\omega) & \approx \sum_{\ell=0}^{\lmax} R_\ell(\varepsilon,\varepsilon') P_\ell(\omega) \\
        &= \frac{1}{2} \sum_{\ell=0}^{\lmax} (2 \ell + 1) \Phi_\ell(\varepsilon,\varepsilon') P_\ell(\omega),
    \end{split}
\end{equation*}
where the coefficients $\Phi_\ell$ are the coefficients of the Legendre series obtained by integration of the kernels. This Legendre expansion of the kernels effectively splits the energy and angular dependence of the kernels. Therefore, the expansion can be interpreted as a sum of separable kernels, which corresponds to Eq.~\eqref{eq:R_sum_separable} with $Q=\lmax+1$.

It will be useful for later developments to express the Legendre polynomial $P_\ell$ in the Legendre expansion as a function of real spherical harmonics $Y_\ell^m$ using the addition formula
\begin{equation}
    R(\varepsilon, \varepsilon',\omega) \approx 2 \pi \sum_{\ell=0}^{\lmax} \Phi_\ell(\varepsilon,\varepsilon') \sum_{m=-l}^{l} Y_\ell^m(\vartheta,\varphi)Y_\ell^m(\vartheta',\varphi'). \label{eq:R_spherical_harmonics}
\end{equation}
In its discretized form, the product $Y_\ell^m(\vartheta,\varphi)Y_\ell^m(\vartheta',\varphi')$ is represented as an outer-product of vectors $y_\ell^m$. More precisely,
\begin{equation}
    Y_\ell^m(\vartheta_i,\varphi_i)Y_\ell^m(\vartheta_j,\varphi_j) = (\vm{y}_\ell^m)(\vm{y}_\ell^m)^T,
\end{equation}
where the indices $i$ and $j$ run over all the $N_a$ angular bins. To write the discretization of Eq.~\eqref{eq:R_spherical_harmonics} more compactly, we introduce the rank-$(2\ell+1)$ matrices of size $N_a \times N_a$
\begin{equation}
    \symvm{\Omega}_\ell = 2\pi \sum_{m=-l}^l  (\vm{y}_\ell^m)(\vm{y}_\ell^m)^T.
\end{equation}
Then, we can write the discretization of Eq.~\eqref{eq:R_spherical_harmonics} as
\begin{equation}
    \vm{R} \approx \sum_{\ell=0}^{\lmax}\symvm{\Phi}_\ell \otimes \symvm{\Omega}_\ell,
\end{equation}
where $\symvm{\Phi}_\ell$ is of size $N_e \times N_e$ and generally full-rank.
Since the rank of a Kronecker product of matrices is the product of the ranks of the individual matrices, and the rank of a sum of matrices is at most the sum of the ranks of the individual matrices, we can evaluate the rank of $\vm{R}$ with
\begin{equation}
    \begin{split}
        \mathrm{rank}(\vm R) &\leq \sum_{\ell=0}^{\lmax} \mathrm{rank}(\symvm{\Phi}_\ell) \mathrm{rank}( \symvm{\Omega}_\ell)\\
        &\leq \sum_{\ell=0}^{\lmax} N_e (2\ell+1) = N_e (\lmax + 1)^2
    \end{split}
\end{equation}
If we assume that the sum over $\ell$ does not reduce the rank, then the rank of $\vm{R}$ is
\begin{equation}
    r \equiv \mathrm{rank}(\vm R) = N_e~\mathrm{min}\big(N_a,~(\lmax + 1)^2 \big).
\end{equation}
Therefore, $\vm{R}$ is of low-rank if the truncation is performed at a small order such that $(\lmax + 1)^2 \ll N_a$, which can be exploited to reduce the cost of matrix operations. However, the truncation at low order also reduces the accuracy with which the kernels are represented.

We have thus shown that the Legendre expansion of the kernels allows us to use the methods described in the previous section, with $\vm{R}_{\varepsilon,l} = \symvm{\Phi}_\ell$, $\vm{R}_{\omega,l} = \symvm{\Omega}_\ell$, $Q = \lmax+1$ and $r_\ell' = (2\ell + 1)$. Note that in spherical symmetry, only the $m=0$ component of the spherical harmonics survives the integration over $\varphi$. Consequently, the rank of $\symvm{\Omega}_\ell$ reduces to $r_\ell'=1$ and the total rank of $\vm{R}$ reduces to $r = N_e~\mathrm{min}\big(N_a,~(\lmax + 1) \big)$. The number of angular bins $N_a$ also generally decreases, since only $\vartheta$ is then discretized.

To apply this formalism to our Eq.~\eqref{eq:implicit_f}, there is another property of the system that we can leverage. The angular part $\symvm{\Omega}_\ell$ of the kernels solely depends on spherical harmonics (hence on propagation angles) and does not depend on the interactions considered. Since the matrices $\vm{A}$ and $\vm{B}$, and vector $\vm{c}$ are constructed from sums of these kernels, the angular parts can be factored out and only the energy part $\symvm{\Phi}_\ell$ is added. More precisely, we can express the matrices $\vm{A}$ and $\vm{B}$ of our system of equations as
\begin{align}
    &\vm{A} = \sum_{\ell=0}^{\lmax} \symvm{\Phi}_{A,\ell} \otimes \vm{U}_\ell \vm{V}_\ell^T,\label{eq:A_kronecker}\\
    &\vm{B} = \vm{D}_{B} + \sum_{\ell=0}^{\lmax} \symvm{\Phi}_{B,\ell} \otimes \vm{U}_\ell \vm{V}_\ell^T, \label{eq:B_kronecker}
\end{align}
where $\vm{D}_B$ is a diagonal matrix of size $N \times N$ arising from isotropic interactions, and $(U_\ell)_{ij} = (y_\ell^{m=j})_i$ and $(V_\ell)_{ij} = (U_\ell)_{ij}\sin\vartheta_i \Delta\vartheta_i \Delta \varphi_i$. Both matrices are of size $N_a \times r'_\ell$. The change from $\symvm{\Omega}_\ell$ to $\vm{U}_\ell \vm{V}_\ell^T$ comes from the fact that the kernels are integrated over angles, which explains the discrete integration factor in the definition of $\vm{V}_\ell$. In other words, the columns of $\vm{U}_\ell$ correspond to discretized spherical harmonics, and the columns of $\vm{V}_\ell$ compute the coefficients of the spherical harmonics expansion. We refer the reader to Appendix~\ref{app:source_terms_implicit} for more details on the expression of $\vm{A}$ and $\vm{B}$. As a result of Eqs.~\eqref{eq:A_kronecker} and \eqref{eq:B_kronecker}, both $\vm{A}$ and $\vm{B} - \vm{D}_B$ have a maximum rank $r = r' N_e = (\lmax + 1)^2 N_e$. The whole discussion in the previous section is thus relevant to our problem, even though some changes are needed because of $\vm{D}_B$.

Finally, we reformulate the Jacobi method such that it can be used when the matrix to invert is not constructed explicitly. Moreover, removing the diagonal from the matrix may also remove the low-rank property.
We therefore modify Eq.~\eqref{eq:Jacobi} to
\begin{equation}
    (\symvm{\Delta} \vm{f})^{(k+1)} = (\symvm{\Delta} \vm{f})^{(k)} -\vm{D}^{-1}\vm{J}(\symvm{\Delta} \vm{f})^{(k)} - \vm{D}^{-1}\vm{h}.
\end{equation}
In the Kronecker-product representation, the Jacobian is obtained by substituting the expression of $\vm{A}$ and $\vm{B}$ in Eqs.~\eqref{eq:A_kronecker} and \eqref{eq:B_kronecker} into Eq.~\eqref{eq:jacobian}.

The theory from Sec.~\ref{sec:separable_kernels} on separable kernels indicates that the Kronecker-product representation enables computations of matrix-vector products at a lower cost. Moreover, the Kronecker-product representation does not require storing all the matrices in their full size $N\times N$. Instead, the matrices are split into $N_e \times N_e$ and $N_a \times r_\ell'$ blocks (since $\symvm{\Omega}_\ell$ is of low rank and need not be stored fully). However, both advantages disappear if the Jacobi method does not converge, since we would then have to construct the Jacobian of size $N\times N$ explicitly and use the LU-decomposition of complexity $O(N^3)$. In the next section, we will show how to circumvent this issue while preserving the advantages of the Kronecker-product representation.

\subsubsection{Dimensionality reduction \label{sec:dimensionality_reduction}}

\noindent We have shown in Sec.~\ref{sec:low_rank_kernels} that our system of equations is in fact low-rank when expanding the kernels in a Legendre series up to order $\lmax$. This low-rank structure hints at the possibility of finding a lower-dimensional representation of the problem. In fact, we can reduce the size of our problem from solving a system of $N$ equations with $N$ unknowns to a system of $r=r' N_e$ equations with $r$ unknowns while maintaining the Kronecker-product properties. As a reminder, $r'=(\lmax+1)^2$ (or $\lmax +1$ in spherical symmetry) is the rank of the angular part of the kernels. We shall see that these $r$ unknowns correspond to the coefficients of the spherical harmonics expansion of the distribution function ($r'$ coefficients) for each of the $N_e$ energy-species bins.

From Eqs.~\eqref{eq:A_kronecker} and \eqref{eq:B_kronecker}, we introduce the following matrices of size $N_e N_a \times r' N_e$:
\begin{align}
    & \bvm{U}_{A/B} \equiv \begin{pmatrix} \symvm{\Phi}_{A/B,0} \otimes \vm{U}_0 & \cdots & \symvm{\Phi}_{A/B,\lmax} \otimes \vm{U}_{\lmax} \end{pmatrix}, \label{eq:U_kronecker}\\
    & \bvm{V} \equiv \begin{pmatrix} \mathbb{1}_{N_e} \otimes \vm{V}_0 & \cdots & \mathbb{1}_{N_e} \otimes \vm{V}_{\lmax} \end{pmatrix} \label{eq:V_kronecker}
\end{align}
We also introduced the $\bvm{\cdot}$ notation for block matrices, and we denote the matrix block as $\bvm{\cdot}_\ell$. For example, the block $\bvm{U}_{A,\ell} = \symvm{\Phi}_{A,\ell} \otimes \vm{U}_\ell$ is of size $N_e N_a \times r_\ell' N_e$, with $r_\ell' = (2\ell +1)$ the number of spherical harmonics of degree $\ell$.
With these definitions, we retrieve the matrices $\vm{A}$ and $\vm{B}-\vm{D}_B$ from the product $\bvm{U}_{A/B}\bvm{V}^T$. Similarly, we rewrite the Jacobian from Eq.~\eqref{eq:jacobian} as
\begin{align}
    &\vm{J} = \vm{D}_J + \vm{U}_J \bvm{V}^T \label{eq:jacobian_matrix}\\
    &\vm{D}_J = \vm{D}_B - \diag{\vm{w}_p} + \diag{\vm{A} \vm{f}} \label{eq:jacobian_matrix_diag}\\
    &\vm{U}_J = \bvm{U}_B + \diag{\mathbf{f}} \bvm{U}_A \label{eq:jacobian_matrix_outer}
\end{align}
As a reminder, the columns of $\vm{U}_\ell$ correspond to the discretization of the spherical harmonics $Y_\ell^m$, whereas $\vm{V}_\ell^T$ projects onto $Y_\ell^m$ and computes the coefficients of the spherical harmonics expansion. We define the coefficients of the spherical harmonics expansion of the distribution function as
\begin{equation}
    \bvm{y} = \bvm{V}^T \vm{f}.
\end{equation}
The resulting vector $\bvm{y}$ is organized as $\lmax+1$ blocks $\bvm{y}_\ell$ of size $r_\ell' N_e$. The block $\bvm{y}_\ell$ therefore contains the spherical harmonics coefficients for $\ell$ and $m\in {-\ell,\dots,\ell}$ for all energy bins.
Substituting that relation into our implicit-step equation~\eqref{eq:implicit_f}, we find
\begin{equation}
    (\bvm{U}_A \, \bvm{y}) \odot \vm{f} + \vm{D}_B \vm{f} + \bvm{U}_B \, \bvm{y} + \vm{c} - \vm{w}_p \odot \vm{f} = 0.
\end{equation}
Using the relation $\vm{a} \odot \vm{b} = \diag{\vm{a}} \vm{b}$ of the Hadamard product, we can express $\vm{f}$ as a function of $\vm{y}$ as
\begin{equation*}
    \vm{f} = -\big(\diag{\bvm{U}_A \, \bvm{y}} + \vm{D}_B  - \diag{\vm{w}_p}\big)^{-1}\left( \vm{c} + \bvm{U}_B \, \bvm{y} \right).
\end{equation*}
One can recognize the diagonal matrix as being the diagonal part $\vm{D}_J$ of the Jacobian from Eq.~\eqref{eq:jacobian_matrix_diag}. The last equation can then be rewritten
\begin{equation}
    \vm{f}(\bvm{y}) = -\vm{D}_J^{-1}\left( \vm{c} + \bvm{U}_B \, \bvm{y} \right).\label{eq:f_of_y}
\end{equation}
Note that $\vm{D}_J$ being diagonal, its inverse is straightforwardly formed and its product with a vector computed as an element-wise product.
Then, we use the function $\vm{f}(\bvm{y})$ to modify our master equation. Instead of solving $\vm{h} = 0$ from Eq.~\eqref{eq:implicit_f}, we now solve
\begin{equation}
    \vm{h}_y \equiv \bvm{y} - \bvm{V}^T \vm{f}(\bvm{y}) = 0. \label{eq:h_of_y}
\end{equation}
This equation ensures that the coefficients $\bvm{y}$ indeed represent the spherical harmonics coefficients of the distribution $\vm{f}$ that solves Eq.~\eqref{eq:implicit_f}. Once the root $\bvm{y}_*$ is found, we can retrieve the corresponding distribution function as $\vm{f}_* = \vm{f}(\bvm{y}_*)$.

We stress that this method is not equivalent to expanding $f$ in spherical harmonics, computing the source terms with the obtained expansion coefficients, and then reconstructing $f$ from the updated coefficients. The main reason is that, even if $\bvm{y}$ itself only contains the harmonic content up to $\lmax$, $f$ itself contains information beyond the maximum order $\lmax$ used in the Legendre expansion of the kernels, and interactions produce higher-order content through the $\vm{D}_J^{-1}$ factor (provided it is not isotropic). The presence of these higher harmonics is the reason why $f$ is reconstructed from $\vm{f}(\bvm{y})$, which contains all the information about the updated $f$, and not from $\sum_\ell \vm{U}_\ell \bvm{y}_\ell$, which only accounts for the harmonics up to $\lmax$.

To use the \ac{NR} method and solve Eq.~\eqref{eq:h_of_y}, we still need to compute its Jacobian $\vm{J}_y$. To this end, we need to compute the derivative of $\vm{f}(\bvm{y})$, which can be done through implicit differentiation of $\vm{D}_J \vm{f}(\bvm{y}) = -\left( \vm{c} + \bvm{U}_B \, \bvm{y} \right)$ from Eq.~\eqref{eq:f_of_y}. These calculations lead to the Jacobian
\begin{equation}
    \vm{J}_y = \mathbb{1}_{r} + \bvm{V}^T\vm{D}_J^{-1}\vm{U}_J, \label{eq:jacobian_of_y}
\end{equation}
where $\vm{U}_J$ is defined in Eq.~\eqref{eq:jacobian_matrix_outer}.
Note that this Jacobian can also be found after using the Woodbury matrix identity on the Jacobian in Eq.~\eqref{eq:jacobian_matrix}. 

Besides the dimensionality reduction, an advantage of this new formulation is that the Kronecker-product structure is retained in $\bvm{V}^T$ and $\bvm{U}_{A/B}$, which allows for more efficient matrix-vector product computations. The details of the implementation of the matrix-vector products in the Kronecker-product representation, as well as their time and memory complexity can be found in Appendix~\ref{app:kronecker_product}.

We summarize here the main results. We reduced the size of the problem from $N = N_e N_a$ to $r = r' N_e$, with $r' = (\lmax+1)^2$ (or $(\lmax+1)$ in spherical symmetry). The use of the LU-decomposition method, which requires explicitly constructing the Jacobian, is of complexity $O(r'^2 N_a N_e + r'^3 N^3_e)$. The factor $N_e^3$ from the second term is a similar asymptotic behaviour as a truncated moment scheme. For $\lmax=1$, this dominating term is the same as in an M1 scheme. For the Jacobi method, the time complexity is $O(K r' N_a N_e + K r' N_e^2)$, where $K$ is the number of iterations in the Jacobi method to converge to a solution. Thus, for all these operations, the main source of acceleration is that the heaviest computations only depend on $N_e$ and $r'$, but not on $N_a$. The cost of the implicit solver may thus effectively depend sublinearly on $N_a$ if $r' N_e^2$ is larger than $N_a$. This method is more efficient if $\lmax$ is small (or $r' = (\lmax+1)^2$ is small compared to $N_a$). As mentioned in Sec.~\ref{sec:source_terms}, our solver currently uses a first-order Legendre expansion of the interaction kernels ($\lmax=1$).

\subsection{Discussion and comparison \label{sec:optimization_discussion}}

\noindent Both the low-rank approach and iterative methods have their advantages and drawbacks.
The low-rank approximation allows one to decouple the energy and angular parts of the kernel to reduce the size of the problem and lower the memory requirements. This however comes at the price of approximating the angular dependence of kernels. Such an approximation is between the M1 scheme, where $f$ itself is also low-rank, and the full Boltzmann solver.

Iterative methods for solving linear systems can be used independently of the low-rank approximation, including in an M1 scheme. In contrast to the low-rank scheme, they do not require any additional approximations in the physics model. Their efficiency depends on the convergence rate of the method and the computational time of an iteration. If the method's convergence rate is small, the total computational time of the implicit time step may increase. The convergence criteria and maximum number of iterations must thus be chosen to minimize this risk. These methods are particularly efficient when the interactions are weak or the time step is small. In our case, we found the Jacobi iterative method to be efficient because it requires few operations per iteration.

We also point out that the complexities mentioned in the previous sections are valid for the \ac{MSMG} mode (see Sec.~\ref{sec:implicit_modes}). For \ac{SSMG}, we have $N_e = N_\varepsilon$ and the solver is used independently for all $N_\mathrm{spec}$. The complexity in $N_\mathrm{spec}$ is thus linear. Similarly, for \ac{SSSG}, $N_e=1$ and the complexity is linear in $N_\mathrm{spec} N_\varepsilon$. Moreover, $\vm{A}= \symvm{0}$ in the \ac{SSSG} mode and the implicit update reduces to solving a linear system. All the methods described in the previous sections can then be applied on this system instead of the \ac{NR} iteration.

We shall now compare these methods quantitatively. We consider the full method (F), where the full system of $N=N_e N_a$ equations is solved directly, and the Low-rank method with Kronecker-product representation (L). For both methods, we also consider the use of the Jacobi iterative method (J) to solve the \ac{NR} iteration. The method is considered to fail if the number of iterations exceeds the maximum number \textit{it\_max}. The method is considered to converge if the relative change in the iterate is smaller than a threshold \textit{tol}. Note that, as explained in Sec.~\ref{sec:iterative}, the method is stopped if the estimated number of iterations before convergence is larger than \textit{it\_max}. If the latter occurs or the method does not converge, the LU method is used to solve the \ac{NR} iteration. We denote the Jacobi method with maximum iteration \textit{it\_max} and tolerance \textit{tol} as J$^{-\log(\textit{tol})}_{\textit{it\_max}}$. For this comparison, we use J$^{10}_{50}$, J$^{10}_{20}$ and J$^{6}_{20}$.

As a representative test, we consider a 1D \ac{CCSN} simulated with M1 radiation transport, as described in \texttt{Gmunu}'s M1 code paper~\citep{Cheong_2023}. We use the SFHo equation of state (EOS) of Ref.~\cite{Steiner_2013} and a neutrino interaction table from \texttt{Weakhub}~\citep{Ng_2024} with $N_\varepsilon=16$ energy bins. We use the conventional set of interactions~\citep{Bruenn_1985}, with nucleon-nucleon bremsstrahlung included as an emissivity. We perform our test from a snapshot at about $20\,$ms after core bounce, and use the zeroth and first moments to reconstruct the distribution function. Then, we time a single implicit step for each method with $\Delta t = 5\times 10^{-4}$ (in code units), including the construction of the matrices that each method needs, as well as the computation time of the source terms. The complexity, however, is dominated by the \ac{NR} method. We only consider the \ac{SSMG} and \ac{MSMG} modes, since \ac{SSSG} does not require a rootfinding method. We repeat the tests $20$ times on the same machine to account for variations in execution times and present the results as averages over these runs. To ensure a fair comparison of the Jacobi method at different resolutions $N_\vartheta$, we enforce the same time step for each resolution. Finally, we stress that the full method F still uses the Legendre expansion of the kernels, but does not exploit the induced low-rank structure.

\begin{figure}
    \centering
    \includegraphics[width=\linewidth]{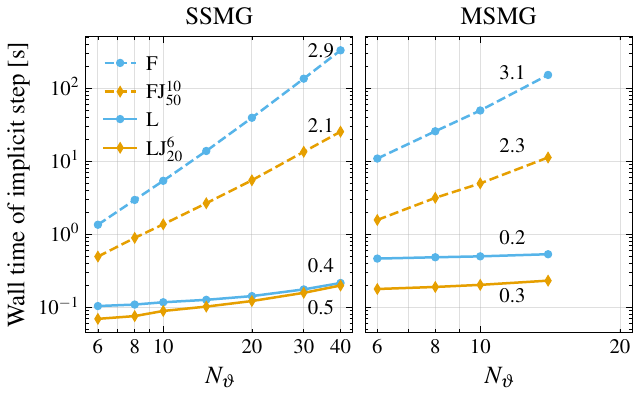}
    \caption{Comparison of the total computation time of the implicit step as a function of $N_\vartheta$ for the full method (F) and the low-rank method (L), for both the \ac{SSMG} (energy coupling, left) and \ac{MSMG} (species and energy coupling, right) modes. The methods FJ$^{-\log(\textit{tol})}_{\textit{it\_max}}$ and LJ$^{-\log(\textit{tol})}_{\textit{it\_max}}$ correspond to the methods F and L where the Newton-Raphson iteration is solved using the Jacobi method with a tolerance \textit{tol} for convergence and a maximum number of allowed iterations \textit{it\_max}. We indicate the approximate slope next to each curve. This slope represents the complexity, which approximately matches the theoretical expectations for F and FJ. For L and LJ, the dependence in $N_\vartheta$ of the theoretical complexity $O(N_a N_e + N_e^3)$ is dominated by the $N_e^3$ ($N_e^2$ for LJ) term, which explains the sublinear dependence. The slope of L and LJ is however expected to approach $1$ as $N_a$ increases.}
    \label{fig:comp_time}
\end{figure}

In Fig.~\ref{fig:comp_time}, we show the scaling of the methods F and L (with and without Jacobi iterative method) as a function of $N_\vartheta$. As expected, the full method F scales cubically in both \ac{SSMG} and \ac{MSMG}. The Jacobi method allows for a better complexity for the F method, although it does not scale exactly quadratically as predicted. This discrepancy is explained by the fact that the iterative method does not converge for all points and is thus a mix between the LU decomposition and the Jacobi method. For the low-rank method, both L and LJ scale sublinearly, even though the theoretical complexity ($O(r'^2 N_e N_a + r'^3 N_e^3)$ for L and $O(r' N_e N_a + r' N_e^2)$ for LJ) includes a term linear in $N_a$. The reason is that this linear term is subdominant compared to the contribution of the $N_e^3$ (or $N_e^2$) at small $N_a$. As $N_a$ grows, the $N_a$ term becomes relatively more important, and we expect the slope to approach 1. This regime is not reached yet in Fig.~\ref{fig:comp_time}, even though we can see the slope increasing in the \ac{SSMG} mode. For \ac{MSMG}, on the other hand, $N_e$ is larger (it contains a factor $N_\mathrm{spec}$), such that this behaviour only occurs at larger $N_a$. For the $N_a$ tested in Fig.~\ref{fig:comp_time}, the computation times are almost constant with increasing $N_a$ in the \ac{MSMG} mode. We can also see that the L method is about $300$ times faster than the F method at $N_\vartheta = 14$ for \ac{MSMG} and at $N_\vartheta = 20$ for \ac{SSMG}. The speed-up provided by the $L$ method goes up to $\sim 10^3$ for $N_\vartheta = 40$ in the \ac{SSMG} mode. When including the Jacobi method, the $LJ$ methods become about $30$ times faster than the FJ method at $N_\vartheta = 14$ for \ac{MSMG} and at $N_\vartheta = 20$ for \ac{SSMG}, and about $100$ times faster at $N_\vartheta = 40$ for the \ac{SSMG} mode.

\begin{figure}
    \centering
    \includegraphics[width=\linewidth]{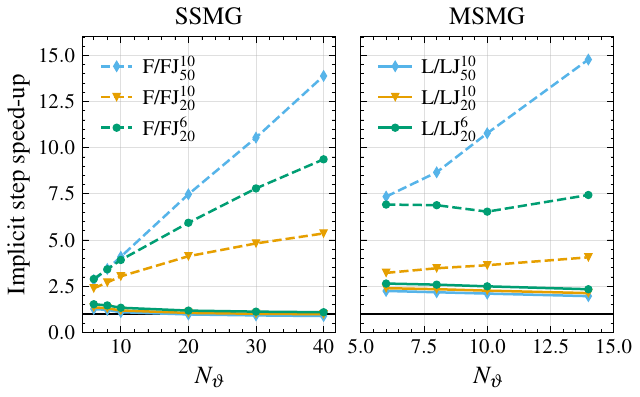}
    \caption{Acceleration provided by the Jacobi method for solving the \ac{NR} iteration for the \ac{SSMG} (left) and \ac{MSMG} (right) modes. The methods FJ$^{-\log(\textit{tol})}_{\textit{it\_max}}$ and LJ$^{-\log(\textit{tol})}_{\textit{it\_max}}$ correspond to the full method (F, dashed lines) and low-rank method (L, solid lines) using the Jacobi method with a tolerance \textit{tol} for convergence and a maximum number of allowed iterations \textit{it\_max}. The black line corresponds to $1$ (no acceleration). For F, the acceleration is expected to scale linearly, since the complexity is quadratic in $N_\vartheta$ for FJ and cubic for F. For L, on the other hand, we expect first the speed-up to decrease with $N_\vartheta$, since the regime at which the $N_\vartheta$ term in the complexity dominates occurs at lower $N_\vartheta$ for LJ than for L. At large $N_\vartheta$, we expect the speed-up to converge to a fixed value, as seen in \ac{SSMG}. }
    \label{fig:comp_jacobi}
\end{figure}

We see from the previous plots that the Jacobi methods provide different acceleration levels for each method and each implicit solver mode. In Fig.~\ref{fig:comp_jacobi}, we show the ratio of the computation time of the FJ (LJ) methods over the F (L) method, respectively, for the \ac{SSMG} mode on the left and \ac{MSMG} mode on the right. Dashed lines represent the F and FJ methods, and solid lines the L and LJ methods. First, we can see that the speed-up scales at most linearly with $N_\vartheta$ for F and FJ. This is expected, since F scales cubically and FJ quadratically in $N_\vartheta$. As for L and LJ, their ratio decreases or levels out with increasing $N_\vartheta$. At smaller $N_\vartheta$,  the ratio L/LJ scales approximately as $r'^2 N_e/K$, where $K$ is the number of iterations required for convergence. In the large $N_\vartheta$ limit, the ratio L/LJ tends to $r'/K$. If $K$ does not depend on $N_a$, the ratio tends to a fixed number in the large $N_a$ limit, which may be above or below $1$ depending on the number of iterations required to converge. Note also that the number of points at which the term linear in $N_\vartheta$ becomes relatively important is smaller for the LJ methods than the L method. This also partially explains why the speed-up of the LJ methods compared to L decreases with $N_\vartheta$.

We also observe different behaviours for different methods and different modes. First, the acceleration tends to be larger for \ac{MSMG} compared to \ac{SSMG}, because the number of elements in the matrices is larger in \ac{MSMG} by a factor $N^2_\mathrm{spec}$. Then, for the FJ methods in \ac{MSMG}, we see that decreasing \textit{it\_max} while keeping the same tolerance significantly decreases the acceleration. This is due to the fact that there are then more occurrences of failed convergence, and hence more use of the more expensive F methods. Similarly, increasing the tolerance increases the number of successful occurrences of the Jacobi method, and/or decreases the number of iterations. The latter behaviour is also observed for the LJ methods, whereas their behaviour differs for the decrease of \textit{it\_max}. We indeed see that the performance is slightly better for LJ$^{10}_{20}$ than LJ$^{10}_{50}$. A possible explanation is that the penalty for a failed Jacobi method is lower than for the F method, and the cost of all failed convergence in LJ$^{10}_{20}$ may be similar to the cost of the additional iterations in LJ$^{10}_{50}$ leading to convergence.

As a reminder, the numbers reported in this section are obtained for a single implicit step, and timings are summed over the whole grid. The resulting acceleration on the total simulation would thus also depend on the number of implicit steps performed in the IMEX scheme, on the tolerance used for the \ac{NR} method and on the number of points in more or less expensive parts of the grid. Moreover, the results presented in this section may not hold at all times of the \ac{CCSN} or in different systems, especially for the Jacobi methods. The convergence and efficiency of this method indeed depend on the strength of the interactions and on the time step. The time step used for these tests was smaller than the ones that would be used in an actual simulation, especially at small $N_\vartheta$. Thus, the acceleration provided by the Jacobi method reported in Fig.~\ref{fig:comp_jacobi} should be considered optimistic. Better criteria for convergence and use of the Jacobi method, for example as a function of the maximum opacity, can be used to minimize the cost of the implicit solver. 

Despite the acceleration provided by the L and LJ methods, simulating \acp{CCSN} with $N_\vartheta=40$ is likely still unaffordable in multiple dimensions, since the global cost of the implicit step remains large ($\sim 0.2$s in 1D \ac{SSMG}) and the maximum time step also depends on the angular resolution in momentum space (see Sec.~\ref{sec:coord}). On the other hand, $N_a = N_\vartheta N_\varphi$ does reach values around $60$ in the lowest resolution simulations. To get an order of magnitude estimate of the acceleration in multiple dimensions, we measure the total computation time of the implicit step for an isolated neutron star (using the same EOS and neutrino interactions) in 2D. We limit the domain to the neutron star itself. For $N_a = 60$ ($N_\vartheta = 10$, $N_\varphi = 6$), we find a speed-up F/L of order $2\times 10^3$ for \ac{SSMG}, which is of the order of what is expected from our 1D tests ($10^3$ for $N_\vartheta = 40$).

Finally, we point out that the Jacobi method favours regions with weak interactions, accelerating the computation in these regions even more than the other regions. As a consequence, the load imbalance between cores may be increased by the use of this method.

\section{Numerical tests \label{sec:tests}}

\noindent Following the description of the solver in all the previous sections, we see that the solver contains multiple features that must be tested. First, there is the advection part. Contrary to classical $3+1$ solvers, our solver must account for advection in momentum space. Moreover, our method is strongly coupled to the velocity field, since we discretize momentum in the fluid rest frame. Second, there is the interaction part. The associated implicit solver requires solving large systems of equations, which may then be prone to numerical errors. We must also retrieve the diffusive limit in regions of large scattering opacities.

In this section, we validate the implementation of our solver on established test cases. First, we analyse the accuracy of the advection part of the equation in both position and momentum spaces in Sec.~\ref{sec:tests_sch}. Then, we introduce interactions between neutrinos and fluid in Sec.~\ref{sec:tests_interaction}, more precisely elastic scattering, emission and absorption. We include a test with shocks to evaluate the performance of the code in the presence of strong velocity gradients, as well as interactions in curved spacetimes, in Sec.~\ref{sec:advection_collisions_curved}.

Unless otherwise specified, the tests that are presented were performed on uniform grids (in position and momentum space) with the HLL Riemann solver, the second-order monotonized central limiter~\citep{Van_Leer_1977,Woodward_1984} and the IMEX-SSP2(2,2,2) time integrator~\citep{Pareschi_2005}. In tests where there are no interactions, Heun's method is used as a time integrator. For all tests, neutrinos are considered massless.

\begin{figure*}
    \centering
    \begin{minipage}{0.48\linewidth}
        \centering
        \includegraphics[width=\linewidth]{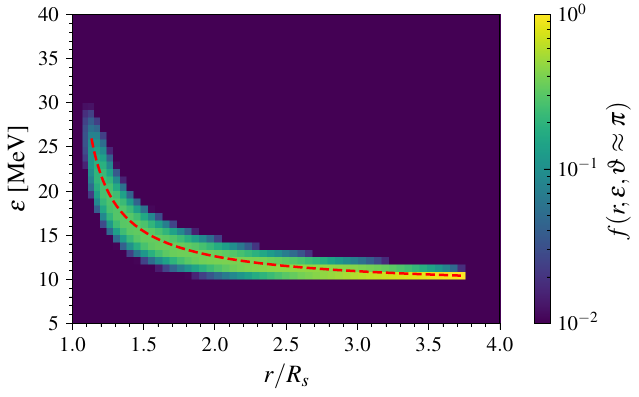}
    \end{minipage}%
    \hfill
    \begin{minipage}{0.48\linewidth}
        \centering
        \includegraphics[width=\linewidth]{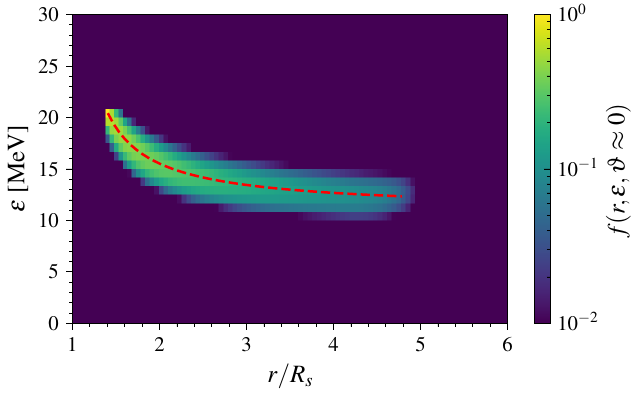}
    \end{minipage}
    \caption{Results of the test for neutrinos going towards a black hole (left) and away from a black hole (right). The analytical trajectory in energy space is shown by a red dashed line. Our solver correctly reproduces the analytical behaviour in both cases.}
    \label{fig:schwarzschild_rad}
\end{figure*}

\subsection{Advection in Schwarzschild spacetime \label{sec:tests_sch}}
\noindent As a first test, we focus on the part that extends the standard spacetime framework: the momentum space. We start our investigation with the simplest setup containing energy and angular advection, \textit{i.e.} empty Schwarzschild spacetime, to ensure that the basic advection is correctly handled by the solver. The following test setups are based on Ref.~\cite{Akaho_2021}. We consider a central black hole of mass $3.62\, M_\odot$, which corresponds to a Schwarzschild radius $R_s \approx 10.69\,$km. All the tests are performed in $1$D in isotropic coordinates. The domain is also fixed for all the tests. The domain is $r \in [1.92,51.92]$ (which approximately corresponds to $[10.7,82.1]\,$km in Schwarzschild coordinates), $\varepsilon\in[0, 50]\,$MeV and $\vartheta\in[0,\pi]$.

\subsubsection{Energy advection in Schwarzschild spacetime \label{sec:tests_sch_rad}}
\noindent In this test, we consider a beam of neutrinos propagating radially towards or away from the black hole. For both tests, we use $N_r=128$, $N_\vartheta=20$ and $N_\varepsilon=60$. Note that the code solves the equation in isotropic coordinates, but we show the results in Schwarzschild coordinates.

First, we consider a source of neutrinos ($f = 1$) emitting continuously outwards from a radius $r_s = 15\,$km and at an energy $\varepsilon_s = 20\,$MeV. We set $f=0$ on the rest of the domain. Neutrinos are thus redshifted as they travel outwards. The neutrino energy (or frequency) measured in the comoving frame in Schwarzschild coordinates is given by (see Eq.~\eqref{eq:eps_curv})
\begin{equation}
    \varepsilon_\mathrm{true}(r) = \sqrt{\left(1-\frac{R_s}{r_s}\right) \left(1-\frac{R_s}{r}\right)^{-1}} \, \varepsilon_s \label{eq:schw_eps}
\end{equation}
Then, we consider a source of neutrinos emitting continuously inwards from $r_s = 40\,$km at an energy $\varepsilon_s = 10\,$MeV. The trajectory in energy space given in Eq.~\eqref{eq:schw_eps} will here correspond to a blueshift. The results of both tests are shown in Fig.~\ref{fig:schwarzschild_rad}. We can see that our solver's prediction aligns with the analytical solution (red dashed line). The width of the distribution function around this line corresponds to numerical diffusion.

\subsubsection{Angular advection in Schwarzschild spacetime \label{sec:tests_sch_ang}}
\noindent Similarly to the two previous tests, we test the propagation of the neutrinos around a black hole. In this case, we focus on the angular advection. We now consider neutrinos emitted with $\vartheta_s = \pi/2$, \textit{i.e.} perpendicular to the radial axis, and $\varepsilon_s = 25\,$MeV. To ensure that we get exactly the cell-centred value $\vartheta = \pi/2$, we only use odd numbers of points in the $\vartheta$ space. We fix the number of points to $N_\varepsilon = 20$ and $N_\vartheta = 41$.

\begin{figure*}
    \centering
    \begin{minipage}{0.48\linewidth}
        \centering
        \includegraphics[width=\linewidth]{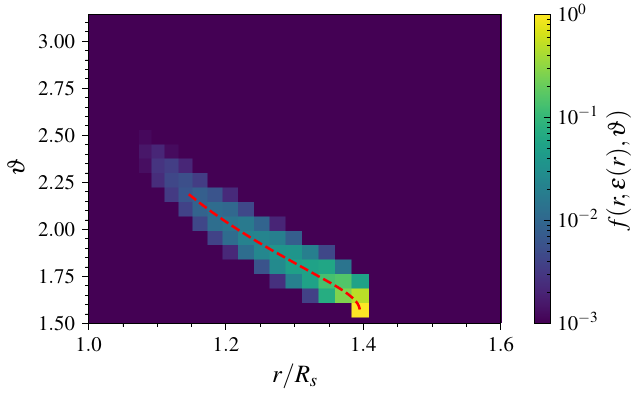}
    \end{minipage}%
    \hfill
    \begin{minipage}{0.48\linewidth}
        \centering
        \includegraphics[width=\linewidth]{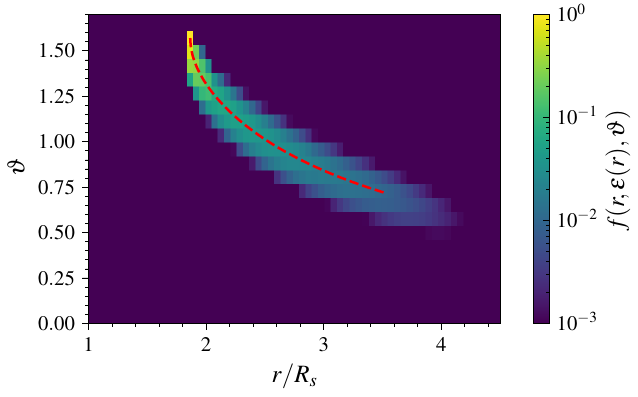}
    \end{minipage}
    \caption{Results of the test for neutrinos emitted at an angle $\vartheta=\pi/2$ inside the photon sphere (left) and outside the photon sphere (right). The analytical trajectory in angular space is shown by a red dashed line. The predictions of our solver match the analytical trajectory.}
    \label{fig:schwarzschild_ang}
\end{figure*}

In the first test, the source of neutrinos is outside the photon sphere $R_p = 3R_s/2 \approx 16\,$km. The emitted neutrinos will thus spiral outwards. We choose the source radius to be $r_s = 20\,$km and a grid with $N_r = 128$. In the second case, the source is located at $r_s = 15 \, \mathrm{km} < R_p$, so that the emitted neutrinos will spiral inwards. Because the motion occurs on a more restricted range of $r$, we use $N_r=256$ for this test. The deflected motion of the neutrinos around the black hole is represented by a change in the propagation direction, hence of $\vartheta$. The results are shown in Fig.~\ref{fig:schwarzschild_ang}. We can see that, in both cases, the change in propagation angle is properly accounted for by the solver.

\subsection{Interaction with a fluid \label{sec:tests_interaction}}
\noindent In this section, we analyse the performance of our solver when including interaction with the fluid. We limit the interactions to absorption, emission and elastic scattering. 

\subsubsection{Shadow casting test}
\noindent We consider a beam of neutrinos propagating through a cylinder of radius $R$ inside which neutrinos are absorbed. The absorption by the cylinder will then leave a shadow behind it. The test setup is the same as in \texttt{Gmunu}'s M1 paper~\citep{Cheong_2023}. We place the cylinder at $(-0.2,0)$ with a radius of $R=0.07$ and let neutrinos propagate from left to right. We use Cartesian coordinates in position space and use the $z$-axis as the reference axis for the momentum space. We therefore set $f=1$ for $\vartheta = \pi/2$ and $\varphi = 0$. For this test, we disable the momentum advection and use a limited number of points in momentum space, since $\tensor{\Gamma}{^\hrho_{\hmu\nu}}=0$. The position space has $256\times 128$ points with coordinates in $[-0.5,0.5]$ and $[-0.25, 0.25]$ for $x$ and $y$, respectively. We set $\kappa_a^* = 10^6$ inside the cylinder to test our solver under stiff conditions.

\begin{figure}[h]
    \centering
    \includegraphics[width=\linewidth]{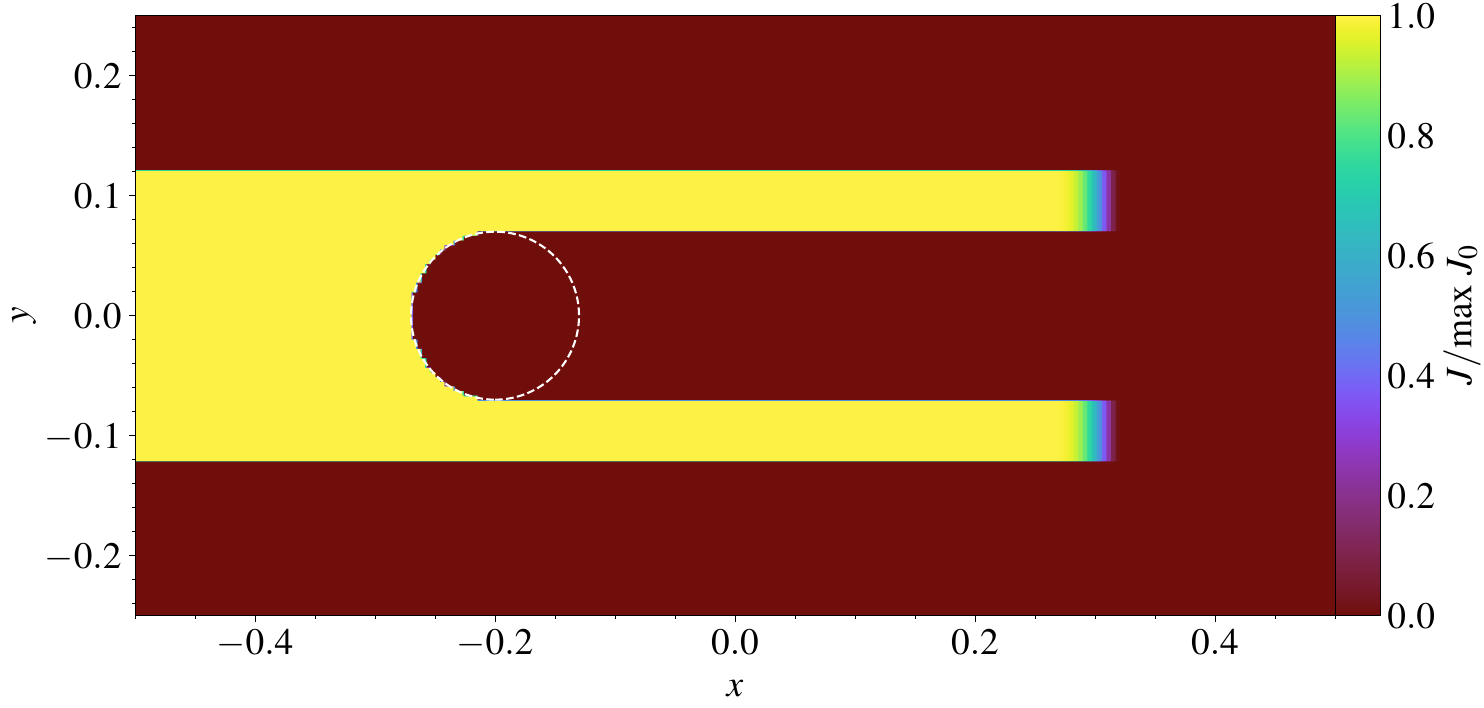}
    \caption{Distribution function at $t=1$, $\vartheta = \pi/2$ and $\phi = 0$ for the shadow casting test from a beam crossing an absorbing cylinder (white dashed line) with $\kappa_a^*=10^6$. The beam propagates from left to right and the shadow is clearly visible behind the cylinder. There is little numerical diffusion on the upper and lower sides of the beam.}
    \label{fig:shadow}
\end{figure}

The results of the test are shown in Fig.~\ref{fig:shadow}, where we plot the distribution function at $t=1$ for $\vartheta = \pi/2$ and $\phi = 0$. The absorption of particles causes a shadow to appear within the beam behind the cylinder, as expected. Contrary to the M1 (see~\cite{Cheong_2023}), there is very little diffusion on the side of the beam because neutrinos propagate almost exactly in the positive $x$ direction. A close look indeed reveals a slight diffusion towards increasing $y$ values only, reflecting the small misalignment of the propagation angle with the $x$ direction. Numerical diffusion would be more significant if the beam propagated diagonally, as will be seen in the next test.

\subsubsection{Homogeneous radiating sphere \label{sec:tests_radiating_sphere}}
\noindent In this test, we consider a spherically symmetric system with a sphere of radius $R$ emitting and absorbing neutrinos. More precisely, we consider $\eta = B \kappa_a$ for $r<R$ and $0 < B \leq 1$. The analytic solution to this problem is derived in Appendix~\ref{app:analytic_sol} and given in Eqs.~\eqref{eq:general_sol_eq} and~\eqref{eq:sol_rad_sphere}.

\begin{figure}[h!]
    \centering
    \includegraphics[width=\linewidth]{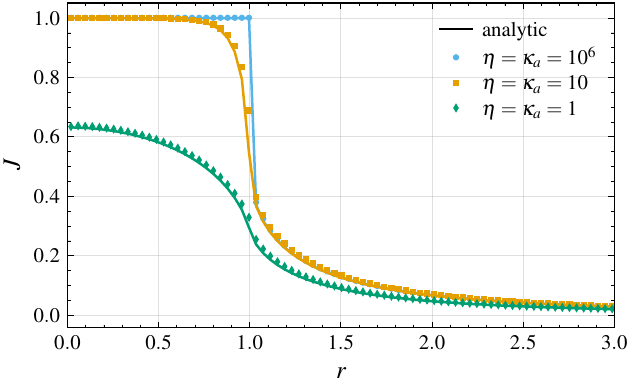}
    \caption{Comparison between the analytical solution and simulation for the homogeneous radiating sphere test. The blue line corresponds to an optically thin medium with $\kappa_a=1$ whereas the orange line ($\kappa_a = 10^6$) represents the optically thick regime. Contrary to the M1 scheme, the Boltzmann solver is able to represent accurately both the optically thin and thick limits.}
    \label{fig:rad_sphere}
\end{figure}

We consider three different cases, from semi-transparent to optically thick regimes with $\eta = \kappa^*_a = 1, 10$ and $10^6$. We consider a spatial grid $N_r = 128$ with $r\in[0,5]$ and a momentum space grid $N_\varepsilon = 1$, since there is no energy coupling or advection, and $N_\vartheta = 20$. 
The simulation is shown in Fig.~\ref{fig:rad_sphere} where we see that the simulation agrees with the analytical solution. Contrary to the M1 (see~\cite{Cheong_2023}), our method properly resolves both optically thin and thick limits, whereas the M1 is less accurate in the optically thin limit.

\subsubsection{Diffusion of a Gaussian profile \label{sec:tests_diffusive}}
\noindent We will now test the validity of the adjustment given in Eq.~\eqref{eq:diffusive} to model the diffusive limit correctly. We consider a spherically symmetric system with a large scattering opacity $\kappa_s = 10^3$. In practice, we set $R_{ES,0} = \kappa_s (4\pi \varepsilon^2)^{-1}$ and $R_{ES,1}=0$, since $\kappa_s$ is defined as in Eq.~\eqref{eq:kappa_s}. We use a spatial grid of $N_r = 256$ points over a domain $r\in[0,4]$. The resulting $\Delta r$ is larger than the mean free path of neutrinos such that the correction from Eq.~\eqref{eq:diffusive} does not vanish. The momentum space grid contains $N_\vartheta=20$ angular bins. Again, since there is no energy coupling or advection, we use $N_\varepsilon=1$. Similarly to Refs.~\cite{Swesty_2009,Sumiyoshi_2012}, we consider an initial Gaussian profile $f(t=0,\vec{r}) = e^{-r^2/d_0^2}$ with $d_0 = 1\,$km and let it evolve until about $t=1\,$ms.

In the limit of $\kappa_s \gg 1$, $f$ is almost isotropic and can thus be approximated by its first two moments $f \approx J + 3 H^{\hmu} l_{\hmu}$. We consider a Minkowski space with no fluid. In that case, it can be shown that the Boltzmann equation becomes equivalent to a diffusion equation for $J$, assuming that the components of $H^{\hati}$ do not change with time~\citep{Bruenn_1985}. The neutrino flux is then given by
\begin{equation}
    H^{\hatj}(\vec{r}) = -\frac{L^{i\hatj}}{3 \kappa_s} \partial_i J  \label{eq:diff_F}
\end{equation}
The analytical solution for the zeroth-order moment is the solution to the diffusion equation and is thus~\citep{Sumiyoshi_2012}
\begin{equation}
    J(t,\vec{r}) = J_0 \, \left(\frac{t_0}{t+t_0} \right)^{3/2} e^{-\frac{\vec{r}^2}{4D(t+t_0)}}, \label{eq:diff_E}
\end{equation}
where $t_0 = d_0^2/4D$ and $D=(3\kappa_s)^{-1}$.
The analytical solution for $f$ in spherical coordinates is therefore given by
\begin{equation}
    f(t,\vec{r},\vartheta) = J(t,\vec{r})~\left( 1 + \frac{3 r \cos\vartheta}{2 (t + t_0)} \right),
\end{equation}
and in Cartesian coordinates by
\begin{equation}
    \begin{split}
        &f(t,\vec{r},\vartheta,\varphi) = J(t,\vec{r}) + \\
        & \frac{3J(t,\vec{r})}{2 (t + t_0)} \left(x\cos\vartheta + y\sin\vartheta \cos\varphi + z \sin\vartheta \sin\varphi \right),
    \end{split}
\end{equation}
where the $x$-axis is the reference axis for the momentum space angles.
Note that the solution for $f$ is still spherically symmetric, but the addition of $\varphi$ in the coordinates is necessary to represent this symmetry in Cartesian coordinates.

\begin{figure}[h]
    \centering
    \includegraphics[width=\linewidth]{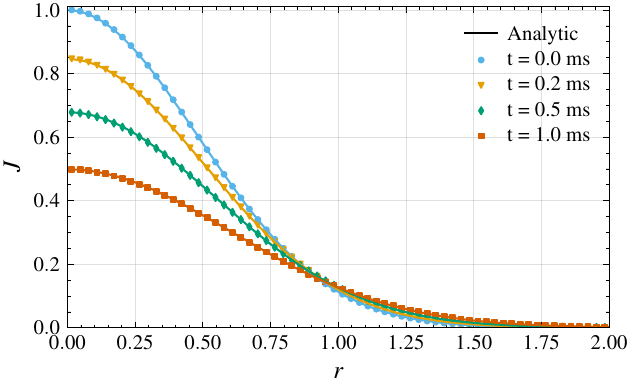}
    \caption{Diffusion of a Gaussian packet in spherical symmetry with $\kappa_s = 10^3$. The simulation (dots) agrees with the analytical solution (solid lines) at different times, validating the correction for the diffusive limit.}
    \label{fig:diffusive}
\end{figure}

The result of the simulation is shown in Fig.~\ref{fig:diffusive}. We plot the zeroth-order moment $J(t,r)$, but the results hold more generally for $f(t,r)$. We can see that the simulation agrees with the analytical solution. We note that the region close to $r=0$ is more sensitive to $N_\vartheta$ than at larger radii, because the momentum advection occurs mainly at small $r$.

We perform a convergence test to ensure that the diffusive limit, the implicit evolution of the source term and the angular advection converge quadratically, as expected with the method used. We do the convergence test in position and momentum space separately. For each of them, we use the highest resolution in the other space to make sure that the error is dominated by the resolution in the considered space. The error considered for the convergence is the difference between the analytical solution and the simulation averaged over $r$ and $\vartheta$, computed as
\begin{equation}
    \begin{aligned}
    &|| f(t,r,\vartheta) - f_{\mathrm{true}}(t,r) || = \\
    &\frac{\sum_{i,j} \left|f(t,r_i,\vartheta_j)-f_{\mathrm{true}}(t,r_i)\right| r_i^2 \sin\vartheta_j \Delta r_i \Delta \vartheta_j}{\sum_{i,j} r_i^2 \sin\vartheta_j \Delta r_i \Delta \vartheta_j}
    \end{aligned}\label{eq:error_ks}
\end{equation}

\begin{figure*}
    \centering
    \includegraphics[width=0.95\linewidth]{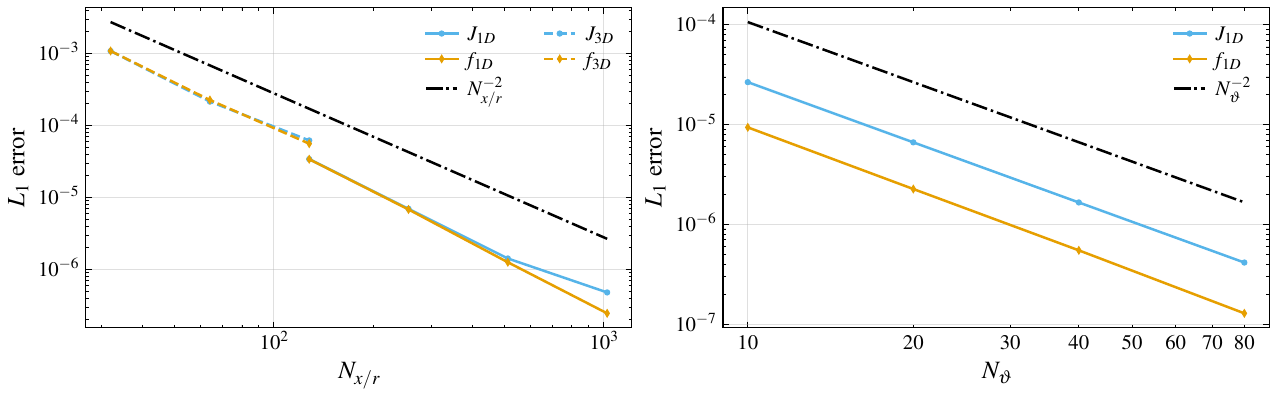}
    \caption{Convergence test for the diffusive limit test case in position space with $N_\vartheta = 80$ (left) and in momentum space with $N_x = 2048$ (right). The error on $f(t,\vec{r},\vec{p})$ and the zeroth-order moment $J$ both converge approximately quadratically. The change in slope for $J$ for the spatial convergence is due to the fact that the error is then dominated by the momentum-space resolution.}
    \label{fig:ks_conv}
\end{figure*}

For the zeroth-order moment, the expression is similar except that we do not average over $\vartheta$. Similarly, we perform a convergence test on the $3$D case in Cartesian coordinates. In the latter case, we use $\kappa_s = 100$, $N_x = 32,64,128$ for each spatial dimension, $N_\varepsilon =1$, $N_\vartheta=20$ and $N_\varphi = 40$. The error is computed in a similar way as Eq.~\eqref{eq:error_ks}, except that the spatial volume element is given by $\Delta x \Delta y \Delta z$ and the (numerical) solution also depends on $\varphi$. We do not study the convergence of angles in momentum space in the Cartesian case because there is no momentum advection in this case. We show the results of the convergence tests in Fig.~\ref{fig:ks_conv}. We can see that the error converges quadratically both in $r$ and $\vartheta$, as expected. The change in slope of $J$ observed for the last point in spatial convergence can be explained by the fact that the error due to the resolution in $\vartheta$ dominates the error due to the resolution in $r$ or $x$.

\subsubsection{Diffusive limit in a moving medium}
\noindent This test extends the previous ones by adding a background velocity field. We choose a setup similar to Refs.~\cite{Radice_2022} and~\cite{Cheong_2023}. We consider a Gaussian pulse
\begin{equation}
    f(x,\vartheta) = e^{-x^2/\sigma^2},
\end{equation}
where we choose $\sigma = 1/3$. This corresponds to a trapped radiation pulse, \textit{i.e.} $\mathcal{H}^\mu=0$. We assume slab geometry and axisymmetric momentum space. We use $N_x = 1024$ spatial points over the domain $x \in [-5,5]$, $N_\varepsilon = 1$, and $N_\vartheta = 20$. Similarly to the previous case, we consider a purely scattering medium with $\kappa_s = 10^3$. The background velocity is set to $v^x = 0.5$.

\begin{figure}[h]
    \centering
    \includegraphics[width=\linewidth]{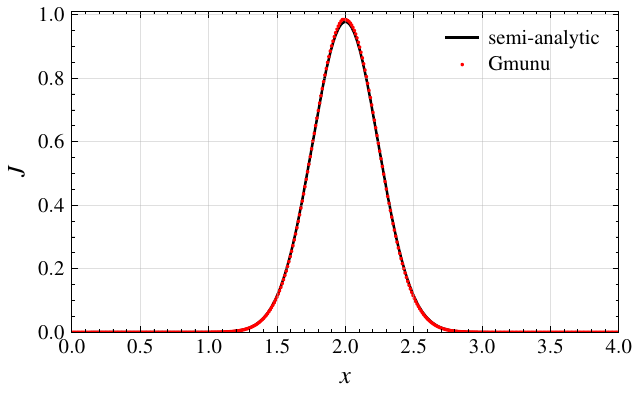}
    \caption{Diffusion of a Gaussian packet in a moving medium in slab geometry with $\kappa_s = 10^3$ and $v_x = 0.5$. The predictions of the solver agree with the semi-analytical solution.}
    \label{fig:diffusive_moving}
\end{figure}

The zeroth-order moment computed in the test is shown in Fig.~\ref{fig:diffusive_moving}. We can see that the predictions of the solver match the semi-analytical solution.

\subsection{Advection and collisions in fixed curved spacetime \label{sec:advection_collisions_curved}}

Because advection is based on comoving-frame momenta, it is crucial to study the behaviour of the solver in the presence of shocks in the fluid. Moreover, collisions typically occur on a curved background.
However, our previous tests have focused on either collisions in flat spacetime or advection in empty curved spacetime. Therefore, we shall now study both collisions and advection in curved spacetime in the presence of matter.

The tests in this section approach more realistic conditions than in the previous sections. To evaluate the performance of the code in more representative setups, we use an energy-grid spacing similar to that of interaction tables. More precisely, we consider a uniform spacing of $\Delta \varepsilon_u = \varepsilon_{\mathrm{max},u}/N_u$, where $\varepsilon_{\mathrm{max},u}$ is the maximum energy for the uniform grid and $N_u$ is the number of energy bins in the uniform grid. The spacing for the next bins is determined through $\Delta \varepsilon_i = a^i \Delta \varepsilon_u$ for $i=1,\dots, N_\varepsilon-N_u$. The scale factor $a$ is determined such that $\varepsilon_{\mathrm{max},u} + \Delta \varepsilon_u \sum_i a^i = \varepsilon_{\mathrm{max}}$, where $\varepsilon_{\mathrm{max}}$ is the maximum interface value of the energy grid. We implicitly assumed that $\varepsilon_\mathrm{min} = 0\,$MeV. This method was directly inspired by the publicly available code \texttt{NuLib}~\citep{O'Connor_2015}. Unless stated otherwise, we use $N_u=2$ uniform bins up to an energy $\varepsilon_{\mathrm{max},u} = 4\,$MeV.

\subsubsection{Core-collapse-supernova environments \label{sec:energy_advection_ccsn}}
In this test, we focus on the energy advection through a shock in a \ac{CCSN} system. We consider a snapshot of a \ac{CCSN} from an M1 simulation. The details of the simulation setup are the same as in \texttt{Gmunu}'s M1 code test in Ref.~\cite{Cheong_2023}. We use a snapshot of the simulation at around $20\,$ms post-bounce, where the amplitude of the discontinuity in velocity at the shock front is around $0.13$. We only use the metric and fluid data from the snapshot and remove all information on neutrinos. We also set $\beta^r = 0$ to ease the computation of a reference solution. Because of the challenging advection through the shock, we use the Piecewise Parabolic Method (PPM) limiter~\citep{Colella_1984} to reconstruct the distribution function at cell interfaces in both position and momentum spaces.

We impose a constant opacity $\kappa_a = 60$ cm$^{-1}$ inside a sphere of radius $R = 10\,$km. The emissivity is determined by Kirchhoff's law with $T = 5\,$MeV. This test was introduced by Ref.~\cite{Muller_2010} and later reused in multiple codes~\citep{O'Connor_2015,Kuroda2016,Cheong_2023}. The difference is that we use a snapshot from a \ac{CCSN} simulation, as opposed to a pre-determined velocity profile. A reference solution for the luminosity and average energy can be found from the stationary Boltzmann equation (see~\cite{Muller_2010}). In the free-streaming regime, the luminosity and average energy satisfy
\begin{align}
    &L_c \equiv \alpha \frac{1+\psi^2 v^r}{1-\psi^2 v^r} L = \alpha^2 \psi^6 r^2 \int \mathcal{H}^r~\varepsilon^2 d\varepsilon d\Omega_p = \mathrm{cst},\\
    &\braket{\varepsilon}_c \equiv \alpha W(1+\psi^2 v^r) \braket{\varepsilon} = \mathrm{cst},
\end{align}
where the average energy $\braket{\varepsilon}$ is defined as
\begin{equation}
    \braket{\varepsilon} \equiv \frac{T^{\hat{0}\hat{0}}}{N^{\hat{0}}} = \frac{\int \varepsilon^3 f d\varepsilon d\Omega_p}{\int \varepsilon^2 f d\varepsilon d\Omega_p},
\end{equation}
that is, the neutrino energy over the neutrino number.
In theory, the new variables $L_c$ and $\braket{\varepsilon}_c$ are constant. This may however not be true in practice. To evaluate how $L_c$ and $\braket{\varepsilon}_c$ deviate from a constant value in the test, we introduce the relative deviations
\begin{align}
    &\delta L_c \equiv \mathrm{abs}\left(\frac{L_c}{\mathrm{mean}(L_c)} - 1\right), \label{eq:rel_L_c} \\
    &\delta \braket{\varepsilon}_c \equiv \mathrm{abs}\left(\frac{\braket{\varepsilon}_c}{\mathrm{mean}(\braket{\varepsilon}_c)} - 1\right) \label{eq:rel_eps_c},
\end{align}
where the $\mathrm{mean}$ operator computes the mean outside of the radiating sphere.

\begin{figure}
    \centering
    \includegraphics[width=\linewidth]{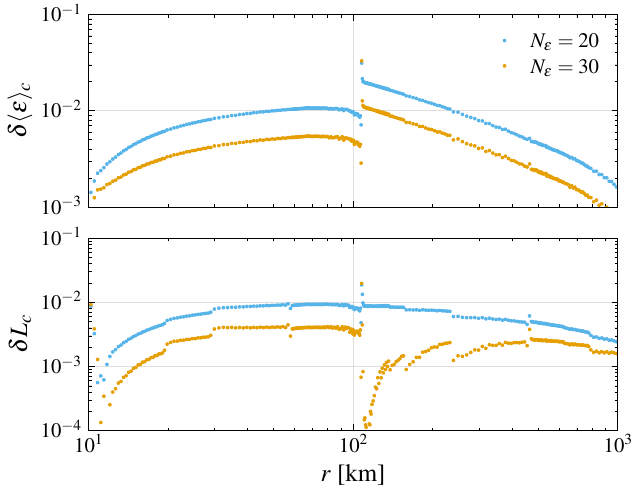}
    \caption{Relative deviations from constant values $\braket{\varepsilon}_c$ and $L_c$ as defined in Eqs.~\eqref{eq:rel_eps_c} and \eqref{eq:rel_L_c} for $N_\varepsilon = 20$ (blue dots) and $N_\varepsilon = 30$ (orange dots) energy bins. The error is below the percent level at all resolutions for the luminosity, except at the shock front, whereas $N_\varepsilon=30$ is needed to reach that accuracy in average energy. The error is dominated by the advection at the shock, as expected from the large velocity gradients.}
    \label{fig:energy_advection_ccsn}
\end{figure}

In Fig.~\ref{fig:energy_advection_ccsn}, we show $\delta L_c$ and $\delta \braket{\varepsilon}_c$ for $N_\varepsilon = 20$ and $N_\varepsilon = 30$. The number of angular bins is fixed to $N_\vartheta = 20$. We can see that the error for the luminosity is below the percent level, except at the shock front. We find that $N_\varepsilon = 30$ is needed for an error on $\braket{\varepsilon}_c$ below $10^{-2}$. Nonetheless, the solver captures the shock-induced and gravity-induced energy advection, since the relative deviations for the average energy are contained below $\sim 2\%$, except at the shock front ($\sim 4\%$), and are even smaller for the luminosity.

\subsubsection{The inhomogeneous radiating sphere \label{sec:inhomogeneous_radiating_sphere}}

\noindent In the previous test, we focused on the free-streaming regime and we computed the error only on energy-integrated variables. In this test, we shall consider solutions in all regimes and study the performance of the solver on individual energy bins. To find a (semi-)analytical solution to compare with in different regimes and energy, we generalize the radiating sphere test case from Sec.~\ref{sec:tests_radiating_sphere} by adding curvature and both position- and energy-dependent opacities.

To ensure a (semi-)analytical solution to the problem, we consider modifications to Lorentzian profiles for both $\alpha$ and $\psi$. More precisely, we introduce $h(r) = \psi^2(r)/\alpha(r)$ and express it as
\begin{equation}
    h(r) = \frac{\psi^2(r)}{\alpha(r)} = \sqrt{\frac{r^2 + A^2}{r^2 + B^2}}, \label{eq:inhomogeneous_sphere_h}
\end{equation}
which corresponds to $\sqrt{1 + L(r)}$, where $L$ is a Lorentzian, and where $A > B$ are real parameters. Then, we consider a similar profile for $\alpha$ of the form
\begin{equation}
    \alpha(r) = \frac{r^2 + A_\alpha^2}{r^2 + B_\alpha^2}, \label{eq:inhomogeneous_sphere_alpha}
\end{equation}
with $A_\alpha < B_\alpha$ real parameters. The remaining profile $\psi(r)$ is deduced from $h$ and $\alpha$. Regarding the opacity, we consider a profile
\begin{equation}
    \kappa^*_a(r,\varepsilon) = \left(\frac{\varepsilon}{\varepsilon_r}\right)^{n_e} \kappa~\mathrm{max}\left[1 - \left(\frac{r}{R} \right)^2, 0 \right]^n H(R-r), \label{eq:inhomogeneous_sphere_kappa}
\end{equation}
where $H(r)$ is the Heaviside step function, $n$ and $n_e$ are integers, and $\kappa$, $\varepsilon_r$ and $R$ are real parameters corresponding to the central reference opacity $\kappa = \kappa^*_a(0,\varepsilon_r)$, reference energy and radius of the inhomogeneous sphere, respectively. For $n_e = 0$ and $n = 0$, we retrieve the opacity profile used for the homogeneous radiating sphere from Sec.~\ref{sec:tests_radiating_sphere}.

\begin{figure}
    \centering
    \includegraphics[width=\linewidth]{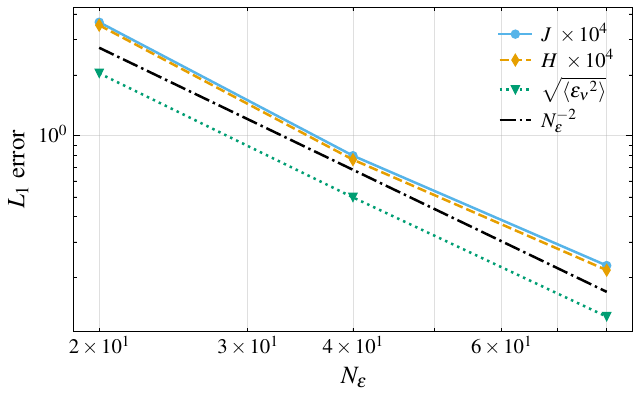}
    \caption{Convergence test for the inhomogeneous radiating sphere with $n=2$ and $n_e=2$. We observe approximately quadratic convergence in energy space for the zeroth moment $\mathcal{J}$ (solid blue line), first moment $\mathcal{H}^r$ (dashed orange line) and root-mean-squared energy (dotted green line), as expected from the scheme used.}
    \label{fig:rad_sphere_conv}
\end{figure}

\begin{figure*}
    \centering
    \includegraphics[width=\linewidth]{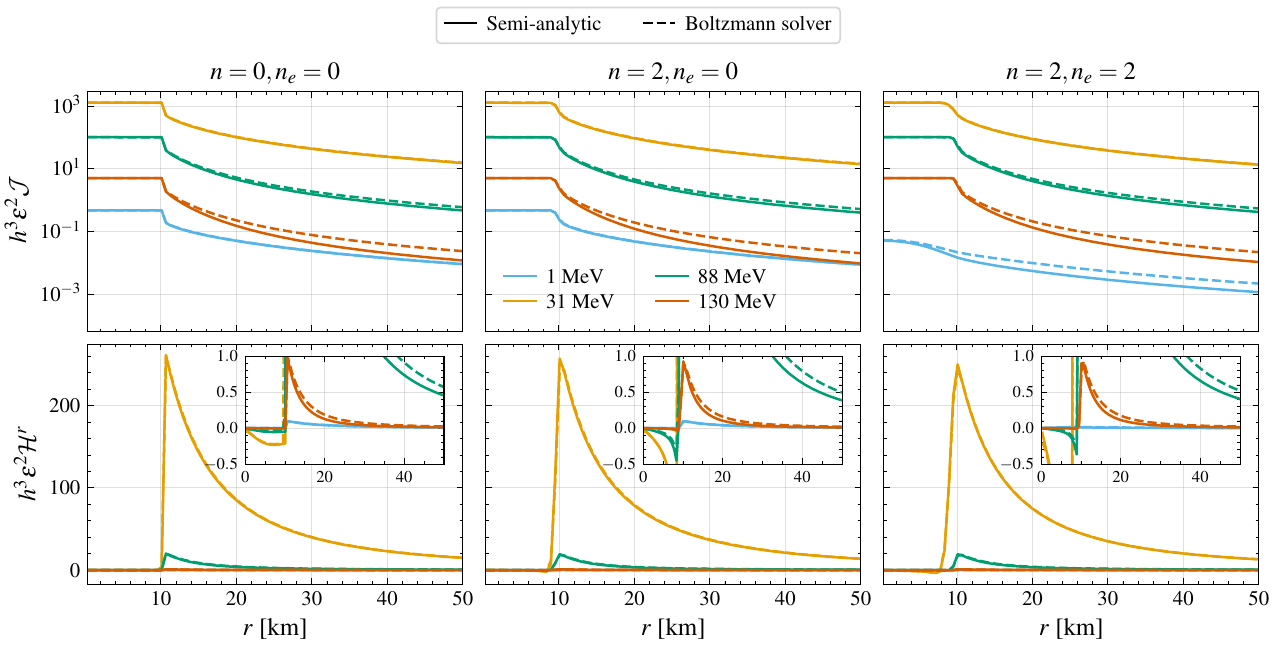}
    \caption{Spectral moments $\mathcal{J}$ (top panels) and $\mathcal{H}^r$ (bottom panels) for different parameters of the inhomogeneous radiating sphere test case. The solid lines correspond to the semi-analytical solution and dashed lines to the predictions of our solver. Different colours represent different energies. We rescale the spectral moments by $\varepsilon^2$, which is the integration weight to compute $J$ and $H^r$. The rescaled moments thus indicate their relative significance in the computation of the luminosity and average energy. At larger energies, the Boltzmann solver's predictions deviate (relatively) more from the analytical solution because of the lower resolution at higher energy. Because the predictions exceed the exact solutions, we expect the solver to overestimate the luminosity and average energy.}
    \label{fig:inhomog_spectral}
\end{figure*}

These profiles are directly inspired by and fitted to a spherically symmetric isolated neutron star profile. The reference parameters are $R = 7$ ($\approx 10\,$km), $A = 9.96989$, $B = 5.47854$, $A_{\alpha} = 6.82503$ and $B_\alpha = 7.98152$. The reference energy $\varepsilon_r$ is chosen to be $20\,$MeV and the reference opacity $\kappa$ will be varied in different tests. The emissivity is given by
\begin{equation}
    \eta(r,\varepsilon) = \frac{1}{1+e^{\varepsilon/T}} \kappa_a^*(r, \varepsilon), \label{eq:inhomog_emissivity}
\end{equation}
where we set $T=10\,$MeV. The semi-analytical solution to the general problem is given in Appendix~\ref{app:analytic_sol}. Note that we do not consider any velocity field in this case. Since we showed in the previous test that the error is dominated by shock-induced advection, we remove this source of error to focus on the behaviour induced by variations in the opacity profile and corresponding free-streaming regime.

First, we perform a convergence test for the energy dimension with the parameters $n_e=2$, $n=2$ and $\kappa=10$. These parameters yield a smoother solution that is more adapted to a convergence test. We fix the number of points $N_r = 512$ and $N_\vartheta=80$ to ensure that the error is dominated by the energy resolution. We also use $N_\varepsilon = 20, 40, 80$ to measure the convergence. In this case, we use a purely log-spaced energy grid with $\varepsilon \in [0.1, 300]$ for consistent refinement of the whole grid as $N_\varepsilon$ increases. We compute the convergence rate based on the zeroth moment $\mathcal{J}$, first moment $\mathcal{H}^r$ and root-mean-squared (RMS) energy
\begin{equation}
    \sqrt{\braket{\varepsilon^2}} = \sqrt{\frac{\int \varepsilon^4 f d\varepsilon d\Omega_p}{\int \varepsilon^2 f d\varepsilon d\Omega_p }}. \label{eq:rms_energy}
\end{equation}
For a quantity $Q$, we compute the $L_1$-error as
\begin{equation}
    || Q - Q_{\mathrm{ref}} ||_1 = \frac{\sum_{i,l,m} |Q(r_i,\varepsilon_l,\vartheta_m) - Q_{\mathrm{ref}}(r_i,\varepsilon_l,\vartheta_m)| \Delta V_{i,l,m}}{\sum_{i,l,m} \Delta V_{i,l,m}}.
\end{equation}
The result of the convergence test is shown in Fig.~\ref{fig:rad_sphere_conv}. The convergence rate is approximately quadratic for all three reference quantities, as expected.

Then, we investigate the performance of the solver on the spectral moments $\mathcal{J}$ and $\mathcal{H}^r$ for three different parameter sets $(n,n_e,\kappa)$: $(0,0,100)$ (constant opacity), $(2,0,100)$ (spatial variation), $(2,2,10)$ (spatial and energy-space variations). We use $N_r = 256$ with $r\in [0, 100]$ and $N_\varepsilon = N_\vartheta = 20$. Note that the energy grid is the default one explained at the beginning of Sec.~\ref{sec:advection_collisions_curved}. 

In Fig.~\ref{fig:inhomog_spectral}, we show the spectral moments rescaled by $\varepsilon^2$, which is the integration weight to obtain the integrated moments $J$ and $H^r$. Each curve thus indicates the relative significance of the energy bin in the computation of the luminosity and average energy. 

We can see that the moments generally agree with the analytical solution at all energies and all cases. However, we do see discrepancies in the free-streaming regime especially at higher energies. In particular, the moments are overestimated by the solver in all three cases. Thus, we expect the luminosities and average energies to be overestimated by the solver. This discrepancy may be explained by the lower resolution at higher energies. In the stationary spherically symmetric Boltzmann equation with no source terms, no velocity and $\beta^i =0$, the radial shape of $\mathcal{H}^r$ is determined by the derivative in energy space through
\begin{equation}
    \partial_r (\alpha \psi^6 r^2 \mathcal{H}^r) = \frac{r^2}{\varepsilon} \partial_\varepsilon (\alpha \psi^6 \varepsilon \mathcal{H}^r \partial_r \ln\alpha ), \label{eq:H_stationary}
\end{equation}
The error on the energy-space derivative would then propagate to the radial profile of $\mathcal{H}^r$ and to the luminosity. Significant errors in the luminosity (and average energy) are only visible if the error is significant around the peak of the spectrum.

Inside the radiating sphere, on the other hand, the solutions and predictions agree in all cases, except for the lowest energy bin for $n=2$ and $n_e=2$. In that energy bin, the opacity is $2.5\times 10^{-2}$, as opposed to $100$ in the other cases. Therefore, the effects of advection (similar to the free-streaming regime) are relatively stronger.

Finally, we note that the analytical solution may contain negative luminosities, as can be seen by negative values of $\mathcal{H}^r$ at the dominating energy bin at $31\,$MeV. These negative values arise because of energy advection, and energy-dependent opacities and emissivities. Inside the sphere, neutrinos are emitted (and absorbed) isotropically in the comoving frame. Neutrinos propagating outwards ($\cos\vartheta > 0$) at a point $r$ and an energy $\varepsilon$ either come from the isotropic source at that point, or from redshifted neutrinos emitted at a smaller radius and a higher energy. Neutrinos propagating inwards, on the other hand, either come from the isotropic source at that point, or from blueshifted neutrinos emitted at larger radii and lower energies. In the case $n=0$ and $n_e=0$, the absorption is constant and the emissivity only depends on energy through the Fermi-Dirac statistics. Therefore, more neutrinos are emitted at lower energies, inducing an excess of ingoing neutrinos at larger energies, and hence a negative $\mathcal{H}^r$ as seen in Fig.~\ref{fig:inhomog_spectral} for the bin at $31\,$MeV (orange line). The luminosity is negative only if this phenomenon occurs around the peak of spectrum (which dominates the luminosity). Similar reasoning can be applied to the other cases. This phenomenon does not occur (equally) at all energies and radii, since it strongly depends on the difference in emissivities and on the spacetime curvature.

\section{Comparison with the M1 scheme \label{sec:comparison_M1}}
In the previous section, we validated the performance of our solver on (semi-)analytical solutions. We now proceed with a comparison to the M1 scheme, which is described in Refs.~\cite{Cheong_2023,Cheong_2024}. The approximations made in the M1 scheme, \textit{i.e.} an approximate closure and discarding higher-order moments, are known to break down in specific cases, such as intersecting beams. Evaluating the impact of these approximations onto the evolution of systems such as \acp{CCSN} is however challenging, since the equations are highly non-linear and many different factors can contribute to the observed discrepancies.

In this section, we compare \texttt{Gmunu}'s M1 solver and our Boltzmann solver first on simplified tests with analytical solutions. These tests allow us to find the possible intrinsic sources of differences. Then, we proceed with a relaxation test case in a hot neutron star environment and a \ac{CCSN} up to core bounce. These last two tests are also used to validate the implementation of inelastic scattering, pair-processes, and number and momentum exchange with the fluid.

We stress that we do not aim to provide a complete comparison between the two schemes, but merely to highlight differences that could impact the evolution of the systems. Moreover, the M1 scheme implements the energy-conservative formulation of the Boltzmann equation, whereas our Boltzmann solver implements the number-conservative formulation. A complete comparison must address these differences and is out of the scope of this paper.

\subsection{The crossing beam problem}
\noindent The M1 scheme is known to fail when two beams cross. Since the M1 closure in the optically thin limit assumes axisymmetry around the direction of propagation, it can only account for a single direction of propagation. The beams therefore merge upon crossing and propagate in the direction corresponding to the average of the two initial beams. We also use this test to see how beams propagate in diagonal directions in Cartesian coordinates with our Boltzmann solver. 

We consider two beams of particles propagating from the upper and lower left corners of the domain. The spatial grid has $256\times 256$ points with both $x$ and $y$ in $[-0.5,0.5]$.  We use the $z$-axis as the reference axis for momentum space and set $f=1$ at $\vartheta=\pi/2$ and $\varphi=-\pi/8$ for the upper beam and $\varphi=\pi/8$ for the lower one. We only consider a limited number of points in momentum space, since no momentum advection is expected. We use $N_\varepsilon =1$, $N_\vartheta =3$ and $N_\varphi = 8$.

\begin{figure}[h]
    \centering
    \includegraphics[width=0.97\linewidth]{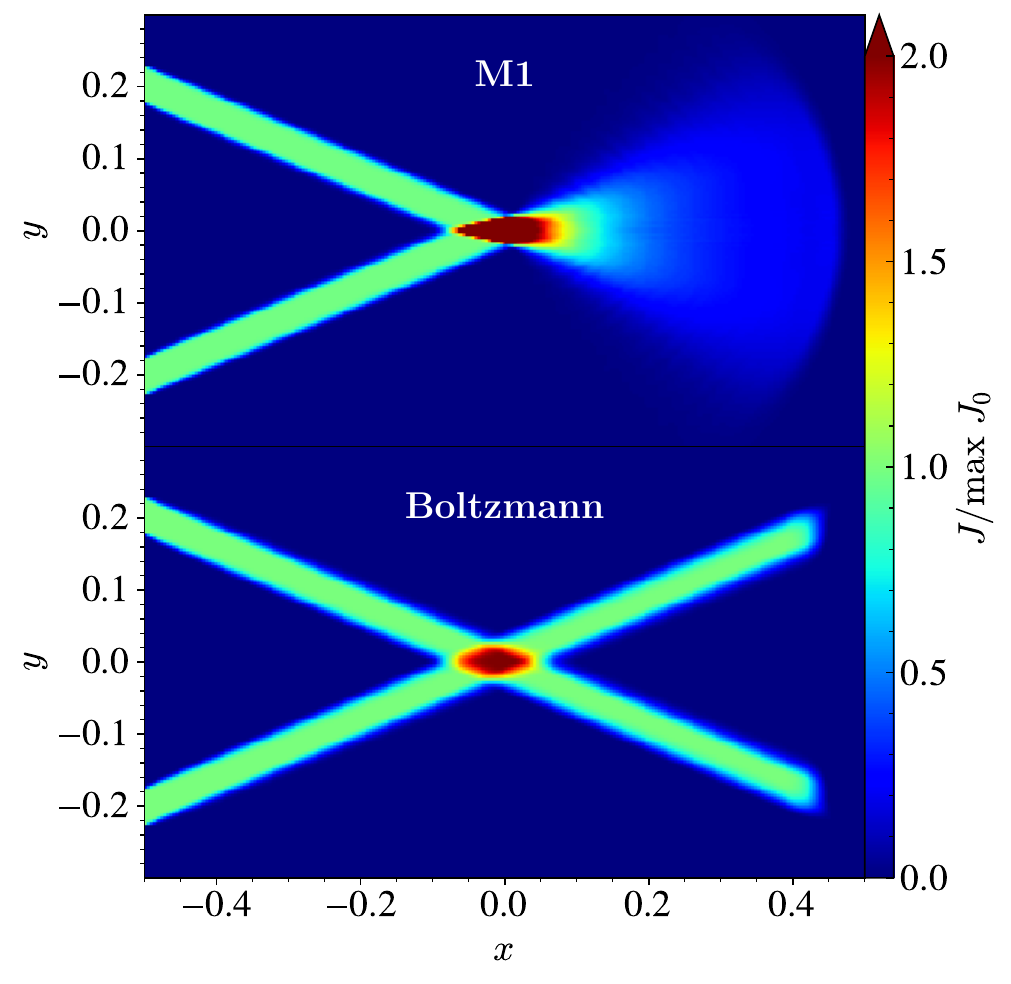}
    \caption{Normalized zeroth moment $J$ of particles for the crossing beam test. Contrary to the M1, the beams cross without interacting, as expected. Numerical diffusion appears on the sides of the beams due to the particles propagating diagonally, that is not perpendicular to any of the surfaces of the cells (in Cartesian coordinates).}
    \label{fig:crossing_beams}
\end{figure}

In Fig.~\ref{fig:crossing_beams}, we show the zeroth moment $J$ normalized to the maximum of its initial value $J_0 = J(t=0)$ for this test case. We also show the prediction of the M1 solver. As expected, the two beams cross without merging with the Boltzmann solver. The M1 scheme, however, fails as predicted: the two beams merge at the crossing point. Contrary to the previous test, numerical diffusion appears on the side and the beams seem to widen slightly. This is due to the fact that the direction of propagation is not perpendicular to any surface of the cubic cell.

\subsection{Inhomogeneous radiating sphere: comparison with M1 \label{sec:inhomogeneous_radiating_sphere_m1}}

\noindent We consider the inhomogeneous radiating sphere in curved spacetime test case from Sec.~\ref{sec:inhomogeneous_radiating_sphere} and compare the results to the M1 scheme's results with the Minerbo closure~\cite{Minerbo_1978}. We use the same setup as in Sec.~\ref{sec:inhomogeneous_radiating_sphere} and the same energy grid for the M1 scheme. We perform simulations with the same parameters as the ones used for Fig.~\ref{fig:inhomog_spectral}. 
\begin{figure*}
    \centering
    \includegraphics[width=\linewidth]{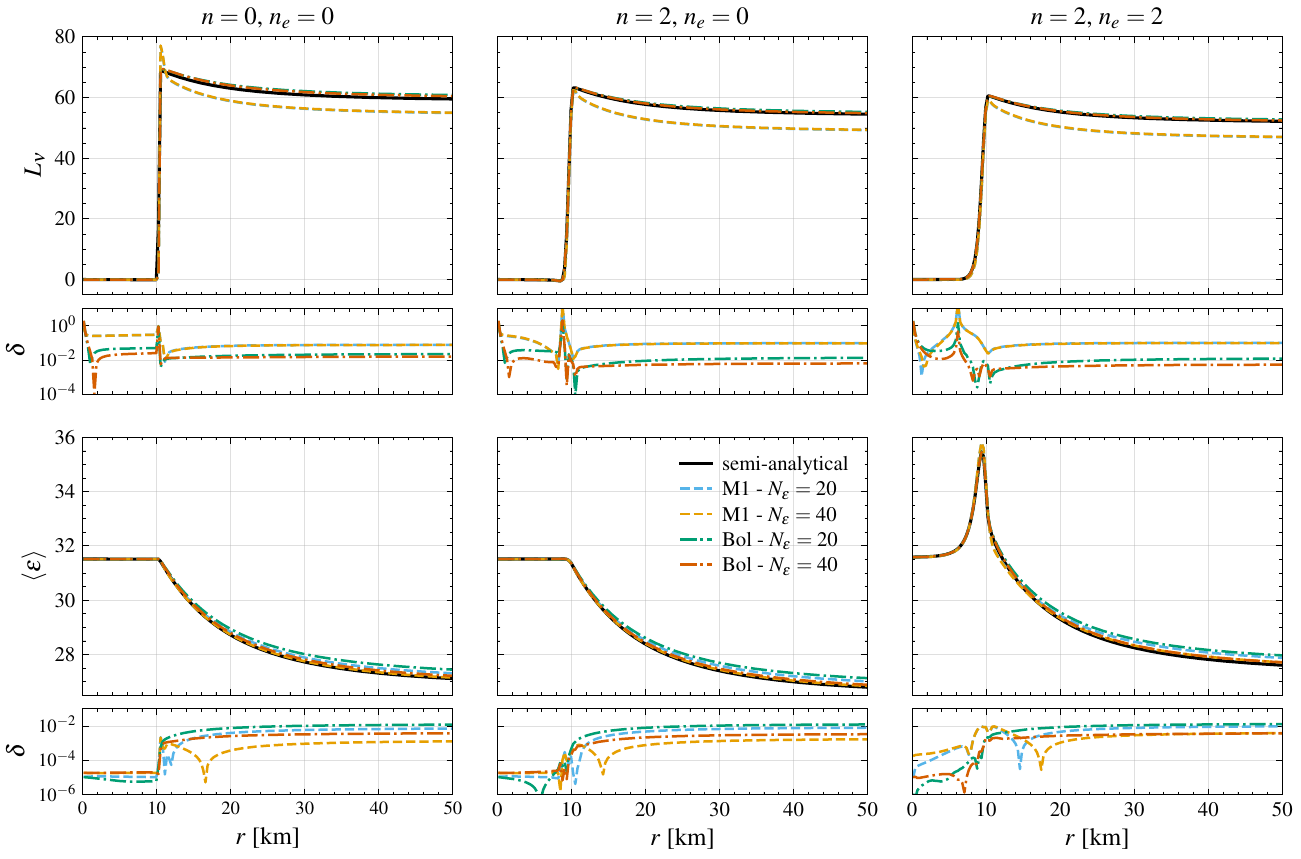}
    \caption{Luminosity, average energy and their relative errors for different number of energy bins with the M1 scheme (dashed lines) and Boltzmann transport (dash-dotted lines) compared to the semi-analytical solution for the inhomogeneous radiating sphere (solid black line, see Sec.~\ref{sec:inhomogeneous_radiating_sphere}). For the average energy, both schemes perform best in the optically thick regime, as expected for M1. For the same number of energy bins, the M1 scheme reconstructs the average energy better than Boltzmann in the free-streaming regime. The M1 performance decreases in the optically thick regime for energy-dependent opacities. For the luminosity, Boltzmann retrieves the luminosity in the free-streaming regime at the percent level in all cases. The error is lower for $n=2$ (smoother transition to optically thin regime). For M1, there is an offset in the luminosity that persists when increasing the number of energy bins. The error at the transition to optically thin regime is also larger for M1 than Boltzmann. }
    \label{fig:comp_inhomog_sphere}
\end{figure*}

We plot the comparison of the average energy $\braket{\varepsilon}$ and the luminosity for each case in Fig.~\ref{fig:comp_inhomog_sphere}. We show the luminosity and average energy, as well as the associated relative errors, for the M1 scheme (dashed lines) and the Boltzmann solver (dash-dotted lines). We plot these quantities for $N_\varepsilon = 20$ and $N_\varepsilon = 40$ energy bins. 

First, we see that the error for the average energy remains below or around the percent level in all cases for both solvers. Contrary to advection through shocks (see Sec.~\ref{sec:energy_advection_ccsn}), $N_\varepsilon=20$ is sufficient to reach this level of accuracy. In the energy-dependent opacity case ($n_e = 2$), the error of the M1 in the optically thick regime is larger, especially at the transition to optically thin regime. The M1 scheme is expected to have poorer performance in this semi-transparent transition regime. In the free-streaming regime, the error of the M1 is smaller than that of the Boltzmann solver, except for energy-dependent opacities, suggesting that the M1 scheme reconstructs the free-streaming average energy better for a given number of energy bins. We point out, however, that we see crossings between the average energy from M1 and the semi-analytical solution on the outside of the star in Fig.~\ref{fig:comp_inhomog_sphere}, indicating different free-streaming behaviour in that region. We also note that the reconstruction scheme of $f$ or moments at the cell interfaces differs between the M1 scheme and the Boltzmann solver. The former uses a prescription introduced by Ref.~\cite{Muller_2010}, whereas the latter uses slope limiters. This difference in methods may play a role in how well the average energy is reconstructed for a given resolution. The investigation of the impact of the reconstruction methods at the interfaces in energy space is left for future work.

Concerning the luminosity, the Boltzmann solver retrieves the free-streaming values up to about the percent level. The error decreases with increasing number of energy bins and is larger for the sharper opacity profile ($n = n_e = 0$). Similarly to the average energy, the performance of the M1 is worse in the transition regime for the luminosity, where we can see larger relative errors when the luminosity starts increasing. In contrast to the average energy, however, the free-streaming region is not recovered accurately, with a relative difference of up to $\sim 10\%$ in the luminosity for $n=2$ and $n_e =2$. Since this error persists when increasing the number of energy bins, the dominating source of error is not the lower resolution affecting the accuracy of the energy-space derivative, contrary to our conclusions in Sec.~\ref{sec:inhomogeneous_radiating_sphere}. 

A possible explanation for the larger error on the luminosity is the approximate closure relation. We start from the stationary equation governing the first moment in Eq.~\eqref{eq:H_stationary}. Integrating the equation with $\int d\varepsilon \varepsilon^2$ and reintroducing the collision term considered here, we find
\begin{equation}
    \partial_r (r^2 \alpha^2 \psi^6 H^i) = r^2 \int d\varepsilon \varepsilon^2 \alpha \psi^6 ( \eta - \kappa_a^* \mathcal{J}),
\end{equation}
where $H^i$ is the energy-integrated first moment. The last equation relies on $\mathcal{H}^i$ vanishing at large energies, which is true in our case given the energy-dependent emissivity (see Eq.~\eqref{eq:inhomog_emissivity}). Integrating over $r'$ up to $r > R$, we then find the constant redshifted luminosity $L_{\mathrm{rs}}$
\begin{equation}
    \alpha(r) L(r) = \int_0^R dr ~ r^2 \int d\varepsilon~ \varepsilon^2 \alpha \psi^6 (\eta - \kappa_a^* \mathcal{J}) \equiv L_{\mathrm{rs}}.
\end{equation}
The value of the luminosity outside of the sphere thus directly depends on $\mathcal{J}$ through the interactions. Any error on the value of $\mathcal{J}$ caused by the closure will thus affect the value of the luminosity. We expect such error to appear in the semi-transparent regime. Note that the error in Fig.~\ref{fig:comp_inhomog_sphere} far away from the sphere, where $\alpha \approx 1$, measures the error on $L_{\mathrm{rs}}$, and hence on the integral of the source terms.

\begin{figure*}
    \centering
    \includegraphics[width=\linewidth]{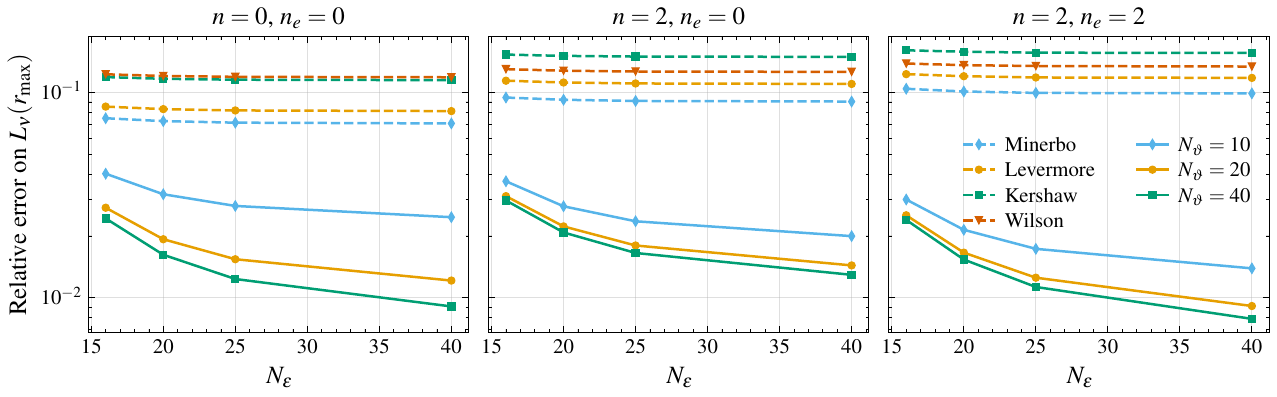}
    \caption{Relative error on the luminosity at the maximum radius for different numbers of angular bins for the Boltzmann solver (solid lines) and different closures for the M1 scheme (dashed lines). For Boltzmann, the predicted luminosity converges in both energy and angle. For M1, increasing the number of energy bins does not significantly lower the error for any closure, as observed in Fig.~\ref{fig:comp_inhomog_sphere} for the Minerbo closure. The error varies more significantly across closures, suggesting the closure is a key source of error for the luminosity.} 
    \label{fig:err_inhomog_sphere}
\end{figure*}

To verify the hypothesis of the closure as the main source of error for the luminosity, we run the same tests for different closures. We use the Wilson closure~\citep{LeBlanc_1970,Wilson_1975}, the Kershaw closure~\citep{Kershaw_1976} and the Levermore closure~\citep{Levermore_1984}. We show the relative error on the luminosity at the maximum radius for the different closures, as well as $N_\vartheta = 10$, $20$ and $40$ angular bins for the Boltzmann solver, in Fig.~\ref{fig:err_inhomog_sphere}. For the Boltzmann solver, the error decreases with increasing number of energy bins and angular bins. We find an error of $\sim 3.5\%$ for $N_\varepsilon = 20$ and $N_\vartheta = 10$ for $n=n_e=0$. For $n_e=2$, the error decreases to $2\%$. For the M1, changing the closure induces more significant changes than increasing the number of energy bins. This suggests that the closure is indeed a dominating factor in the persistent error on the luminosity. Moreover, the error is smaller for $n=0$ than $n=2$, which is consistent with the expected error from the closure in the semi-transparent regime. For $n=2$, the transition from optically thick to free-streaming is smoother, hence extending the semi-transparent region and contributing more significantly to the error.

A more robust proof may only be obtained by designing a closure specifically for this test based on the analytical solution and using it for the M1 evolution. This method is out of the scope of this paper and is left for future comparisons.

\subsection{Relaxing neutrino radiation in a 1D hot neutron star}\label{sec:1dns}
\begin{figure*}
    \centering
    \includegraphics[width=\linewidth]{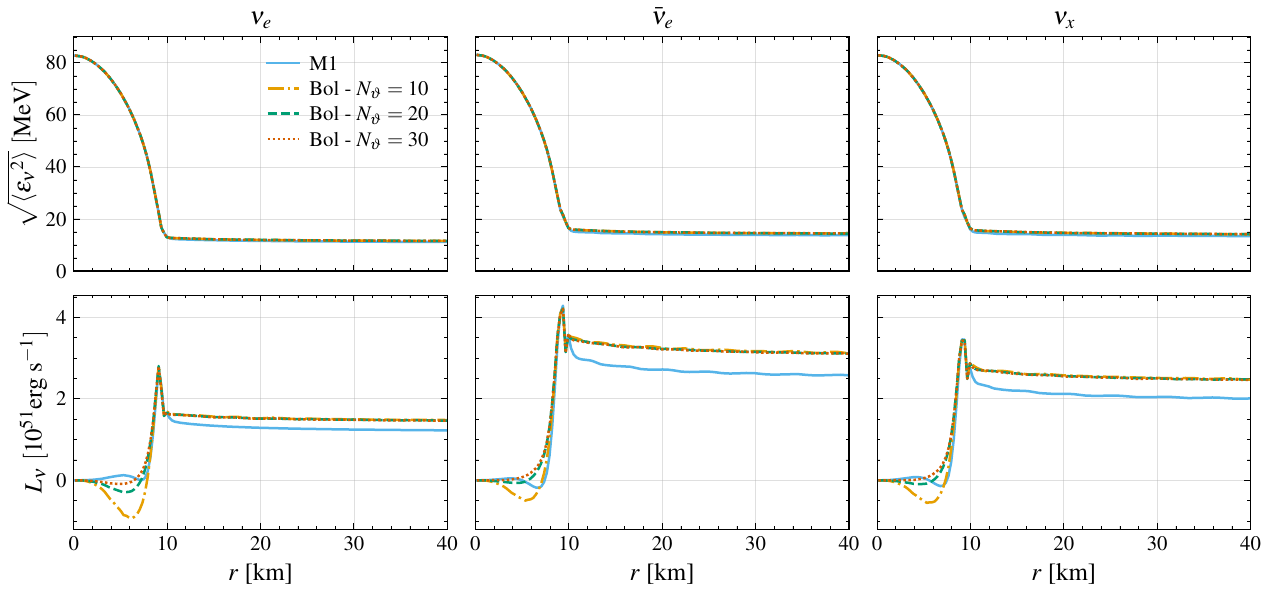}
    \caption{Root-mean-squared (RMS) energy and luminosities of neutrinos in the fluid rest frame for the hot neutron star relaxation test case. The solid blue line corresponds to the M1 simulation, while the Boltzmann results are represented by
the orange dash-dotted ($N_\vartheta =10$), the green dashed ($N_\vartheta
=20$) and the red dotted ($N_\vartheta = 30$) lines. For all three species, the RMS energy agrees between M1 and
Boltzmann transports. Luminosities at large radii differ by $\sim 10\%$ for
each species, which is consistent with observations in Sec.~\ref{sec:inhomogeneous_radiating_sphere_m1}. Luminosities also differ inside of the neutron star, where we observe a significant difference between different $N_\vartheta$. }
    \label{fig:hot_ns}
\end{figure*}

In this test, we consider a hot neutron star in spherical symmetry and 1D
without evolving the fluid or the spacetime, but only the neutrino distribution functions.

The initial data are computed with XNS~\citep{Bucciantini:2014zca} using the
SFHo EOS~\citep{Steiner_2013}, with a central density of
$6.7\times 10^{14}\,\mathrm{g\,cm^{-3}}$, in neutrinoless $\beta$-equilibrium 
and with a
constant entropy per baryon of $1,k_{\mathrm{B}}$/baryon. The central temperature is
approximately $23\,\mathrm{MeV}$.  The initial value of the distribution function is $0$ everywhere.

The tabulated neutrino interactions are generated with
\texttt{Weakhub}~\citep{Ng_2024}. We employ the elastic approximation for
charged-current interactions, while the beta process involving
heavy nuclei is treated following Ref.~\cite{Bruenn_1985}. Electron--positron pair
annihilation and nucleon--nucleon bremsstrahlung, as well as inelastic
neutrino--electron scattering, are treated using interaction kernels. Elastic
scattering is also treated in kernel form, with weak-magnetism and recoil
corrections for neutrino--nucleon scattering following Ref.~\cite{Horowitz2002}.
Details of the elastic-scattering kernels are given in
Appendix~\ref{app:kernels_ES}. We evolve three neutrino species, $\nu_e$,
$\bar{\nu}_e$, and one effective heavy-lepton species $\nu_x$, which represents
any one of $\nu_\mu$, $\bar{\nu}_\mu$, $\nu_\tau$, and $\bar{\nu}_\tau$. All
energy- and species-coupling interactions are solved implicitly with the
\ac{MSMG} mode (see Sec.~\ref{sec:implicit_modes}).
We use $N_r=256$ for $r\in[0, 400]$ with four levels of refinement and
$N_\varepsilon=20$. We consider $N_\vartheta=10$, $20$ and $30$ for the Boltzmann
solver to assess the dependence on the angular resolution.

Figure~\ref{fig:hot_ns} shows the root-mean-squared (RMS) neutrino energies and
luminosities in the fluid rest frame for the M1 and Boltzmann schemes when the stationary state is reached close to the star's surface. The RMS
energies agree well for all three species and show only a weak dependence on $N_\vartheta$. The agreement between M1 and Boltzmann is expected from the inhomogeneous radiating sphere test in the previous section. There is a difference of about $10\%$ between the Boltzmann and the M1 luminosities at $r=40\,$km, which is also expected from the previous test. The suggested explanation of the closure as a source for this discrepancy still holds in this context. 

Both the M1 and the Boltzmann solver predict significant negative luminosities. For the M1, the negative part is located close to the star's surface, where the transition from optically thick to optically thin regimes starts and the closure plays a more significant role. For the Boltzmann solver, the luminosity converges towards positive luminosities as $N_\vartheta$ increases. A higher number of angular bins may be needed in this test for the luminosity to converge, but this number may depend on the slope limiter considered.

Another possible explanation for the discrepancies deeper in the star is the different treatments for the diffusive limit employed in the two schemes, since these differences occur in a region of strong scattering. As in the previous test, a deeper study on the closure is needed to assess its relevance in explaining the differences, compared to, \textit{e.g.}, the diffusive limit treatment.

Overall, considering realistic opacities, and including inelastic scattering and pair processes does not change the main conclusion from the inhomogeneous radiating sphere toy model. M1 and Boltzmann agree on the RMS (or average) energy, but differ significantly in luminosities. The agreement between M1 and Boltzmann (accounting for the expected luminosity difference in the free-streaming regime) provides evidence for the correct implementation of all the source terms.

\subsection{Core-collapse of a $15~M_{\odot}$ star in 1D
\label{sec:test_ccsn}}
\begin{figure*}
    \centering
    \includegraphics[width=\linewidth]{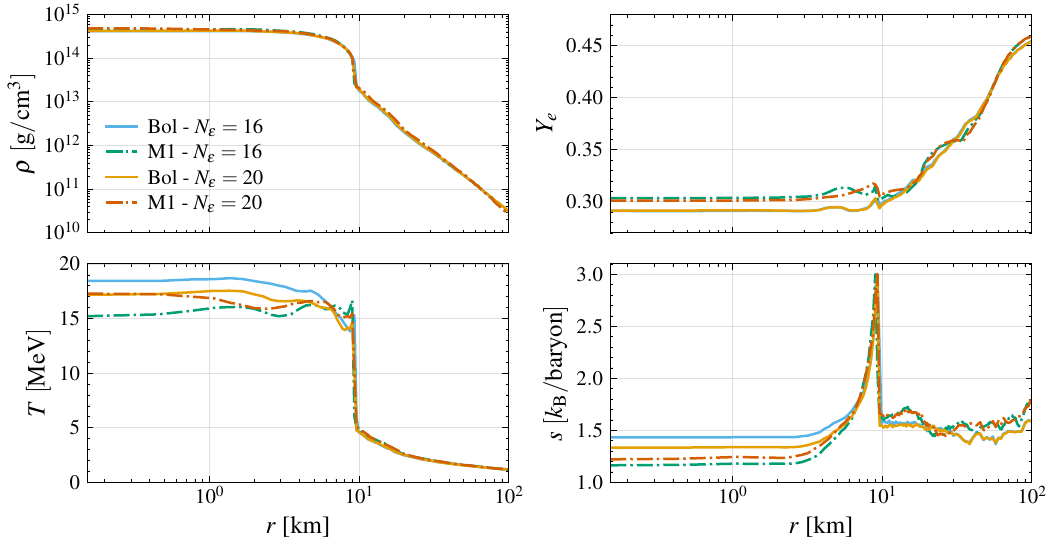}
    \caption{ Radial profiles of the rest-mass density $\rho$, electron fraction $Y_e$, matter temperature $T$, and entropy per baryon $s$ at core bounce for the Boltzmann and M1 transport schemes with $N_\varepsilon=16$ and $20$. The two schemes give nearly identical density profiles and closely agree in $Y_e$ and the bounce-shock structure. The larger differences in the inner-core temperature and entropy for $N_\varepsilon=16$ are substantially reduced for $N_\varepsilon=20$, indicating improved agreement with increasing energy-space resolution.
    }
    \label{fig:ccsn_bounce}
\end{figure*}

As a final and more stringent test of the Boltzmann solver, we compare it
directly with the M1 transport scheme in a 1D core-collapse star 
simulation. We adopt a numerical setup similar to Sec.~5.2 of Ref.~\cite{Cheong_2023}. 
The M1 implementation in \texttt{Gmunu} was previously benchmarked against
different radiation transport codes, showing good quantitative agreement.

We evolve the $15\,M_\odot$ progenitor \textit{s15s7b2} of Ref.~\cite{Woosley_1995}.
We use the same EOS and weak interaction set as Sec.~\ref{sec:1dns}.
The only difference is that we employ the \ac{SSMG} implicit solver and 
pair processes mentioned in Sec.~\ref{sec:1dns} are included only for $\nu_x$. For this test, we use the PPM limiter in both position and momentum spaces to reconstruct the distribution function at the cell surface.

The computational domain is $r\in[0,10^4]\,\mathrm{km}$ with $N_r=128$. We employ \ac{AMR} with a maximum refinement level $l_{\mathrm{max}}=9$. We consider $N_\varepsilon=16$ and $20$. The first energy group is centred at $1\,{\mathrm{MeV}}$ with a width of $2\,{\mathrm{MeV}}$, while the remaining group centres are distributed logarithmically up to $\sim 300\,{\mathrm{MeV}}$, and $N_{\vartheta}$ is set to $20$. Other details of the numerical setup, such as the \ac{AMR} setting, stage-dependency, \ac{CFL} condition and metric solver, are described in Ref.~\cite{Cheong_2023}.

% fig29
Figure~\ref{fig:ccsn_bounce} compares the radial profiles of the matter quantities at core bounce. The Boltzmann and M1 calculations produce nearly identical density profiles and very similar electron-fraction profiles, including the location and structure of the bounce shock for both energy-bin resolutions. The largest differences occur in the temperature and entropy in the inner core for $N_\varepsilon=16$. These differences are substantially reduced for $N_\varepsilon=20$, for which the temperature profiles from the two transport
schemes become very similar. The common dependence on $N_\varepsilon$ suggests
that a substantial part of the discrepancy originates from the discretization
of the neutrino energy spectrum rather than from the angular treatment alone.
Since the neutrino--matter source terms depend strongly on neutrino energy,
insufficient spectral resolution can bias the integrated energy exchange and
accumulate into differences in the matter temperature and entropy. Increasing
$N_\varepsilon$ better resolves these source terms and correspondingly improves
the agreement between the two schemes.

Increasing $N_\varepsilon$ also decreases the amplitude of the energy-conservation violations. We remind the reader that the M1 scheme implements the four-momentum-conservative formulation of the Boltzmann equation, whereas the Boltzmann solver implements the number-conservative formulation. Any violations of the conservation laws may affect the exchange with the fluid, as pointed out in Refs.~\cite{Nagakura_2017} and~\cite{Nagakura_2019}. In Sec.~\ref{sec:conservation_laws_spherical}, we show that energy advection at small radii induces energy non-conservation. Even though we only discuss velocity-induced advection, we expect similar effects with gravity-induced advection. Since these violations decrease with $N_\varepsilon$, we expect the effect of conservation violations on the temperature profile to decrease as well. On the other hand, we cannot evaluate the actual significance of the effect of violations of the conservation laws in these simulations without an improved discretization scheme. A complete comparison between M1 and the Boltzmann solver must therefore use such a discretization for the Boltzmann scheme to separate fully the effects of conservation violations from those of the closure and moment truncation.

This CCSN test is considerably more demanding and more non-linear than the idealized tests considered above, challenging the stiff interaction solver, momentum space discretization  and the treatment of radiation across different optical regimes. The close agreement of the profiles for $N_\varepsilon=20$ with the M1 calculations provides a validation of the Boltzmann solver in \texttt{Gmunu}.

\section{Conclusion \label{sec:conclusion}}

\noindent We presented the implementation of a multidimensional general-relativistic Boltzmann solver in the code \texttt{Gmunu}. We described the generalizations needed with the addition of a momentum space for the \ac{FV} method, the boundary conditions and the discretization. We also discuss the different impacts of the choice of coordinates in position space and of reference axis to measure the momentum space angles. In particular, we show that the maximum time step allowed by the \aclu{CFL} condition can be more constrained in spherical coordinates than in cylindrical coordinates, and more constrained in cylindrical coordinates than in Cartesian coordinates, because of geometrical advection terms. Moreover, the choice of reference axis in cylindrical coordinates also influences the maximum allowed time step. 

Then, we developed a discretization that ensures consistency between the number-conservative and non-conservative formulations for homogeneous distribution functions in empty flat spacetime. A naive discretization induces significant local divergences between the true solution and the solver's prediction. Our discretization is designed to suppress inconsistencies arising from geometrical advection terms with homogeneous distribution functions and retrieves the discretization scheme introduced in Ref.~\cite{Mezzacappa_1993} in spherically-symmetric empty flat spacetime. Our discretization generalizes that of Ref.~\cite{Mezzacappa_1993} to higher dimensions and different coordinate systems, under the same assumptions. Generalizations of our discretization to suppress exactly gravity-induced and velocity-induced inconsistencies are left for future work. Another crucial improvement for the future is the development of a discretization that ensures simultaneous conservation of number and four-momentum.

Our discretization scheme does not ensure conservation of four-momentum. We analyse the amplitude of energy-conservation violations in the presence of a varying velocity background and find two main contributions. In Cartesian coordinates, discontinuity in the spectrum of $f$ is a dominating source of violations at high energy resolutions. More realistic profiles that are smoother show better conservation properties. This is consistent with our discretization being designed for homogeneous distribution functions. In spherical coordinates, the dominating term is the energy advection itself at small radii. Our analysis highlights these sources of violations for future improvements of the discretization.

Our work also addresses matter-radiation interactions. We described the formalism and implementation of the implicit solver used for the stiff source terms. Instead of the full angular dependence of interaction kernels, we use a first-order Legendre expansion as used by the M1. Thus, our solver lies at an approximation level between the M1 scheme and full Boltzmann transport. 

Since the cost of the implicit evolution of source terms scales as the cube of the number of points in the momentum grid for the \aclu{NR} method, we introduce a method that leverages the low-rank structure of the truncated Legendre expansion of the kernels to reduce the complexity to $O((\lmax+1)^4 N_a N_e + (\lmax+1)^6 N_e^3)$, where $N_e$ is the number of energy bins times the number of neutrino species, $N_a$ is the total number of angular bins, and $\lmax$ is the order at which the Legendre expansion is truncated. Note that for $\lmax = 1$, the second term in the complexity of our method is the same as that of the M1 scheme. We measured the speed-up provided by our method in a snapshot of a \aclu{CCSN} $20\,$ms after bounce with $N_\varepsilon=16$. When including energy-coupling interactions, we find a factor $300$ speed-up at $N_a = 20$ and $10^3$ at $N_a = 40$. When including species-coupling interactions, we find a factor $300$ speed-up at $N_a = 14$. In 2D, the minimum number of angular bins used is typically $N_a = 6 \times 10$. Although our method uses a truncated Legendre expansion, other expansions of the kernels or approximation techniques may also provide a low-rank structure that can be leveraged. We also discuss the use of the iterative Jacobi method to solve the \aclu{NR} iteration, even though this method only works in regions of weak interactions or for small time steps. Other iterative methods will be investigated.

To test each part of the implementation, we benchmarked our solver against standard test cases. We find that a higher energy resolution ($N_\varepsilon=30$) than typically used in simulations is required to reach percent-level accuracy of neutrino average energy in the presence of velocity shocks. We find quadratic convergence in propagation angles and energy, as expected from the schemes used.

Finally, we compare our solver to \texttt{Gmunu}'s M1 solver. We compare the luminosities and average (or root-mean-squared) energy for an inhomogeneous radiating sphere (position- and energy-dependent opacities in curved spacetime) and in a more realistic relaxation test case in a hot neutron star environment. Both show significant differences in the free-streaming regime luminosity (about $10\%$ difference) which may be explained by the closure relation. We stress, however, that this $10\%$ difference was obtained in simplified stationary test cases and cannot be straightforwardly transposed to complex dynamical simulations. The average and root-mean-squared energy agree in both cases, and the M1 scheme reconstructs the average energy in the free-streaming region with a generally better accuracy than our Boltzmann solver in the inhomogeneous radiating sphere test case in Sec.~\ref{sec:inhomogeneous_radiating_sphere_m1}. Designing test-specific closures would confirm the role of the closure in the luminosity discrepancies. We also find that our Boltzmann solver may require a large number of angular bins to converge in the hot neutron star test case. This number may however depend on the numerical scheme and on the slope limiters. Additionally, we compare the solvers on a \aclu{CCSN} test case and show good agreement in density and temperature profiles at core bounce for a sufficiently large ($N_\varepsilon = 20$) number of energy bins. We stress that a complete comparison should account for the possible conservation law violations to disentangle the effects of these conservation violations from those of the closure and moment truncation.

\begin{acknowledgements}
We thank all the people who have indirectly contributed to this work throughout its development, including, but not limited to, R. Akaho, M. Bhattacharyya, T. Colemont, F. De Ceuster, M. De Haes, F. Foucart, H.-T. Janka, H. Nagakura, D. Radice, S. Richers and S. Yamada.

A.O. and T.G.F.L. are supported by the Research Foundation Flanders (FWO) research project G086722N and FWO International Research Infrastructure (IRI) projects I002123N and I000725N. A.O. also acknowledges support from FWO grant K212426N.

H.H.Y.N. is supported by the Croucher Postdoctoral Fellowship by the Croucher Foundation. 

P.C.-K.C. acknowledges support from NSF Grant PHY-2020275 (Network for Neutrinos, Nuclear Astrophysics, and Symmetries (N3AS)). 

A.M. is partially supported by the National Science Foundation Gravitational Physics Theory Program through grant PHY-2409148 and by the National Science Foundation Cyberinfrastructure for Sustained Scientific Innovation Program through grant OAC-2513245.
\end{acknowledgements}

\onecolumngrid

\appendix
\section{Comoving tetrad \label{app:comoving_tetrad}}

\noindent We choose the spatial components of ${L^\mu}_{\hmu}$ to be such that ${\ell^i}_{\hati}$ is triangular. We define $a^i$, $b^i$ and $c^i$ such that ${\ell^i}_{\hati} = (a^i,b^i,c^i)$. We can then obtain a system of equations to determine those three vectors, while imposing $a^2=a^3=b^3=0$. We then introduce
\begin{equation}
    \Upsilon_{ij} = \gamma_{ij} - v_i v_j
\end{equation}
The equations in Appendix B of Ref.~\cite{Cardall_2013} can then be rewritten
\begin{align}
    &\Upsilon_{ij} a^i a^j = 1 & &\Upsilon_{ij} b^i b^j = 1 & &\Upsilon_{ij} c^i c^j = 1\\
    &\Upsilon_{ij} a^i b^j = 0 & &\Upsilon_{ij} a^i c^j = 0 & &\Upsilon_{ij} b^i c^j = 0
\end{align}
The solution to these equations including the additional constraints we imposed can be found analytically. The solutions are
\begin{align}
    &a^i = \frac{1}{\sqrt{\Upsilon_{11}}}(1,0,0)^T ,\\[1.5ex]
    &b^i = \frac{1}{\sqrt{\Upsilon_{22} + \Upsilon_{12}\hat{\lambda}}}(\hat{\lambda},1,0)^T ,\\[1.5ex]
    &c^i = \frac{1}{\sqrt{\Upsilon_{33} + \Upsilon_{23}\hat{\beta} + \Upsilon_{13}\hat{\sigma}}}(\hat{\sigma},\hat{\beta},1)^T ,
\end{align}
where we introduced
\begin{align*}
    &\hat{\lambda} = -\frac{\Upsilon_{12}}{\Upsilon_{11}} &  &\hat{\alpha} = -\frac{\Upsilon_{13}}{\Upsilon_{11}} & &\hat{\beta} = -\frac{\Upsilon_{23}\Upsilon_{11}-\Upsilon_{13}\Upsilon_{12}}{\Upsilon_{11}\Upsilon_{22} - (\Upsilon_{12})^2} & & \hat{\sigma} = \hat{\alpha} + \hat{\lambda} \hat{\beta}
\end{align*}

In our case where $\gamma_{ij}$ is diagonal, we can rewrite these expressions in a more concise way. First, we introduce the Lorentz factor along the first axis $W_1 \equiv W(v^1, v^2=0, v^3=0)$. Then, we introduce the Lorentz factor on the $12$-plane $W_2 \equiv W(v^1,v^2, v^3=0)$. With this notation, we can rewrite $a^i$, $b^i$ and $c^i$ as
\begin{align*}
    &a^i = \frac{W_1}{\sqrt{\gamma_{11}}}(1,0,0)^T ,&&b^i = \frac{W_2}{\sqrt{\gamma_{22}}} (W_1 \gamma_{22} v^1 v^2 , \frac{1}{W_1},0)^T , &&c^i = \frac{W}{\sqrt{\gamma_{33}}}(W_2 \gamma_{33} v^1 v^3,W_2 \gamma_{33} v^2 v^3,\frac{1}{W_2})^T.
\end{align*}
It then becomes apparent that the comoving frame coincides with the orthonormal frame when $v^i=0$. The expression of the full tetrad $\tensor{L}{^\mu_\hmu}$ for a diagonal spatial metric $\gamma_{ij}$ is then
\begin{equation}
    \tensor{L}{^\mu_\hmu} = 
        \begin{bmatrix}
            \frac{W}{\alpha} & \frac{W_1}{\alpha} \sqrt{\gamma_{11}} v^1 & \frac{W_1 W_2}{\alpha} \sqrt{\gamma_{22}} v^2 & \frac{W_2 W}{\alpha} \sqrt{\gamma_{33}} v^3\\
            \frac{W}{\alpha}(\alpha v^1 - \beta^1) & \frac{W_1}{\alpha\sqrt{\gamma_{11}}}(\alpha - \gamma_{11}\beta^1 v^1) & \frac{W_1 W_2}{\alpha}(\alpha v^1 - \beta^1) \sqrt{\gamma_{22}} v^2 & \frac{W_2 W}{\alpha}(\alpha v^1 - \beta^1) \sqrt{\gamma_{33}} v^3\\
            \frac{W}{\alpha}(\alpha v^2 - \beta^2) & -\frac{W_1}{\alpha} \sqrt{\gamma_{11}} v^1 \beta^2 & \frac{W_2}{\alpha\sqrt{\gamma_{22}}}(\frac{\alpha}{W_1} - W_1 \gamma_{22}\beta^2 v^2) & \frac{W_2 W}{\alpha}(\alpha v^2 - \beta^2) \sqrt{\gamma_{33}} v^3\\
            \frac{W}{\alpha}(\alpha v^3 - \beta^3) & -\frac{W_1}{\alpha} \sqrt{\gamma_{11}} v^1 \beta^3  & -\frac{W_1 W_2}{\alpha} \sqrt{\gamma_{22}} v^2 \beta^3 & \frac{W}{\alpha\sqrt{\gamma_{33}}}(\frac{\alpha}{W_2} - W_2 \gamma_{33}\beta^3 v^3)
        \end{bmatrix}.
\end{equation}
Note that in our framework, we have $\gamma_{ij} = \psi^4 \hat{\gamma}_{ij}$.

\section{Area and volume elements in momentum space \label{app:area_and_volume}}
\noindent As mentioned in the main text, we want to ensure that parts of the momentum-space flux cancel with parts of the flux in position space, as expressed by Eq.~\eqref{eq:discr}. To this end, we need to compute the area and volume elements using Eqs.~\eqref{eq:Vx} to~\eqref{eq:Ap}. The difference with the spatial elements in spherical coordinates is that we divide by $\po$. For massless particles, we have $\po = \varepsilon$ and
\begin{align}
    &\Delta V_p = \left( \varepsilon \Delta \varepsilon  \right) \, \left( 2 \sin \frac{\Delta \vartheta }{2} \, \sin \vartheta \right) \, \Delta \varphi \label{eq:nm_Vp}\\
    &\Delta A_{p,\varepsilon} = \left( \varepsilon \pm \frac{\Delta \varepsilon}{2}  \right) \, \left( 2 \sin \frac{\Delta \vartheta }{2} \, \sin \vartheta \right) \, \Delta \varphi \label{eq:nm_Ap}\\
    &\Delta A_{p,\vartheta} = \left( \varepsilon \Delta \varepsilon  \right) \, \sin \left( \vartheta \pm \frac{\Delta \vartheta }{2} \right) \, \Delta \varphi \label{eq:nm_Ath}\\
    &\Delta A_{p,\varphi} = \left( \varepsilon \Delta \varepsilon  \right) \, \left( 2 \sin \frac{\Delta \vartheta }{2} \, \sin \vartheta \right) \label{eq:nm_Aphi}
\end{align}

\section{Calculation of kernels for elastic scattering} \label{app:kernels_ES}
\noindent As discussed in Sec.~\ref{sec:source_terms}, the Boltzmann solver requires the
use of kernels $R_\mathrm{ES}$ rather than the opacity $\kappa_s$ to describe
ES interactions between neutrinos and nucleons, light
clusters, and heavy nuclei in dense matter. Here we continue to assume
isoenergetic scattering, i.e., $\varepsilon = \varepsilon^{\prime}$.  

The kernel representation for ES is not commonly adopted in neutrino transport.
Although simplified expressions for ES kernels were derived decades
ago~\citep{Bruenn_1985, Rampp2002}, we reformulate the kernels directly from the
opacities using more accurate prescriptions.  

\subsection{Neutrino--nucleon elastic scattering: $\nu+N \leftrightarrow
\nu+N$}\label{sec:nu_s_Nscat}

\noindent For this interaction, the kernel-form ES treatment must incorporate several
corrections, including weak magnetism, nucleon recoil, the strange-quark
contribution to the nucleon spin, and the vector and axial response factors
arising from density and spin fluctuations of the
medium~\citep{Horowitz2002,Hobbs2016,Horowitz2017}.  We provide a method to
derive the corresponding kernels with these corrections, starting from the
scattering opacity.  

The differential cross section for scattering off a nucleon $N$ (either proton
$p$ or neutron $n$) is given by Ref.~\cite{Burrows_2006} \begin{equation}
\begin{aligned} \frac{d \sigma_N}{d \Omega_p} = & \, \frac{\sigma_0}{16 \pi}
\left(\frac{\varepsilon}{m_e c^2}\right)^2 \left[ C_V^2(1+\omega) S_V +
C_A^2(3-\omega) S_A \right] W_{\mathrm{M}, \nu}^{\mathrm{NC}} \,
W_{\mathrm{R}, \nu}^{\mathrm{NC}}, \end{aligned} \end{equation} where
$W_{\mathrm{R}, \nu}^{\mathrm{NC}}$ and $W_{\mathrm{M}, \nu}^{\mathrm{NC}}$ are
the neutral-current correction factors for nucleon recoil and weak magnetism,
respectively, and depend on the neutrino species~\citep{Horowitz2002}.  Here,
$m_e$ denotes the electron rest mass.  
The factors $S_V$ and $S_A$ are the vector and axial response functions,
respectively, which quantify the medium’s response to density and spin
fluctuations (see~\cite{Horowitz2017,Ng_2024}).  The reference cross
section is defined as \[ \sigma_0 := \frac{4 (m_e c^2 G_F)^2}{\pi (\hbar c)^4}
= 1.761 \times 10^{-44}~\mathrm{cm}^2, \] where $G_F$ is the Fermi constant.  
Using the Weinberg angle $\theta_W$, with $\sin ^2 \theta_W=0.22290$, 
the vector coupling constants $C_V$ are $0.5 - 2 \sin^2\theta_W$ for protons
and $-0.5$ for neutrons, while the axial-vector coupling constants $C_A$ are
$\tfrac{1}{2}(g_A - g_A^s)$ for protons and $-\tfrac{1}{2}(g_A + g_A^s)$ for
neutrons.  Here, $g_A = 1.27$ is the axial-vector coupling constant and $g_A^s
= -0.1$ is the nucleon’s strange helicity~\citep{Hobbs2016}.

It is straightforward to separate the expression into two terms: one
proportional to $\omega$ and one independent of $\omega$. These two
contributions are linearly independent.  Therefore, after computing the
transport (momentum-transfer) cross section, the result naturally splits into
parts with and without $\omega$ dependence.  

Following Ref.~\cite{Burrows_2006}, we obtain the transport cross section
$\sigma_N^t$ as a function of the total cross section $\sigma_N$ and the
angular coefficient $\delta_N$: \begin{equation}\label{eq:sigmatr} \sigma_N^{t}
= \int \frac{d \sigma_N}{d \Omega_p} (1-\omega) \, d \Omega_p = \sigma_N \left(1 -
\tfrac{1}{3} \delta_N \right), \end{equation} where the total cross section is
\begin{equation} \sigma_N = \frac{\sigma_0}{4} \left(\frac{\varepsilon}{m_e
c^2}\right)^2 \left( C_V^2 S_V + 3 C_A^2 S_A \right)
W^{\mathrm{NC}}_{\mathrm{M}, \nu} \, W^{\mathrm{NC}}_{\mathrm{R}, \nu},
\end{equation} and $\delta_N = \frac{C_V^2 S_V - C_A^2 S_A}{C_V^2 S_V + 3 C_A^2 S_A}$.

To derive the Legendre coefficients, we first express the scattering opacity as
\begin{equation}\label{eq:ES_kappa_s_form2} 
\kappa_s(\varepsilon) = \eta_{NN} \sigma^t_N(\varepsilon)= \eta_{NN}\sigma_N \left(1 - \tfrac{1}{3} \delta_N \right),
\end{equation} where $\eta_{NN}$ accounts for nucleon phase-space blocking~\citep{Ng_2024}.  

By comparing Eqs.~\eqref{eq:ES_kappa_s_form2} and \eqref{eq:kappa_s}, one finds
\begin{equation} \begin{aligned} R_{\mathrm{ES},0} &= \frac{\eta_{NN} \,
\sigma_N}{4 \pi \varepsilon^2}, & \quad R_{\mathrm{ES},1} &= \frac{\eta_{NN} \,
\sigma_N \delta_N}{4 \pi \varepsilon^2}, \\[6pt] \Phi_{\mathrm{ES},0} &=
\frac{\eta_{NN} \, \sigma_N}{2 \pi \varepsilon^2}, & \quad \Phi_{\mathrm{ES},1} &=
\frac{\eta_{NN} \, \sigma_N \delta_N}{6 \pi \varepsilon^2}, \end{aligned}
\end{equation} with units of $\mathrm{MeV^{-2}~cm^{-1}}$.  We note that
$\Phi_{\mathrm{ES},0/1}$ carry different units than the inelastic scattering
kernels $\Phi_{\mathrm{IS},0/1}$, which have dimensions of
$\mathrm{cm^{-3}~s^{-1}}$.

\subsection{Neutrino--nucleus elastic scattering: $\nu+A_{\mathrm{nuc}} \leftrightarrow \nu+A_{\mathrm{nuc}}$}\label{sec:nu_s_Hscat}

\noindent In Eq.~(35) of Ref.~\cite{Ng_2024}, the differential cross section takes a
non-trivial form, since both the corrections and the coupling-constant terms
depend explicitly on $\omega$.  To simplify the treatment, we follow the
approximate expression of Ref.~\cite{Bruenn_1985}, including a factor $\langle
S(\varepsilon) \rangle$ to account for the ion--ion correlation function, which
incorporates the effects of Coulomb interactions between
nuclei~\citep{Horowitz1997, Rampp2002}.

The resulting Legendre coefficients are
\begin{equation}\label{eq:heavy_nuclei_ES} \begin{aligned} \Phi_{\mathrm{ES},0} =&
\frac{\sigma_0}{16 \pi m_e^2 c^4} \, A_{\mathrm{nuc}}^2 n_H \left[ C_{V0} -
\tfrac{1}{2}\!\left(1 - \tfrac{2Z_{\mathrm{nuc}}}{A_{\mathrm{nuc}}}\right) C_{V1} \right]^2 \times \\
& \frac{2y - 1 +
e^{-2y}}{y^2} \, \langle S(\varepsilon) \rangle, \\[6pt] \Phi_{\mathrm{ES},1} =&
\frac{\sigma_0}{16 \pi m_e^2 c^4} \, A_{\mathrm{nuc}}^2 n_H \left[ C_{V0} -
\tfrac{1}{2}\!\left(1 - \tfrac{2Z_{\mathrm{nuc}}}{A_{\mathrm{nuc}}}\right) C_{V1} \right]^2 \times \\
& \frac{2 - 3y +
2y^2 - (2+y)e^{-2y}}{y^3} \, \langle S(\varepsilon) \rangle, \end{aligned}
\end{equation} where $y = 4b \varepsilon^2$ with $b \approx 3.70 \times 10^{-6}
A_{\mathrm{nuc}}^{2/3}~\mathrm{MeV}^{-2}$~\citep{Ng_2024}.  The coupling constants are $C_{V0}
= -\sin^2\theta_{W}$ and $C_{V1} = 1 - 2 \sin^2\theta_W$~\citep{Bruenn_1985}.
Here $n_H$, $A_{\mathrm{nuc}}$, and $Z_{\mathrm{nuc}}$ denote the number density of heavy nuclei (excluding
$\alpha$-particles and other light clusters), the mass number, and the charge
number, respectively.  

We note that nuclear form-factor and electron-polarization corrections are
neglected in this treatment.  Furthermore, Eqs.~(C44, C45) in Ref.~\cite{Bruenn_1985} contain a typo in the coupling constant. Specifically,
the factor $ \left(C_{V0} + \tfrac{1}{2}\tfrac{N-Z_{\mathrm{nuc}}}{A_{\mathrm{nuc}}} C_{V1}\right)^2 $
should be replaced with $\left[C_{V0} - \tfrac{1}{2}\!\left(1 -
\tfrac{2Z_{\mathrm{nuc}}}{A_{\mathrm{nuc}}}\right) C_{V1}\right]^2 $ (also corrected in Ref.~\cite{Kuroda2016}).

\subsection{Neutrino--light cluster elastic scattering: $\nu+A_{\mathrm{light}} \leftrightarrow \nu+A_{\mathrm{light}}$}\label{sec:nu_s_light_cluster_scat}
\noindent Light clusters considered here include $\alpha$-particles, deuterons ($^2\mathrm{H}$), tritons ($^3\mathrm{H}$), and helions ($^3\mathrm{He}$).
When focusing on neutrino--$\alpha$ elastic scattering, it is important to note
that treating the $\alpha$-particle as a heavy nucleus in ES (using
Eq.~\eqref{eq:heavy_nuclei_ES}, as done for example in Refs.~\cite{Rampp2002,Kuroda2016}) leads to a slight
overestimate of the scattering opacity before core bounce.  

Analogous to Sec.~\ref{sec:nu_s_Nscat}, the differential cross section can be
separated into $\omega$-dependent and $\omega$-independent contributions, which
are linearly independent.  Starting from Eq.~\eqref{eq:sigmatr}, 
$\sigma^t_{\mathrm{light}} = \sigma_{\mathrm{light}} \left(1 -
\tfrac{1}{3}\delta_{\mathrm{light}}\right)$,
and the relation for cross-section~\citep{Burrows_2006},
$\sigma^t_{\mathrm{light}} =
\tfrac{2}{3}\,\sigma_{\mathrm{light}}$, 
we can obtain $\delta_{\mathrm{light}} = 1$.

The cross section is given by \begin{equation} \begin{aligned} \sigma_{\mathrm{light}}
&= \frac{\sigma_0}{4} \left(\frac{\varepsilon}{m_e c^2}\right)^2 A_{\mathrm
light}^2 \, \bigg[ C_A - 1 + \frac{Z_{\mathrm{light}}}{A_{\mathrm{light}}}\left(2 - C_A -
C_V\right) \bigg]^2, \end{aligned} \end{equation} where $C_A = 0.5$, $C_V = 0.5
+ 2 \sin^2\theta_W$, and $A_{\mathrm{light}}$ and $Z_{\mathrm{light}}$ are the mass and
charge numbers of the light cluster.  

Therefore, the corresponding kernels are \begin{equation} 
\Phi_{\mathrm{ES},0} = \frac{n_{\mathrm{light}} \, \sigma_{\mathrm{light}}}{2 \pi
\varepsilon^2},\,\,\,\,\,\, \Phi_{\mathrm{ES},1} = \frac{n_{\mathrm{light}} \, \sigma_{\mathrm
light} \delta_{\mathrm{light}}}{6 \pi \varepsilon^2}, \end{equation}
where $n_{\mathrm{light}} = n_b \, \frac{X_{\mathrm{light}}}{A_{\mathrm{light}}}$ is the
number density of a particular light cluster species with mass fraction $X_{\mathrm
light}$, and $n_b$ is the baryon number density.

\section{Implicit evolution of source terms} \label{app:source_terms_implicit}
\noindent We will derive the expression of the discretization of the source terms and show that it can be written as indicated in Eq.~\eqref{eq:discr_source}:
\begin{equation*}
    \vm{s}_{\mathrm{rad}} = \alpha \psi^6 \sqrt{\frac{\bar{\gamma}}{\hat{\gamma}}} \left[\left( \tilde{\vm{A}} \vm{f} \right) \odot \vm{f} + \tilde{\vm{B}} \vm{f} + \tilde{\vm{c}}\right].
\end{equation*}

We consider the vector $f_I = f_s(\textbf{p}_{l,m,n})$, where $I=g(l,m,n,s)$ with $g$ a mapping from the $4$D indices (momentum and species) to a $1$D index, and $f$ is the distribution function at a given point in space. For simplicity, we will omit the spatial dependence in the notation.

We start with the discretization of each source term. We compute the integral of a function $h$ as
\begin{equation}
    \int h(\vm{p}) d^3p \approx \sum_I h_I \, \varepsilon^2_I \, \sin\left(\vartheta_{I} \right) \Delta\varepsilon_I \Delta \vartheta_{I} \Delta \varphi_{I}.
\end{equation}
In the following, we will also omit the $\po$ term and reintroduce it at the end.
We can then write the discretization of the absorption and emission source term of Eq.~\eqref{eq:ae} as
\begin{equation}
    C_{\mathrm{E/A}, I} = \eta_I - \kappa^*_{a,I} f_I,
\end{equation}
where $\eta_I = \eta(\varepsilon_I)$ and $\kappa^*_{a,I} = \kappa^*_a(\varepsilon_I)$. In vector notation, this reads
\begin{equation}
    \vm{C}_{\mathrm{E/A}} = \symvm{\eta} - \symvm{\kappa}^*_{a} \odot \vm{f} = \symvm{\eta} - \mathrm{diag}\left( \symvm{\kappa}^*_{a} \right) \vm{f},\label{eq:discr_ae_mat}
\end{equation}
where $\odot$ denotes the Hadamard, or element-wise, product and \textit{diag}$(\cdot)$ returns the diagonal matrix whose diagonal elements are the elements of the argument. For elastic scattering, Eq.~\eqref{eq:es} becomes
\begin{equation}
    C_{ES,I} = \varepsilon^2_I \sum_J \sin(\vartheta_{J}) \Delta\vartheta_J \Delta\varphi_J R^{\mathrm{iso}}_{IJ} \left( f_J - f_I \right), \label{eq:discr_es}
\end{equation}
where $R^{\mathrm{iso}}_{IJ} = R^{\mathrm{iso}}(\varepsilon_I, \omega_{IJ})$ with 
\begin{equation}
    \omega_{IJ} = \cos\vartheta_I \cos\vartheta_J + \sin\vartheta_I \sin\vartheta_J \cos(\varphi_I - \varphi_J).
\end{equation}
We introduce the matrix $G^\mathrm{iso}_{IJ} \equiv \varepsilon_I^2 \sin(\vartheta_{J}) \Delta\vartheta_J \Delta\varphi_J R^{\mathrm{iso}}_{IJ}$ to write Eq.~\eqref{eq:discr_es} in the matrix form
\begin{equation}
    \vm{C}_{ES} = \vm{G}^{\mathrm{iso}} \vm{f} - (\vm{G}^{\mathrm{iso}} \vm{1}) \odot \vm{f} = \vm{G}^{\mathrm{iso}} \vm{f} - \mathrm{diag}(\vm{G}^{\mathrm{iso}} \vm{1}) \vm{f},\label{eq:discr_es_mat}
\end{equation}
where $\vm{1}$ is a vector of ones. $\vm{G}^{\mathrm{iso}} \vm{1}$ thus represents the sum of columns of $\vm{G}^{\mathrm{iso}}$, \textit{i.e.} in this case the integral of $R^{\mathrm{iso}}$ over the solid angle in momentum space.
We proceed in a similar manner for the inelastic scattering and pair processes and write Eqs.~\eqref{eq:is} and \eqref{eq:pp} as
\begin{align}
    &C_{IS,I} = \sum_J \Delta \varepsilon_J \Delta \vartheta_J \Delta \varphi_J \varepsilon^2_J \sin\vartheta_J \left[R^{\mathrm{in}}_{IJ} f_J (1-f_I) - R^{\mathrm{out}}_{IJ} (1-f_J) f_I \right], \label{eq:discr_is}\\[2ex]
    &C_{PP,I} = \sum_J \Delta \varepsilon_J \Delta \vartheta_J \Delta \varphi_J \varepsilon^2_J \sin\vartheta_J \left[R^{\mathrm{pro}}_{IJ} (1-f_J) (1-f_I) - R^{\mathrm{ann}}_{IJ} f_J f_I \right],\label{eq:discr_pp}
\end{align}
where $R^{\mathrm{in/out/pro/ann}}_{IJ} = R^{\mathrm{in/out/pro/ann}}(\varepsilon_I, \varepsilon_J, \omega_{IJ})$.
Again, we introduce matrices
\begin{equation}
    G^{\mathrm{in/out/pro/ann}}_{IJ} = \Delta \varepsilon_J \Delta \vartheta_J \Delta \varphi_J \varepsilon^2_J \sin\vartheta_J R^{\mathrm{in/out/pro/ann}}_{IJ}
\end{equation}
to write Eqs.~\eqref{eq:discr_is} and \eqref{eq:discr_pp} in the matrix form
\begin{align}
    &\vm{C}_{IS} = \left[\left( \vm{G}^{\mathrm{out}} - \vm{G}^{\mathrm{in}}\right) \vm{f}  \right] \odot \vm{f} + \left[ \vm{G}^{\mathrm{in}} - \mathrm{diag}\left(\vm{G}^{\mathrm{out}} \vm{1} \right) \right] \vm{f}, \label{eq:discr_is_mat}\\[2ex]
    &\vm{C}_{PP} = \left[\left( \vm{G}^{\mathrm{pro}} - \vm{G}^{\mathrm{ann}}\right) \vm{f}  \right] \odot \vm{f} - \left[ \vm{G}^{\mathrm{pro}} + \mathrm{diag}\left(\vm{G}^{\mathrm{pro}} \vm{1} \right)\right] \vm{f} + \vm{G}^{\mathrm{pro}} \vm{1}.\label{eq:discr_pp_mat}
\end{align}
Note that the components $G^{\mathrm{pro}}_{II'}$ and $G^{\mathrm{ann}}_{II'}$ with $I=g(l,m,n,s)$ and $I'=g(l',m',n',s')$ are 0 if $s = s'$ (except for the heavy-lepton species $\nu_x$), since pair processes couple a particle and its antiparticle, whereas for the other interactions $G_{II'}$ are 0 if $s\neq s'$.
Gathering Eqs.~\eqref{eq:discr_ae_mat}, \eqref{eq:discr_es_mat}, \eqref{eq:discr_is_mat} and \eqref{eq:discr_pp_mat} and reintroducing $\po$, we find
\begin{align*}
    \vm{s}_{\mathrm{rad}} &= \alpha \psi^6 \sqrt{\frac{\bar{\gamma}}{\hat{\gamma}}} \Big(\vm{C}_{\mathrm{E/A}} + \vm{C}_{ES} + \vm{C}_{IS} + \vm{C}_{PP}\Big)\\
        &= \alpha \psi^6 \sqrt{\frac{\bar{\gamma}}{\hat{\gamma}}} \vm{p}^{\hat{0}} \odot \left[\left( \tilde{\vm{A}} \vm{f} \right) \odot \vm{f} + \tilde{\vm{B}} \vm{f} + \tilde{\vm{c}}\right]
\end{align*}
where we identify (no sum over repeated indices)
\begin{align}
    &\Tilde{A}_{IJ} = G^{\mathrm{out}}_{IJ} + G^{\mathrm{pro}}_{IJ} - G^{\mathrm{in}}_{IJ} - G^{\mathrm{ann}}_{IJ},\\
    &\Tilde{B}_{IJ} =  G^{\mathrm{iso}}_{IJ} + G^{\mathrm{in}}_{IJ} - G^{\mathrm{pro}}_{IJ} - \delta_{IJ}\Big[ \kappa^*_{a,I} + \sum_K \big( G^{\mathrm{iso}}_{IK} + G^{\mathrm{out}}_{IK} + G^{\mathrm{pro}}_{IK}  \big) \Big] ,\\
    &\Tilde{c}_{I} = \eta_I + \sum_K G^{\mathrm{pro}}_{IK} .
\end{align}
We will now derive the expression of the equation to solve for the implicit evolution. We assume that $\psi$ does not change through neutrino interactions with the fluid. From Eqs.~\eqref{eq:implicit} and \eqref{eq:q}, and accounting for the fact that $\bar{\gamma} = \hat{\gamma}$, we have
\begin{equation}
    \psi^6 \exppar{\left(\mathcal{L}_{\hmu} \vm{p}^{\hmu}\right)}{n+1} \odot \exppar{\vm{f}}{n+1} - \psi^6 \exppar{\left(\mathcal{L}_{\hmu} \vm{p}^{\hmu}\right)}{n} \odot \exppar{\vm{f}}{n} - \Delta t \alpha \psi^6 \vm{p}^{\hat{0}} \odot \left[ \left( \tilde{\vm{A}} \exppar{\vm{f}}{n+1} \right) \odot \exppar{\vm{f}}{n+1} + \tilde{\vm{B}} \exppar{\vm{f}}{n+1} + \tilde{\vm{c}}\right] = 0 
\end{equation}
Using the element-wise division, we can then identify $\vm{A}$, $\vm{B}$ and $\vm{c}$ from Eq.~\eqref{eq:implicit_f} 
\begin{align}
    &A_{IJ} = \alpha \Delta t \Tilde{A}_{IJ} ,\\[1.5ex]
    &B_{IJ} = \alpha \Delta t \Tilde{B}_{IJ},\\[1.5ex]
    &c_{I} = \alpha \Delta t \Tilde{c}_I + \exppar{(w_p)_I}{n} \exppar{f_I}{n},
\end{align}
where $\vm{w}_p$ is defined in Eq.~\eqref{eq:w_p}.
Under the assumption that the fluid velocity does not change significantly through neutrino interactions, we have 
\begin{equation}
    \exppar{\vm{w}_p}{n} = \exppar{\vm{w}_p}{n+1},
\end{equation}
which we use in the solver.

\section{Matrix-vector products in the Kronecker-product representation \label{app:kronecker_product}}
\noindent In Sec.~\ref{sec:dimensionality_reduction}, we described how to reduce the dimensionality of the implicit solver problem from a size $N\equiv N_e N_a$, where $N_e = N_\varepsilon N_\mathrm{spec}$ and $N_a = N_\vartheta N_\varphi$, into a problem of size $r = r' N_e$, with $r' = (\lmax +1)^2$ and $\lmax$ the highest order considered in the Legendre expansion of the interaction kernels. This reduction is performed with the matrices $\bvm{V}$ and $\bvm{U}$, defined in Eqs.~\eqref{eq:V_kronecker} and~\eqref{eq:U_kronecker}, respectively. Both matrices can be expressed in terms of Kronecker products. This structure enables more efficient computations of matrix-vector products. In this section, we describe how to perform these operations efficiently in our specific formalism and discuss their complexity.

We remind the reader that the Kronecker-product formalism does not require building any matrices explicitly, except for the Jacobian $\vm{J}_y$ in case of failure of the Jacobi method. Instead, we only implement the functions that compute matrix-vector products from the individual matrices that are used in the Kronecker products.

There are two main types of matrix-vector products that we have to compute. First, the projections $\bvm{V}^T \vm{z}$, where $\vm{z}$ is a vector of size $N$.
For a block $\ell$, we have
\begin{equation}
   \bvm{V}_\ell^T \mathrm{vec}(\vm{Z}) = \mathrm{vec}\big(\vm{V}^T_\ell \vm{Z} \mathbb{1}_{N_e}\big) = \mathrm{vec}\big(\vm{V}^T_\ell \vm{Z} \big),
\end{equation}
where $\vm{Z} \in \mathbb{R}^{N_a \times N_e}$ is a matrix representation of the vector $\vm{z}$ (\textit{i.e.} $\vm{z}= \mathrm{vec}(\vm{Z})$). As a reminder, the block $\bvm{V}_\ell = \mathbb{1}_{N_e} \otimes \vm{V}_\ell$ and $\vm{V}_\ell$ is defined below Eq.~\eqref{eq:B_kronecker}. Thus, the complexity of the matrix multiplication for a block $\bvm{V}_\ell$ of size $N_e N_a \times N_e r_\ell'$ is $O(r_\ell' N_a N_e)$, with $r_\ell' = (2\ell +1)$. Since this operation must be performed for $(\lmax+1)$ blocks, computing $\bvm{V}^T \vm{z}$ is of complexity $O(r' N_a N_e)$, with $r' = (\lmax + 1)^2$. In comparison, the naive implementation of this operation using full-size vectors and matrices is $O(r' N_a N^2_e)$. 

The second type of matrix-vector product is $\bvm{U}_{A/B} \, \bvm{x}$ for a vector $\bvm{x}$ of size $r = r' N_e$, where $\bvm{U}_{A/B}$ is defined in Eq.~\eqref{eq:U_kronecker}. Because both $\bvm{U}_{A/B}$ and $\bvm{x}$ are organized by block, we can perform the matrix multiplication block-by-block. For each block $\ell$, we have
\begin{equation}
    \bvm{U}_{A/B,\ell} \, \bvm{x}_\ell = \mathrm{vec}( \vm{U}_\ell  \, \vm{X}_\ell  \, \symvm{\Phi}^T_{A/B,\ell} ),
\end{equation}
where $\bvm{x}_\ell = \mathrm{vec}(\vm{X}_\ell)$ and $\vm{X}_\ell$ is of size $r_\ell' \times N_e$. The cost of the multiplication for each block $\ell$, neglecting the \enquote{vec} operation, is of complexity $O(r_\ell' N_a N_e +r_\ell' N_e^2)$ if one computes the product $\vm{X}_\ell \, \symvm{\Phi}^T_{A/B,\ell}$ first. The total cost of the product $\bvm{U}_{A/B} \, \bvm{x}$ is $O(r' N_a N_e +r' N_e^2)$, whereas the complexity for the naive full-size implementation is $O(N_a N_e^2 r')$. Both types of products can then be used in the Jacobi method to compute $\vm{J}_{y} \vm{h}$ efficiently. Note that computing the product $\vm{U}_J \, \bvm{x}$, with $\vm{U}_J$ defined in Eq.~\eqref{eq:jacobian_matrix_outer}, requires computing $\vm{U}_A  \, \bvm{x}$ and $\vm{U}_B \, \bvm{x}$ separately. The result of $\vm{U}_A \, \bvm{x}$ is then multiplied by $\diag{\vm{f}}$ on the left, which is effectively performed with a Hadamard product.  

Concerning memory requirements, this formulation only needs to form $\vm{D}_B$, $\symvm{\Phi}_{A/B,\ell}$, $\vm{U}_\ell$ and $\vm{V}_\ell$. In the worst case, the largest matrix to store is the Jacobian $\vm{J}_y$ of size $r' N_e \times r' N_e$. In total, the memory requirement scales approximately as $O(r' N_a + N_e N_a + r'^2 N_e^2)$.

Part of the cost of the LU decomposition comes from the fact that we have to construct the Jacobian. It can nonetheless be more efficiently constructed in $O(r'^2 N_e N_a + r'^2 N_e^2)$ using Kronecker product properties. Since the Jacobian is of size $r \times r$, the complexity of the LU decomposition is $O(r^3)$, or $O(r'^3 N_e^3)$. The Jacobi method, on the other hand, only requires matrix-vector products and, therefore, scales as $O(K r' N_a N_e + K r' N_e^2)$, where $K$ is the number of iterations before convergence.

Note that, in practice, we never apply the \enquote{vec} operator. Instead, we store all vectors as matrices. For example, $\bvm{y}$ is stored as a matrix of size $r' \times N_e$ and $\vm{f}$ as a matrix of size $N_a \times N_e$. We also construct the Jacobian as a set of $N_e^2$ blocks of size $r' \times r'$ instead of $(\lmax+1)^2$ blocks of size $N_e r_\ell' \times N_e r_\ell'$. When reconstructed for the LU-method, the vector $\bvm{y}$ must be accordingly reshaped into $N_e$ blocks of size $r'$, \textit{i.e.} grouping all spherical harmonics coefficients of a given energy bin. This representation is more suitable for the construction of the Jacobian and the memory organization in \texttt{Fortran}. Note that this does not affect any other variables in the system of reduced dimensionality, since the matrices $\bvm{U}$ and $\bvm{V}$ are never constructed explicitly, and the matrix representation of vectors is not sensitive to such block organization. This discussion is only relevant when flattening into 1D vectors is required.

\section{Analytical solutions of the Boltzmann equation in the comoving frame} \label{app:analytic_sol}

\noindent The Boltzmann equation can be solved analytically in simple cases, for example when the source terms vanish or only contain absorption and emission. The presence of the velocity in Eq.~\eqref{eq:cons_bol_ref}, with the definition of the fluxes in Eqs.~\eqref{eq:flux_x} and \eqref{eq:flux_p} and conserved variable in Eq.~\eqref{eq:q}, makes solving the equation more tedious. The homogeneous equation can be solved using the method of characteristics, expressing each coordinate along a trajectory parametrized by a variable $s$. One therefore finds a set of trajectories $t(s), \vec{x}(s), \vec{p}(s), \po(s)$ that describes the motion of neutrinos in phase space. Since the solution is constant along the characteristic curves that are found,  the solution at a time $t$ and position $(\vec{x}, \vec{p})$ can be expressed by going back in time along the characteristic to the initial time $t_0$, where the solution is given by the initial condition. In other words, we have $f(t,\vec{x}, \vec{p}) = f(t_0, \vec{x}_0, \vec{p}_0)$, where $(\vec{x}_0, \vec{p}_0)$ is the initial position in phase space for a particle that is at position $(\vec{x}, \vec{p})$ at time $t$.

In the comoving frame, the presence of the velocity makes the use of this method more involved, since it requires computing integrals involving the velocity and its derivatives. We can however use the fact that the solution in the orthonormal frame is not influenced by the velocity (provided the collision terms vanish) and can be mapped to the solution in the comoving frame. We introduce indices adorned with a bar to denote quantities in the local orthonormal frame and introduce the tetrad $\tensor{e}{^{\mu}_{\bar{\mu}}}$. We then use the fact that
\begin{equation}
    p^{\mu} =\tensor{L}{^\mu_\hmu}p^{\hmu} = \tensor{e}{^{\mu}_{\bar{\mu}}} p^{\bar{\mu}}, \label{eq:p_com_ortho}
\end{equation}
such that the solution in the comoving frame $p^{\hmu}(s)$ can be expressed as a function of the solution in the local orthonormal frame $p^{\bar{\mu}}(s)$
\begin{equation}
    p^{\hmu}(s) = \tensor{L}{^\hmu_\mu} \tensor{e}{^{\mu}_{\bar{\mu}}} p^{\bar{\mu}}(s).
\end{equation}
In the following, we consider the particles to be massless, \textit{i.e.} $\po = \varepsilon$.
We can simplify Eq.~\eqref{eq:p_com_ortho} in a spherically symmetric configuration with $v(r) \equiv v^r$, $\beta \equiv \beta^r(r)$ and we find, omitting the dependence on $r$ in the notation,
\begin{align}
    p^t &= \frac{\Bar{\varepsilon}}{\alpha} = \frac{\hat{\varepsilon}}{\alpha} \frac{1 + \psi^2 \hat{\mu} v}{\sqrt{1-\psi^4 v^2}},\\
    p^r &= \Bar{\varepsilon} \left(-\frac{\beta}{\alpha} + \frac{\Bar{\mu}}{\psi^2} \right) = \frac{\hat{\varepsilon}}{\alpha\sqrt{1-\psi^4 v^2}} \left[ \alpha v - \beta + \left( \frac{\alpha}{\psi^2} - \beta \psi^2 v\right) \hat{\mu} \right],
\end{align}
where $\mu=\cos\vartheta$ and quantities adorned with a bar are considered in the local orthonormal frame, whereas hatted quantities are considered in the fluid rest frame. We can thus use these relations to find the solution in the comoving frame without explicitly solving the more complex equation in the comoving frame. Note that the solution for $r(s)$ and $t(s)$ must be the same for the momentum in the comoving or orthonormal frame, since the spatial and temporal coordinates are described in the lab frame and thus only depend on $p^\mu$. We then invert the last system of equations and find the trajectory in phase space with the momentum in the comoving frame
\begin{align}
    \hat{\varepsilon}(s) &= \Bar{\varepsilon}(s) \: \frac{1 - \psi^2\big[r(s)\big] v\big[r(s)\big]\, \Bar{\mu}(s)}{\sqrt{1-\psi^4\big[r(s)\big]v^2\big[r(s)\big]}}, \label{eq:eps_com}\\
    \hat{\mu}(s) &= \frac{\psi^2\big[r(s)\big]v\big[r(s)\big] - \Bar{\mu}(s)}{\psi^2\big[r(s)\big]v\big[r(s)\big] \, \Bar{\mu}(s) - 1}. \label{eq:mu_com}
\end{align}
We can see that in the limit $v =0$, we find the equality between the quantities in the comoving frame and in the orthonormal frame, as expected.

We consider a spherically symmetric, conformally flat and stationary space with $\beta^r = 0$ and $v^r=0$. Using the method of characteristics, we find
\begin{align}
    &\bar{\varepsilon}(r) = \bar{\varepsilon}(r_0) \frac{\alpha(r_0)}{\alpha(r)}, \label{eq:eps_curv}\\
    &\bar{\mu}(r) = \pm \frac{\sqrt{h(r)^2 r^2 - h(r_0)^2 r_0^2(1-\bar{\mu}_0^2)}}{r h(r)}, \label{eq:mu_curv}\\
    & h(r) = \frac{\psi(r)^2}{\alpha},
\end{align}
with $r_0 = r(s=0)$ and $\bar{\mu}_0 = \bar{\mu}(s=0) = \bar{\mu}(r_0)$. The sign of $\bar{\mu}(r)$ may change with time upon crossing the point $\bar{\mu} = 0$. Accounting for these changes of sign, we can express $r(t)$ by solving
\begin{equation}
    \sum_{i=1}^N \int_{t_{i-1}}^{t_i} dt' = \sum_{i=1}^N \int_{r_{i-1}}^{r_i} \mathrm{sign}(\bar{\mu}) \frac{h(r')^2 r'}{\sqrt{h(r')^2 r'^2 - h(r_0)^2 r_0^2(1-\bar{\mu}_0^2)}} dr', \label{eq:r_curv}
\end{equation}
where we split the total integral at the points $r_i \equiv r(t_i)$ at which $\bar{\mu}$ changes sign. The time $t_0$ is the initial time and $t_N \equiv t$.

In the specific case where $h(r) = 1$, \textit{i.e.} a conformal spacetime with $\alpha = \psi^2$ (including flat spacetime $\alpha = \psi = 1$), the solutions for $r$ and $\bar{\mu}$ can be written explicitly. These solutions as a function of $t$ are
\begin{align}
    r(t) &= \sqrt{(t + r_0 \Bar{\mu}_0 )^2 + r_0^2 (1-\Bar{\mu}_0^2)}, \label{eq:r_of_t}\\
    \Bar{\mu}(t) &= \frac{t + r_0 \Bar{\mu}_0}{r(t)}, \label{eq:mu_of_t}
\end{align}
where $r_0$, $\Bar{\mu}_0$ refer to the initial position and momentum-space cosine, respectively. From these expressions, we can express $r_0$, $\Bar{\mu}_0$ and $\varepsilon_0$ as a function of $r,\Bar{\mu},t$ and express the solution to the homogeneous Boltzmann equation as
\begin{equation}
    \begin{aligned}
        &f(t,r,\bar{\varepsilon},\bar{\vartheta}) = f(0,r_0, \bar{\varepsilon}_0, \bar{\vartheta}_0), &&r_0 = \sqrt{r^2 + t^2 - 2r\bar{\mu} t}, \\
        &\bar{\varepsilon}_0 = \bar{\varepsilon} \, \alpha(r) \, \alpha(r_0)^{-1}, &&\bar{\vartheta}_0 = \mathrm{arccos}\left(\frac{r \cos(\bar{\vartheta}) - t}{r_0} \right).
    \end{aligned} \label{eq:coord_0_conformal}
\end{equation}
When adding a moving fluid to the system, we use Eqs.~\eqref{eq:eps_com} and \eqref{eq:mu_com} to find the momentum space trajectory. The corresponding solutions, along with the ones for $\Bar{\mu}$, $\Bar{\varepsilon}$ and $r$ were used to compute the analytical trajectories for the tests considered in Sec.~\ref{sec:tests}.

We now present the solution to the Boltzmann equation when the source terms are non-vanishing. More precisely, we consider the case with both absorption and emission. The method of characteristics can still be used, giving the same solution for the trajectory in phase space. The difference is that there is an additional equation to be solved coming from the source terms given in Eq.~\eqref{eq:ae}:
\begin{equation}
    \frac{df}{ds} = \hat{\varepsilon}(s) \left( \eta(s) - \kappa^*_a(s) f \right).
\end{equation}
We can rewrite it as a function of $r$ 
\begin{equation}
    \frac{df}{dr} = \left(\frac{dr}{ds}\right)^{-1} \hat{\varepsilon}(r) \left( \eta(r) - \kappa^*_a(r) f \right),
\end{equation}
where $dr/ds$ is given by $\tensor{L}{^r_\hmu} p^{\hmu}$. The general solution to this equation is
\begin{align}
    &f(t(r),r,\hat{\varepsilon}(r),\hat{\mu}(r)) = e^{-g(r)} \left(\int^{r}_{r_0} e^{g(r')}\eta\big[r',\hat{\varepsilon}(r')\big] \, \hat{\varepsilon}(r') \, \left(\frac{dr}{ds}\right)^{-1} \, dr' + f(0,r_0,\hat{\varepsilon}_0,\hat{\mu}_0)  \right), \label{eq:sol_kappa_eta}\\
    &g(r) = \int^{r}_{r_0} \hat{\varepsilon}(r')\,\kappa^*_a\big[r',\hat{\varepsilon}(r')\big] \, \left(\frac{dr}{ds}\right)^{-1} dr' . \label{eq:sol_kappa_eta_exp}
\end{align}
The solution shows that the distribution is increased by the accumulated emission along the considered trajectory, whereas it is decreased by the absorption accumulated along the trajectory.

In the case where $\eta = B \kappa^*_a$, the integral in Eq.~\eqref{eq:sol_kappa_eta} can be computed directly by noting that the integrand is of the form $\frac{dg(r)}{dr} e^{g(r)}$, where $g(r)$ is given in Eq.~\eqref{eq:sol_kappa_eta_exp}. The solution is then given by
\begin{equation}
    f(r) = B\left(1- e^{-g(r)} \right)+ f(0,r_0,\hat{\varepsilon}_0,\hat{\mu}_0) e^{-g(r)} \, . \label{eq:general_sol_eq}\\
\end{equation}
The full solution $f(t,r,\hat{\varepsilon},\mu)$ is obtained by expressing all the occurrences of $r_0$, $\hmu_0$ and $\hat{\varepsilon}_0$ as a function of $t$, $r$, $\hmu$ and $\hat{\varepsilon}$, similarly to Eq.~\eqref{eq:coord_0_conformal}.

Finally, we use the previous results to derive the solutions used for the test in Sec.~\ref{sec:tests_radiating_sphere} and Sec.~\ref{sec:inhomogeneous_radiating_sphere}. For simplicity, we will drop the hat notation for momentum-space variables in the comoving frame. We consider a curved spacetime, emission and absorption coefficients of the form $\eta = B \kappa^*_a$, $\kappa^*_a = K(r)~H(R-r)$, respectively, where $H(x)$ is the Heaviside function and $K(x)$ is an arbitrary function. For the initial condition, we use $f(t=0,r,\varepsilon,\mu) = 0$. We introduce the impact parameter $r_1$ such that $\mu(r_1) = 0$, where $\mu(r)$ is given in Eq.~\eqref{eq:mu_curv}. The time at which the particle reaches $r_1$ is $t_1 \equiv t(r_1)$. We assume that the gravitational field is such that there is only one real positive solution for the equation $\mu(r) = 0$.  We also assume that $t \gg 1$, such that the system reaches an equilibrium and becomes stationary. We introduce the indefinite integral
\begin{equation}
    I(r) = \int^r K(r')~\frac{\alpha(r') h(r')^2 r'}{\sqrt{h(r')^2 r'^2 - h(r_0)^2 r_0^2 (1-\mu^2_0)}} dr', \label{eq:opacity_integral}
\end{equation}
which corresponds to the integrand of $g(r)$ in Eq.~\eqref{eq:sol_kappa_eta_exp}, without the Heaviside function $H(R-r)$ from the opacity and using $\mu(r)$ from Eq.~\eqref{eq:mu_curv}. Because of the change of sign of $\mu$ before and after reaching $r_1$, and because of the Heaviside function, we can split the solution $g(r)$ into multiple cases:
\begin{equation}
    g(r) = \begin{cases}
        -I(r) + I(R) & \textrm{ if } \mu_0 < 0,~ r_1 < r < R < r_0,~ t(r) < t_1 \\
        I(r) + I(R) - 2 I(r_1) & \textrm{ if } \mu_0 < 0,~ r_1 < r < R < r_0,~ t(r) > t_1 \\
        2 I(R) - 2 I(r_1) & \textrm{ if } \mu_0 < 0,~ r_1 < R < r_0,~ R < r,~ t(r) > t_1 \\
        ... & 
    \end{cases}
\end{equation}
where we used the $t \gg 1$ assumption and the limited domain to neglect some cases (for example with $r_0 < R$). We also note that, for $t > t_1$, the integral in $g(r)$ is split as $\int_{0}^{s} \dots ds' = -\int_{r_0}^{r_1} \dots dr' + \int_{r_1}^r \dots dr'$, where the $-$ and $+$ signs come from the sign of $\mu$ in each portion of the trajectory.
The final solution is then found by replacing $r_0$ and $\mu_0$ by their expression as a function of $r$, $\mu$ and $t$. 

Let us first consider the test in Sec.~\ref{sec:tests_radiating_sphere}, \textit{i.e.} flat spacetime ($h(r) =1$, $\alpha(r)=1$) and $K(r) = \kappa$. In this case, we have $t = r\mu - r_0\mu_0$, $r_1 = r_0\sqrt{1-\mu_0^2}$ and $t_1 = -r_0 \mu_0$. We also have that $I(r_1) = 0$. We then find the solution presented in Ref.~\cite{Smit_1997}
\begin{equation}
    g(r,\mu) = \kappa \, 
    \begin{cases}
    r \mu  + R\sqrt{1-\left(\frac{r}{R}\right)^2 (1-\mu^2)} & \textrm{if } r < R \textrm{ and } -1 \leq \mu \leq 1\: ,\\
    2 R\sqrt{1-\left(\frac{r}{R}\right)^2 (1-\mu^2)} & \textrm{if } r \geq R \textrm{ and } \sqrt{1-\left(\frac{R}{r}\right)^2} \leq \mu \leq 1\: ,\\
    0 & \textrm{otherwise.}
    \end{cases} \label{eq:sol_rad_sphere}
\end{equation}

For the profiles considered in Sec.~\ref{sec:inhomogeneous_radiating_sphere} (see Eqs.~\eqref{eq:inhomogeneous_sphere_h}, \eqref{eq:inhomogeneous_sphere_alpha} and \eqref{eq:inhomogeneous_sphere_kappa}), the solution is only semi-analytical. Because $K(r) \propto \varepsilon(r)^{n_e}$, after expressing $\varepsilon_0$ and $r_0$ as a function of $\varepsilon$ and $r$ the integral $I(r)$ from Eq.~\eqref{eq:opacity_integral} becomes
\begin{equation}
    I(r,\varepsilon) = \left[\varepsilon \alpha(r)\right]^{n_e} ~ \int^r ~\frac{\alpha(r')^{1-n_e} \left[1-\left(\frac{r'}{R}\right)^{2}\right]^n h(r')^2 r'}{\sqrt{h(r')^2 r'^2 - h(r)^2 r^2 (1-\mu^2)}} dr'
\end{equation}
where we used $\varepsilon \alpha(r) = \varepsilon_0 \alpha(r_0)$ and $h(r)^2 r^2 (1-\mu^2) = h(r_0)^2 r_0^2 (1-\mu^2_0)$ (from Eq.~\eqref{eq:mu_curv}). We also still find that $I(r_1,\varepsilon) = 0$. We also need to use the more general solution from Eq.~\eqref{eq:sol_kappa_eta}, because we use $\eta(r,\varepsilon) = f_\mathrm{eq}(\varepsilon) \kappa_a^*(r,\varepsilon)$, where $f_\mathrm{eq}$ is the Fermi-Dirac distribution function in Eq.~\eqref{eq:f_eq} (with $T=10\,$MeV and $\mu_\nu =0$). Integrating by parts and splitting the integrals similarly to the previous case, we find
\begin{equation}
    f(r,\varepsilon,\mu) =
    \begin{cases}
    \begin{aligned}
        f_{\mathrm{eq},x}(r) - f_{\mathrm{eq},x}(R)  e^{I(r,\varepsilon) - I(R,\varepsilon)} + \int_{r}^R e^{I(r,\varepsilon) - I(r',\varepsilon)} \frac{df_{\mathrm{eq},x}}{dr'} dr' 
        \end{aligned} & \textrm{if } r < R \textrm{ and } -1 \leq \mu < 0 \: ,\\[2ex]
    \begin{aligned}
       f_{\mathrm{eq},x}(r) &- f_{\mathrm{eq},x}(R)  e^{-I(r,\varepsilon) - I(R,\varepsilon)} - \int_{r_1}^r e^{I(r',\varepsilon) - I(r,\varepsilon)} \frac{df_{\mathrm{eq},x}}{dr'} dr' \\
       & + \int_{r_1}^R e^{-I(r',\varepsilon) - I(r,\varepsilon)} \frac{df_{\mathrm{eq},x}}{dr'} dr'
       \end{aligned}  & \textrm{if } r < R \textrm{ and } 0 \leq \mu \leq 1 \: ,\\[2ex]
    \begin{aligned}
       f_{\mathrm{eq},x}(R) &( 1 -  e^{-2 I(R,\varepsilon)}) - \int_{r_1}^R e^{I(r',\varepsilon) - I(R,\varepsilon)} \frac{df_{\mathrm{eq},x}}{dr'} dr' \\
       & + \int_{r_1}^R e^{-I(r',\varepsilon) - I(R,\varepsilon)} \frac{df_{\mathrm{eq},x}}{dr'} dr'
       \end{aligned} & \textrm{if } r \geq R \textrm{ and } \sqrt{1-\left(\frac{R h(R)}{r h(r)}\right)^2} \leq \mu \leq 1\: ,\\
    0 & \textrm{otherwise,}
    \end{cases} \label{eq:sol_inhomogeneous_rad_sphere}
\end{equation}
where we used $f_{\mathrm{eq},x}(r') \equiv f_{\mathrm{eq}}(\varepsilon \alpha(r)/\alpha(r'))$ (we omit the $r$ and $\varepsilon$ dependence to ease the notation).
We stress that the condition on $\mu$ for $r \geq R$ is only valid for the $h(r)$ profile we consider. We do not write the full expression of $h(r)$ to ease the notation. For these specific $h(r)$ and $\alpha(r)$ profiles, one can express the integral in $I(r,\varepsilon)$ as a function of elliptic integral functions. The other integrals must be computed numerically.

\begin{acronym}
    \acro{GRMHD}[GRMHD]{general-relativistic magneto-hydrodynamics}
    \acro{CCSN}[CCSN]{core-collapse supernova}
    \acroplural{CCSN}[CCSNe]{core-collapse supernovae}
    \acro{AMR}[AMR]{adaptive mesh refinement}
    \acro{FV}[FV]{finite volume}
    \acro{FD}[FD]{finite difference}
    \acro{CFL}[CFL]{Courant-Friedrichs-Lewy}
    \acro{MC}[MC]{Monte Carlo}
    \acro{MSMG}[MSMG]{Multi-Species Multi-Group}
    \acro{SSMG}[SSMG]{Single-Species Multi-Group}
    \acro{SSSG}[SSSG]{Single-Species Single-Group}
    \acro{NR}[NR]{Newton-Raphson}
\end{acronym}

\twocolumngrid

\bibliography{bibliography}{}

\end{document}